\documentclass[3p,times]{elsarticle}
\usepackage{amssymb}

\usepackage{caption}
\usepackage{subfigure}  
\usepackage{amsmath}
\usepackage{bm}
\usepackage{color}
\usepackage[colorlinks=True]{hyperref}
\begin{document}
\begin{frontmatter}
\title{Is a direct numerical simulation (DNS) of Navier-Stokes equations with small enough grid spacing and time-step  definitely reliable/correct?}

\author[label2]{Shijie Qin}
\author[label2]{Yu Yang}
\author[label3]{Yongxiang Huang}
\author[label4]{Xinyu Mei}
\author[label4]{Lipo Wang}
\author[label1,label2]{Shijun Liao \corref{cor1}}

\cortext[cor1]{sjliao@sjtu.edu.cn}

\address[label1]{State Key Laboratory of Ocean Engineering, Shanghai 200240, China}
\address[label2]{School of  Ocean and Civil Engineering, Shanghai Jiao Tong University, Shanghai 200240, China} 
\address[label3]{State Key Laboratory of Marine Environmental Science,  College of Ocean and Earth Sciences, Xiamen University, Xiamen 361102, China}
\address[label4]{UM-SJTU Joint Institute, Shanghai Jiaotong University, Shanghai 200240, China}

\begin{abstract}
Turbulence  is strongly associated with the vast majority of fluid flows in nature and industry.  Traditionally, results given by the direct numerical simulation (DNS) of Navier-Stokes (NS) equations that relate to a famous millennium problem  are widely regarded as `reliable' benchmark solutions of turbulence, as long as grid spacing is fine enough (i.e. less than the minimum Kolmogorov scale) and time-step is small enough, say, satisfying the Courant-Friedrichs-Lewy condition (Courant number $<$ 1).   Is this really true?        
In this paper a two-dimensional  sustained turbulent Kolmogorov flow  driven by an external body force governed by the NS equations under an initial condition with a spatial symmetry  is investigated numerically by the two numerical methods with detailed comparisons:  one is the traditional DNS, the other is the `clean numerical simulation' (CNS).   In theory, the exact solution must have a kind of spatial symmetry since its initial condition is spatially symmetric.   However, it is found that numerical noises of the DNS are quickly enlarged to the same level as the `true' physical solution, which finally destroy the spatial symmetry of the flow field.   In other words, the DNS of the turbulent Kolmogorov flow governed by the NS equations are badly polluted mostly.  On the contrary, the numerical noise of the CNS is much smaller than the `true' physical solution of turbulence in a long enough interval of time so that the CNS result is very close to the `true' physical solution and thus  can remain symmetric, which can be used as a benchmark solution for comparison.   Besides,     
it is found that numerical noise as a kind of artificial tiny disturbances can lead to huge deviations at large scale on the two-dimensional Kolmogorov turbulence governed by the NS equations, not only quantitatively (even in statistics) but also qualitatively (such as spatial symmetry of flow).  
 This highly suggests that  fine enough spatial grid spacing with small enough time-step alone could not guarantee the validity of the DNS of the NS equations: it is only a necessary condition but not sufficient.  All of these findings might challenge some of our general beliefs in turbulence: for example, it might be wrong in physics to neglect the influences of small disturbances to NS equations.  
    Our results suggest that,  from physical point of view, it should be better to use the Landau-Lifshitz-Navier-Stokes (LLNS) equations, which consider the influence of unavoidable thermal fluctuations, instead of the NS equations, to model turbulent flows.
\end{abstract}	


\end{frontmatter}

\section{Introduction}

Turbulence  is strongly associated with the vast majority of fluid flows in nature and industry \cite{frisch1995turbulence, PeterSmith1998Explaining, bohr2005dynamical}. Traditionally, results given by the direct numerical simulation (DNS) of Navier-Stokes equations under various initial/boundary conditions are widely regarded as `reliable' benchmark solutions of turbulence, as long as spatial grid spacing is fine enough (i.e. less than the minimum Kolmogorov scale) \cite{pope2001turbulent} and time-step is small enough, say, satisfying the Courant-Friedrichs-Lewy condition (Courant number $<$ 1)  \cite{Courant1928}.   Is this definitely true?    

It is widely believed that turbulence flow, as a typical spatio-temporal chaotic system \cite{frisch1995turbulence, pope2001turbulent, bohr2005dynamical}, has the famous  `butterfly effect' \cite{lorenz1963deterministic, GRChen2014}: microscopic disturbances in turbulent flows, caused by either thermal fluctuations or environmental noises that are small-scale but unavoidable in practice, might increase exponentially (and quickly)  to macroscopic levels \cite{leith1972predictability, boffetta2001predictability, boffetta2017chaos, li2020superfast}.
Recently, Qin and Liao \cite{qin_liao_2022} provided a rigorous evidence that numerical noise as a kind of tiny artificial stochastic disturbances can indeed increase quickly to the same order of magnitude of the `true' physical solution of a two-dimensional sustained turbulent Rayleigh-B{\' e}nard convection, and in addition can have quantitative and qualitative large-scale influences in statistics, and even might lead to different types of flow.  

Note that numerical noises, i.e. truncation error and round-off error, always exist and are practically unavoidable in each  result given by the `direct numerical simulation' (DNS).    
Unfortunately, as mentioned above, due to the chaotic property of turbulence  \cite{platt1991investigation,feudel1995bifurcations},  the tiny numerical noises as a kind of micro-level disturbances increase exponentially (and thus quickly) to a macroscopic level, therefore each DNS result of turbulent flow is a kind of mixture of `true' physical solution (denoted by $p$) and `false' numerical noise (denoted by $\delta$) that are mostly at the {\em same} order of magnitude, i.e. $p \sim \delta$,  as currently illustrated by Qin and Liao \cite{qin_liao_2022}  using a two-dimensional Rayleigh-B{\'e}nard turbulent flow as an example by means of the so-called `clean numerical simulation' (CNS) \cite{Liao2009, Liao2023book, lin2017origin, hu2020risks, qin2020influence, qin_liao_2022}.  For investigations on turbulence,  statistic properties are more important than spatio-temporal trajectory of numerical simulation.  It is a pity that, traditionally,  the `true' solution $p$  is  unknown so that  one had to use the the badly polluted simulation, i.e. the above-mentioned mixture $p+\delta$, to gain statistic results $ \langle p + \delta \rangle$, where $\langle \; \rangle$ denotes an operator of statistics.  Unfortunately, this is based on such a `hypothesis' that    $\langle p + \delta \rangle = \langle p \rangle$, say, the statistics is {\em stable} under  disturbances, where $p$ denotes the `true' physical solution.  But, as illustrated by Liao \cite{Liao2023book}, this hypothesis is {\em not} definitely true for  many chaotic systems \cite{hu2020risks, qin2020influence} and even some turbulent flows: the DNS results of a kind of two-dimensional Rayleigh-B{\'e}nard turbulent flow are quite different from these given by CNS   not only in statistics but also even in type of flow  \cite{qin_liao_2022}.    

Here,  let us briefly describe the basic idea of the CNS.  
In order to obtain a `convergent' (or reproducible) numerical simulation (i.e. a result very close to `true' solution $p$) of chaotic dynamical systems \cite{lorenz1963deterministic,  PeterSmith1998Explaining, sprott2010, lee2014wind, gao2018flow}, Liao \cite{Liao2009} proposed a new numerical strategy, namely the `clean numerical simulation' (CNS) \cite{ Liao2009, Liao2013, Liao2014, Liao2023book, lin2017origin, hu2020risks, qin2020influence,  qin_liao_2022}, to reduce the background numerical noise (i.e. the truncation error and round-off error) so greatly that numerical noise is negligible (say, the `false' numerical noise $\delta$ is much smaller than the `true' physical solution $p$, i.e. $|\delta| \ll |p|$) during a {\em finite} but long enough interval of time $t\in[0,T_c]$, where $T_c$ is the so-called `critical predictable time'.  
In the frame of the CNS, temporal and spatial truncation errors can be decreased to a required small level by means of the Taylor expansion with a high enough order in the temporal dimension and a fine enough discretization in the spatial dimension (such as the high-order spatial Fourier expansion), respectively.
Besides, all of the physical/numerical variables and parameters must be  in the multiple precision (MP) \cite{oyanarte1990mp} with a large enough number of significant digits so that the round-off error can be decreased to a required small level.
Moreover, an additional numerical simulation with the even smaller numerical noise of the same chaotic system is required to determine the value of $T_c$ via a comparison with the previous CNS result so that the corresponding numerical noise of the simulation could be negligible and thus this computer-generated simulation is convergent/reproducible within the whole spatial domain in the temporal interval $[0,T_c]$. In this way, unlike other traditional numerical algorithms using double precision, the CNS can give reliable numerical simulation of a chaotic system in a {\em finite} but long {\em enough} interval of time, say, the CNS result is so close to the `true' physical solution $p$ that one can gain its statistics $\langle p \rangle$ accurately.  
Here, it should be emphasized that the so-called  `critical predictable time'  denoted by $T_{c}$ is a very important concept in the frame of the CNS: unlike DNS results whose interval of time can be as long as one would like {\em without} considering the influences of the exponential increase of numerical noises, CNS can give a reliable result only in a {\em finite} but long {\em enough} interval of time, i.e. $t\in[0,T_{c}]$.     
So, for the first time, the CNS provides us a tool to carefully check the hypothesis $\langle p + \delta \rangle = \langle p \rangle$ for chaotic systems and turbulent flows, where  $p+\delta$ on the left hand side denotes numerical simulations given by a traditional numerical method using double precision (such as DNS), and $p$ on the right-hand side can be accurately gained by means of the CNS, respectively.    

Up to now, the CNS has been successfully applied to many chaotic dynamical systems, such as Lorenz system \cite{Liao2009, LIAO2014On}, the famous three-body problem \cite{Liao2014, liao2015inherent,  Li2017More, li2018over, Liao2022NA}, H\'{e}non-Heiles chaotic system \cite{Liao2013}, hyperchaotic R\"{o}ssler system \cite{AAMM-14-799, AAMM-15-1191}, a chaotic free-fall disk \cite{xu2021accurate},  Arnold-Beltrami-Childress (ABC) flow \cite{qin_liao_2023}, the complex Ginzburg-Landau equation (CGLE) \cite{hu2020risks}, the damped driven sine-Gordon equation (SGE) \cite{qin2020influence},  the two-dimensional turbulent Rayleigh-B{\'e}nard convection \cite{qin_liao_2022}, and so on.
It was found that the hypothesis $\langle p + \delta \rangle = \langle p \rangle$ indeed holds for {\em many} chaotic systems and turbulent flows, say, their statistics are stable, which are called `normal-chaos'.  But, it was found that the hypothesis   $\langle p + \delta \rangle = \langle p \rangle$ does {\em not} hold for {\em some} chaotic systems and turbulent flows, say, their statistics are {\em unstable} to small disturbances, which are called `ultra-chaos' \cite{AAMM-14-799, qin_liao_2023, YANG2023113037, ZHANG2023133886}.   The ultra-chaos as a brand-new concept  might greatly deepen and enrich our understandings about chaos as well as turbulence.

In this paper, let us further consider a two-dimensional turburlent flow of viscous fluid, driven by an external body forcing that is stationary, monochromatic, and sinusoidally varying in space. This kind of flow was first introduced by Kolmogorov in 1959 \cite{arnold1960seminar}, 
called  `Kolmogorov flow' and usually regarded as a mathematically and experimentally tractable flow model whose instability \cite{meshalkin1961investigation, beaumont1981stability, thess1992instabilities, balmforth2002stratified}, global stability \cite{marchioro1986example}, chaotic property \cite{platt1991investigation, feudel1995bifurcations}, and turbulent state \cite{green1974two, sivashinsky1985weak, chandler2013invariant} have been investigated in details.  The Kolmogorov flow can be practically realized in the laboratory by means of the magnetohydrodynamics (MHD) \cite{bondarenko1979laboratory, obukhov1983kolmogorov, sommeria1986experimental} or in a soap film \cite{burgess1999instability}. Furthermore, due to some additional physical factors such as the Coriolis term (i.e. the effect of the planetary rotation) \cite{lorenz1972barotropic, kazantsev1998unstable}, the bottom friction \cite{manfroi1999slow, tsang2008energy, tsang2009forced}, and the stratification \cite{balmforth2002stratified, balmforth2005stratified}, Kolmogorov flows have also been widely studied in geophysical fluid dynamics.

In this investigation, we compare the CNS result of the two-dimensional Kolmogorov turbulence with that given by a traditional DNS using double precision.  In this way, we can check the hypothesis $\langle p + \delta \rangle = \langle p \rangle$ in details, where the left-hand side is given by the DNS, and the right-hand side is given by the CNS, respectively.   The mathematical model is described in \S~2, the detailed comparisons between CNS and DNS result are given in \S~3, and the discussions and conclusions are given in \S~4, respectively.    As reported in \S~3,  numerical noises as  a kind of small-scale disturbances have huge influences on large-scale properties of the two-dimensional  Kolmogorov turbulence not only in spatio-temporal trajectories and its spatial symmetry but also in various statistics.   
Especially, it is found that the hypothesis $\langle p + \delta \rangle = \langle p \rangle$ does {\em not} stand up for the two-dimensional Kolmogorov turbulence under consideration,  say, the fine enough spatial grid spacing with small enough time-step  {\em cannot} guarantee the validity of the DNS: it is only a necessary condition but {\em not} sufficient.     

\section{Mathematical model and numerical strategy}

\subsection{Governing equation and initial/boundary conditions \label{geq-initial}}

Consider an incompressible flow in a domain $[0,L]\times[0,L]$ under the so-called `Kolmogorov forcing', which is stationary, monochromatic, and sinusoidally varying in space, with an integer $n_K$ describing the forcing scale and $\chi$ representing the corresponding forcing amplitude per unit mass of fluid \cite{chandler2013invariant}.
Using the length scale $L/2\pi$ and the time scale $\sqrt{L/2\pi\chi}$, the non-dimensional governing equation of this two-dimensional Kolmogorov flow in the form of stream function reads
\begin{align}
& \frac{\partial}{\partial t} \left( \nabla^{2}\psi\right)+\frac{\partial(\psi,\nabla^{2}\psi)}{\partial(x,y)}-\frac{1}{Re}\nabla^{4}\psi+n_K\cos(n_Ky)=0,       \label{eq_psi}
\end{align}
where $x$ and $y$ are horizontal and vertical coordinates with $x,y\in[0,2\pi]$, $t$ denotes the time, the stream function $\psi$ is defined by 
\begin{align}
& u=-\frac{\partial\psi}{\partial y}, \hspace{1.0cm} v=\frac{\partial\psi}{\partial x},       \label{psi}
\end{align} 
 $u$ and $v$ represent horizontal and vertical velocities,  $Re$ is the Reynolds number  defined by
\begin{align}
& Re=\frac{\sqrt{\chi}}{\nu}\left(\frac{L}{2\pi}\right)^{\frac{3}{2}},       \label{Re}
\end{align}
$\nu$ denotes the kinematic viscosity,
$\nabla^{2}$ is the Laplace operator with the definition $\nabla^{4}=\nabla^{2}\nabla^{2}$, and 
\begin{align}
& \frac{\partial(a,b)}{\partial(x,y)}=\frac{\partial a}{\partial x}\frac{\partial b}{\partial y}-\frac{\partial b}{\partial x}\frac{\partial a}{\partial y}       \label{Jacobi}
\end{align}
is the Jacobi operator, respectively.  Note that the stream-function $\psi$ satisfies the periodic condition on the boundary
\begin{equation}
\psi (0,y,t) = \psi (2\pi,y,t), \;\;\; \psi(x,0,t) = \psi(x,2\pi,t).  \label{psi-bc-periodic}
\end{equation}

Following Chandler and Kerswell \cite{chandler2013invariant}, we choose the physical parameters in  (\ref{eq_psi}) as follows:
\begin{align}
& n_K=4, \hspace{1.0cm} Re=40,       \label{parameter_values}
\end{align}
with the initial condition
\begin{align}
& \psi(x,y,0)=-\frac{1}{2}\Big\{ \cos(x+y)+\cos(x-y)\Big\},       \label{initial_condition}
\end{align}
corresponding to a sustained turbulent Kolmogorov flow.

Note that there exists a kind of spatial symmetry in the initial condition (\ref{initial_condition}).    According to the governing equation (\ref{eq_psi}), the periodic boundary condition (\ref{psi-bc-periodic}) and the initial condition   (\ref{initial_condition}) with the spatial symmetry,  it is straight forward that the solution of the corresponding NS equations should be in the form of a series
\begin{equation}
\psi(x,y,t) = \sum_{m=0}^{+\infty}\sum_{n=1}^{+\infty}\left\{a_{m,n}(t)\cos(m x + ny) + b_{m,n}(t)\cos(m x - ny)\right\}, \label{psi:series-expression}
\end{equation}    
where $a_{m,n}(t)$ and $b_{m,n}$ are unknown coefficients dependent upon the time $t$.  Therefore,  the exact solution of the  Navier-Stokes equations  have the  {\em spatial symmetry} 
\begin{equation}
\psi(x,y,t) = \psi(-x,-y,t) = \psi(2\pi-x,2\pi-y,t),  \label{symmetry}
\end{equation} 
although the coefficients $a_{m,n}(t)$ and $b_{m,n}$ in (\ref{psi:series-expression}) are unknown.  
Thus, the loss of such kind of spatial symmetry clearly indicates the large deviation from the true solution of the Navier-Stokes equations.  This is exactly the reason why we choose the periodic boundary condition (\ref{psi-bc-periodic}) and the initial condition  (\ref{initial_condition}) with the spatial symmetry!     

\subsection{Strategy of CNS \label{CNS-strategy}}
First of all, let us simply describe the strategy of CNS.   The CNS is used to greatly decrease the background numerical noise, i.e. the truncation and round-off errors, to such a required tiny level that the corresponding `false' numerical noise of the simulation is much smaller than, and thus negligible compared with, the `true' physical solution of the considered Kolmogorov flow  in an interval of time that is long enough for statistics. In this way, a convergent (reproducible) and reliable numerical result of the two-dimensional Kolmogorov turbulence can be obtained, which is used here as the `clean' and `true' benchmark solution. 
Briefly speaking, to decrease the spatial truncation error of the simulation to a small enough level, we discretize the spatial domain of the flow field by a uniform mesh $N_x \times N_y = 256 \times 256$, and adopt the Fourier spectral method for spatial approximation with the $3/2$ rule for dealiasing.  The corresponding spatial resolution is high enough for the Kolmogorov flow under consideration: the grid spacing is less than the minimum instantaneous Kolmogorov scale \cite{pope2001turbulent}, as shown later in \S~3.5.  Besides, to decrease the temporal truncation error to a required small enough  level,  we use the 60th-order (i.e. $M = 60$) Taylor expansion with a time-step $\Delta t=1\times10^{-3}$.  Furthermore, in order to decrease the round-off error to a required small enough level, we use multiple precision with 200 significant digits (i.e. $N_s = 200$) for {\em all} physical/numerical variables and parameters. In addition, the self-adaptive CNS strategy \cite{AAMM-15-1191} is adopted  to dramatically reduce the calculation amount.  
Similarly, we run another CNS using the {\em same} Fourier spectral method with the {\em same} uniform spatial mesh but having the even smaller temporal truncation error and round-off error by means of a higher-order (i.e. $M = 62$) Taylor expansion with the same time-step ($\Delta t=1\times10^{-3}$) and the higher multiple precision with more significant digits (i.e. $N_s = 205$).
Comparing these two CNS results by means of the so-called  `spectrum-deviation' $\delta_s(t)$ whose definition is given in \S~2.3, it is found that $\delta_s < 10^{-19}$ is satisfied throughout the whole simulation $0\leq t \leq 1500$, say, these two CNS results have no distinct difference in an interval of time $0\leq t \leq 1500$ that is long enough for calculating statistics. This verifies the convergence and reliability of our CNS result in $t \in [0, 1500]$ given by means of $M = 60$, $\Delta t=10^{-3}$ and $N_s = 200$, which can be regarded as a `clean' benchmark result, since it is very close to the `true' physical solution of the two-dimensional Kolmogorov turbulence under consideration.
 For details about the CNS algorithm of the Kolmogorov turbulence, please refer to Appendix~A.

On the other hand, in the case of the {\em same} initial condition (\ref{initial_condition}) and the {\em same}  physical parameters (\ref{parameter_values}), the same governing equation (\ref{eq_psi}) is solved numerically in $t \in [0, 1500]$ by a traditional DNS algorithm, i.e. the fourth-order Runge-Kutta's method with the double precision using the time-step $\Delta t=10^{-4}$, whose numerical noise increases exponentially (and quickly) up to the {\em same} order of magnitude of the `true' physical solution and thus can not be negligible thereafter.  Therefore, the corresponding DNS result becomes a mixture of the `true' physical solution and the `false' numerical noise, which are mostly at the same order of magnitude.   Then,  comparing this DNS result with the CNS benchmark solution, we can investigate the influence of numerical noises  as tiny artificial stochastic disturbances on the two-dimensional Kolmogorov turbulence in details.  

For both of DNS and CNS,  the  grid spacing is fine enough, i.e. satisfying the spacing criterion  \cite{pope2001turbulent},   and the  time-steps are small enough, i.e. satisfying the Courant-Friedrichs-Lewy condition \cite{Courant1928}, as shown in \S~3.5 in details.   

\subsection{Some measures of the flow}    \label{Key_measures}

For the sake of simplicity, the definitions of some statistic operators are briefly described below.   The spatial average is defined by
\begin{align}
& \langle\,\,\rangle_A=\frac{1}{4\pi^2}\int^{2\pi}_0\int^{2\pi}_0 dxdy       \label{average_A}
\end{align}
and the spatiotemporal average (along the $x$ direction) is defined by
\begin{align}
& \langle\,\,\rangle_{x,t}=\frac{1}{2\pi (T_2-T_1)}\int^{2\pi}_0\int^{T_2}_{T_1} dxdt,       \label{average_xt}
\end{align}
respectively, where $T_1=200$ and $T_2=1500$ are chosen in \S~\ref{Influence} corresponding to a long enough temporal interval, as well as a relatively stable state of the flow. Besides, the spatiotemporal average (over the whole field) is defined by
\begin{align}
& \langle\,\,\rangle=\frac{1}{4\pi^2 (T_2-T_1)}\int^{2\pi}_0\int^{2\pi}_0\int^{T_2}_{T_1} dx dy dt,       \label{average_all}
\end{align}
where different values of $T_1$ and $T_2$ correspond to different integral intervals of time.

For the two-dimensional Kolmogorov turbulence under consideration, vorticity is expanded as the Fourier series
\begin{align}
& \omega(x,y,t)=\nabla^{2}\psi(x,y,t)=\sum^{\lfloor N_x/3 \rfloor}_{\,m=-\lfloor N_x/3 \rfloor}\sum^{\lfloor N_y/3 \rfloor}_{\,n=-\lfloor N_y/3 \rfloor}\Omega_{m,n}(t) \exp(\mathbf{i}\,mx)\exp(\mathbf{i}\,ny),       \label{vorticity}
\end{align}
where $m$, $n$ are integers, $\lfloor\,\,\rfloor$ stands for a floor function, $\mathbf{i}=\sqrt{-1}$ denotes the imaginary unit, and for dealiasing $\Omega_{m,n}=0$ is imposed for wavenumbers outside the above domain $\sum$, respectively.  Note that, when $\omega$ is a real number, $\Omega_{-m,-n}=\Omega^*_{m,n}$ must be satisfied, where $\Omega^*_{m,n}$ is the conjugate of $\Omega_{m,n}$. The enstrophy spectrum is defined by
\begin{align}
& B_l(t)=\sum_{l-1/2 \leq \sqrt{m^2+n^2} < l+1/2}\left| \Omega_{m,n}(t) \right|^2,       \label{enstrophy_spectrum}
\end{align}
where the wave number $l$ is a non-negative integer.  In addition, for two different numerical results that give different enstrophy spectrum  $B_l(t)$ and $B'_l(t)$, respectively, where $B'_l(t)$ corresponds to the simulation with smaller numerical noise,  we define the so-called `spectrum-deviation'
\begin{align}
& \delta_s(t) = \frac{\sum\limits_{l=0}\Big|B_l(t)-B'_l(t)\Big|}{\sum\limits_{l=0}B_l'(t)}    \label{delta_s}
\end{align}
to measure the deviation of the simulation corresponding to $B_l(t)$,  from the other corresponding to $B'_l(t)$,  at a given time $t$.

In \S~3, we will focus on the kinetic energy
\begin{align}
& E(x,y,t) = \frac{1}{2}[u^2(x,y,t)+v^2(x,y,t)]    \label{kinetic_energy}
\end{align}
and
the kinetic energy dissipation rate
\begin{align}
& D(x,y,t)=\frac{1}{2Re}\sum_{ij}\big [ \partial_iu_j(x,y,t)+\partial_ju_i(x,y,t) \big ]^2,    \label{dissipation_rate}
\end{align}
where $i,j=1,2$, $u_1=u$, $u_2=v$, $\partial_1=\partial /\partial x$, and $\partial_2=\partial /\partial y$, respectively.

\section{Influence of numerical noises as tiny disturbances}    \label{Influence}

As mentioned in \S~\ref{CNS-strategy}, comparing the DNS result, which is a mixture of the `true' physical solution and the `false' numerical noise that are mostly at the {\em same} order of magnitude, with the CNS benchmark solution, whose `false' numerical noises are {\em negligible} and thus which is very close to its `true' physical solution in a finite but long enough interval of time, we can investigate in details the influence of numerical noise as a kind of tiny artificial stochastic disturbances on the two-dimensional Kolmogorov turbulence under consideration.  Some results of detailed comparisons are shown in this section. 

\subsection{Trajectory, spatial symmetry  and statistics}

\begin{figure}
    \begin{center}
        \begin{tabular}{cc}
             \includegraphics[width=2.2in]{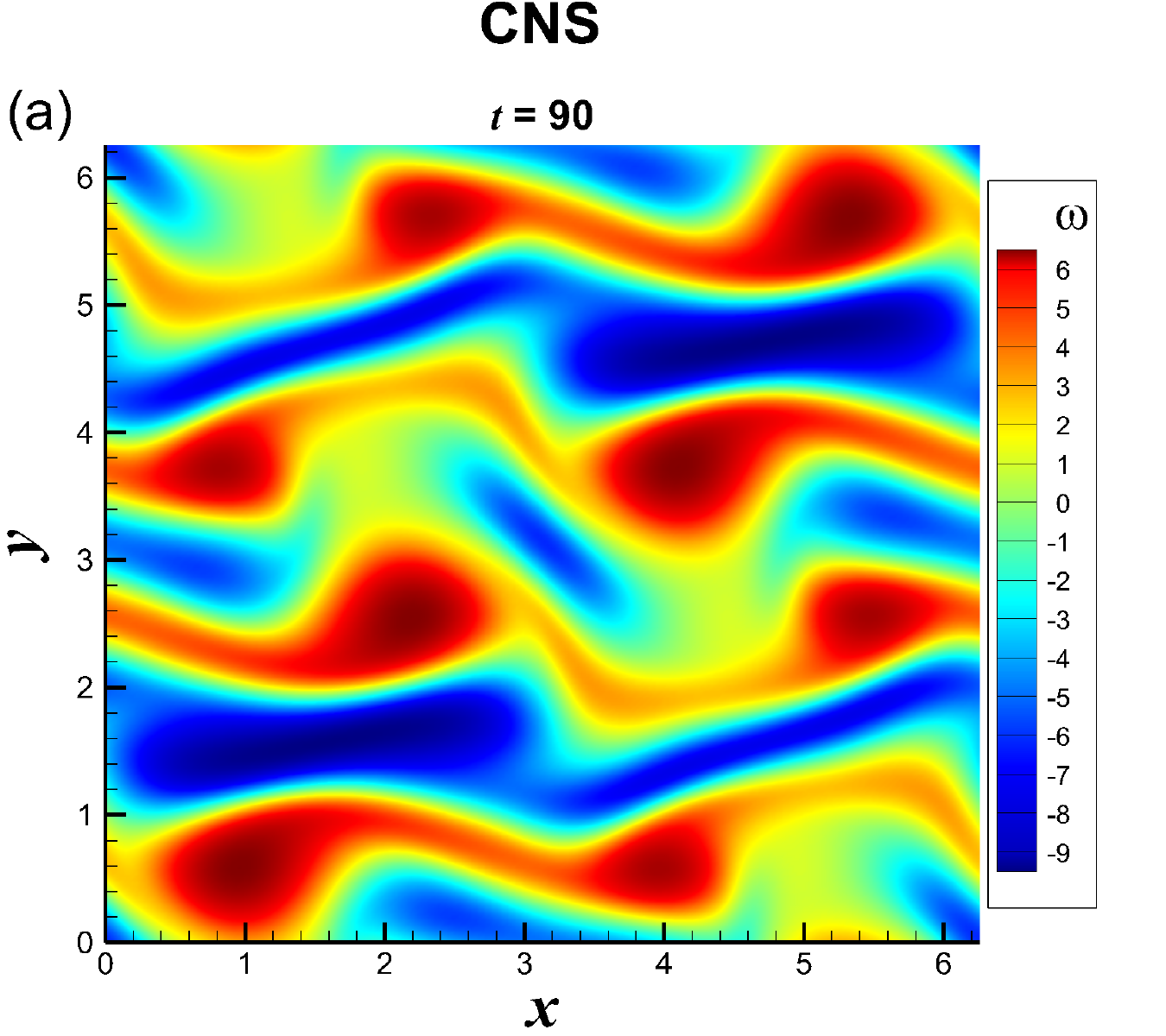}
             \includegraphics[width=2.2in]{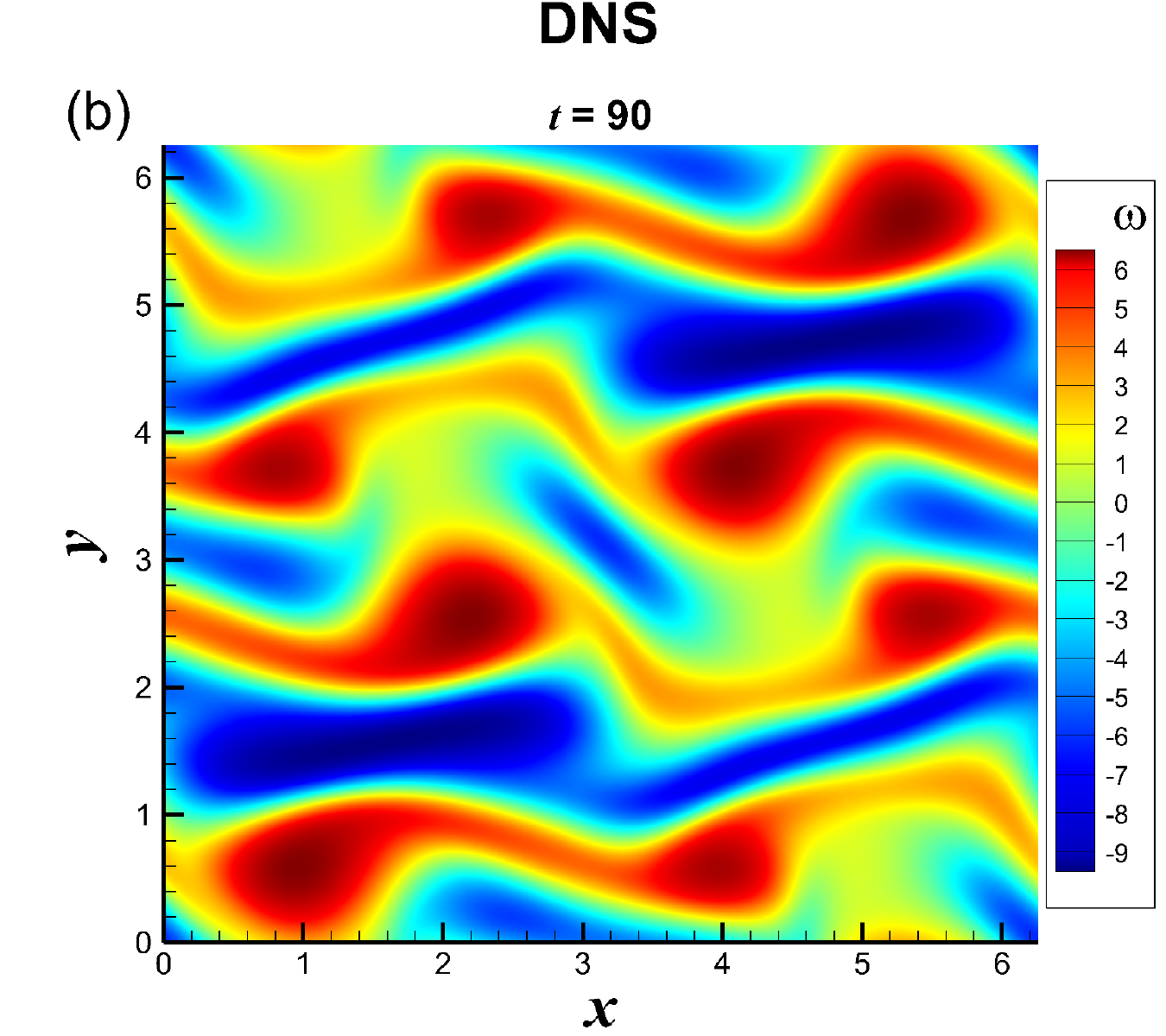}    \\
             \includegraphics[width=2.2in]{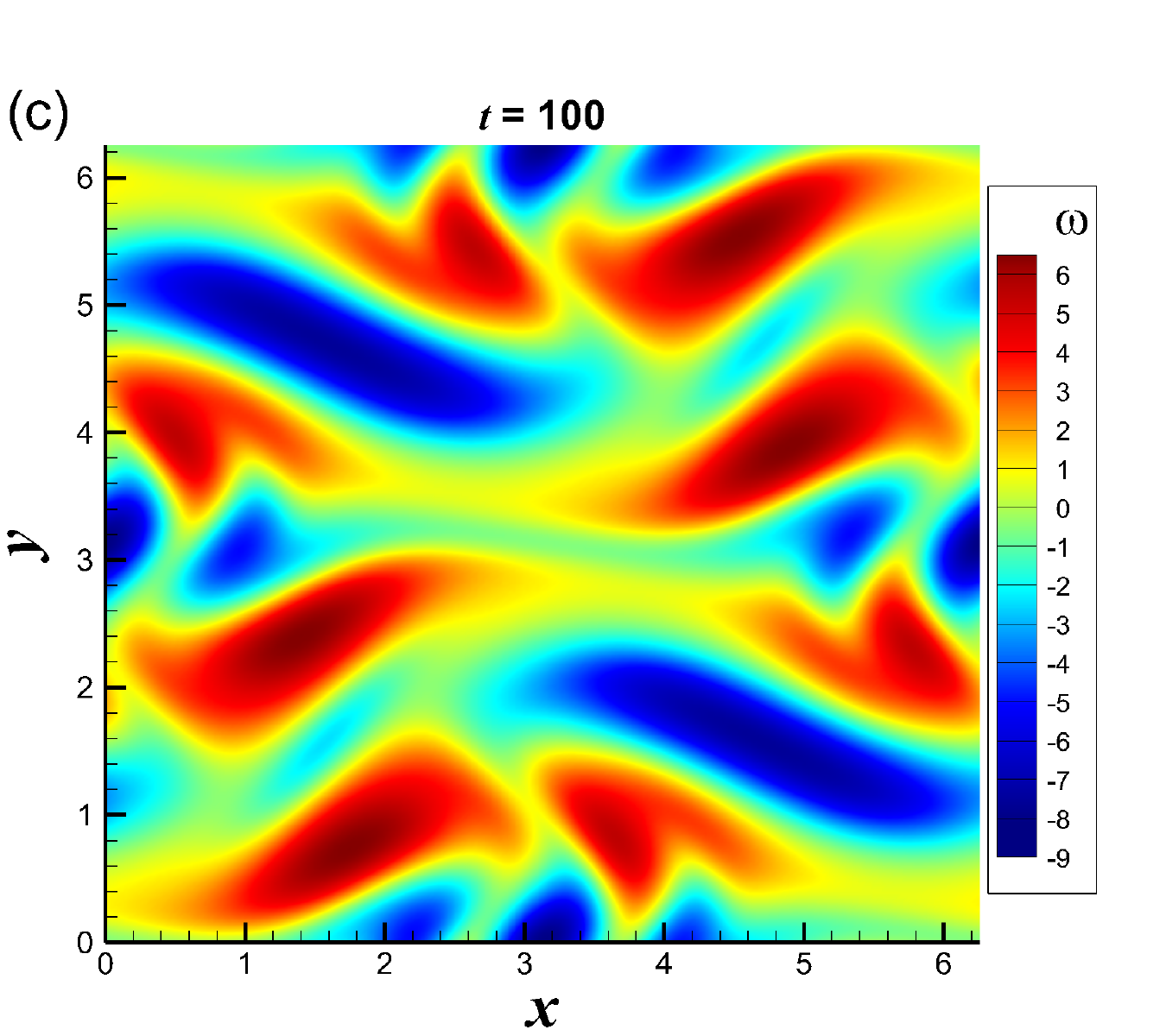}
             \includegraphics[width=2.2in]{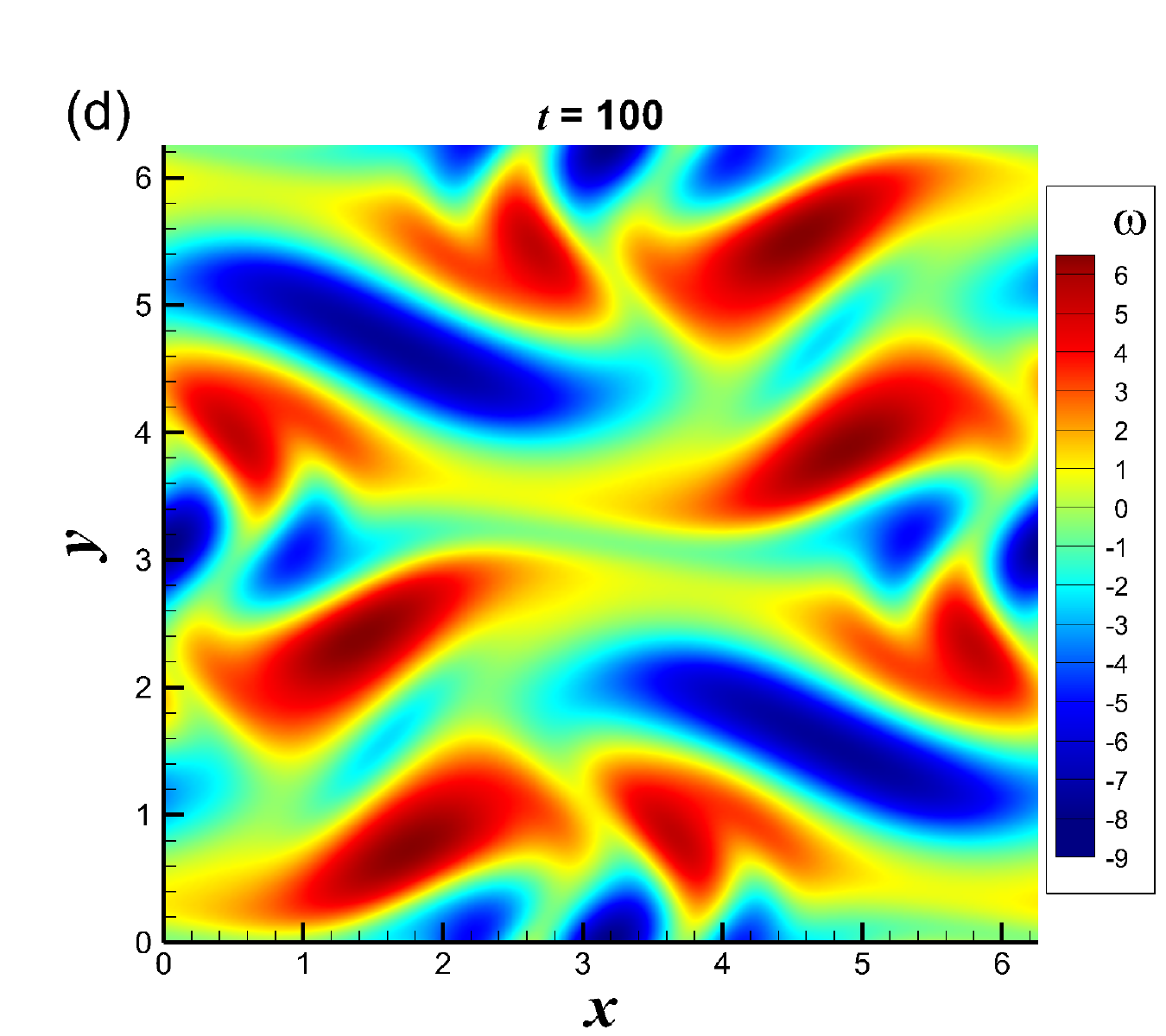}    \\
             \includegraphics[width=2.2in]{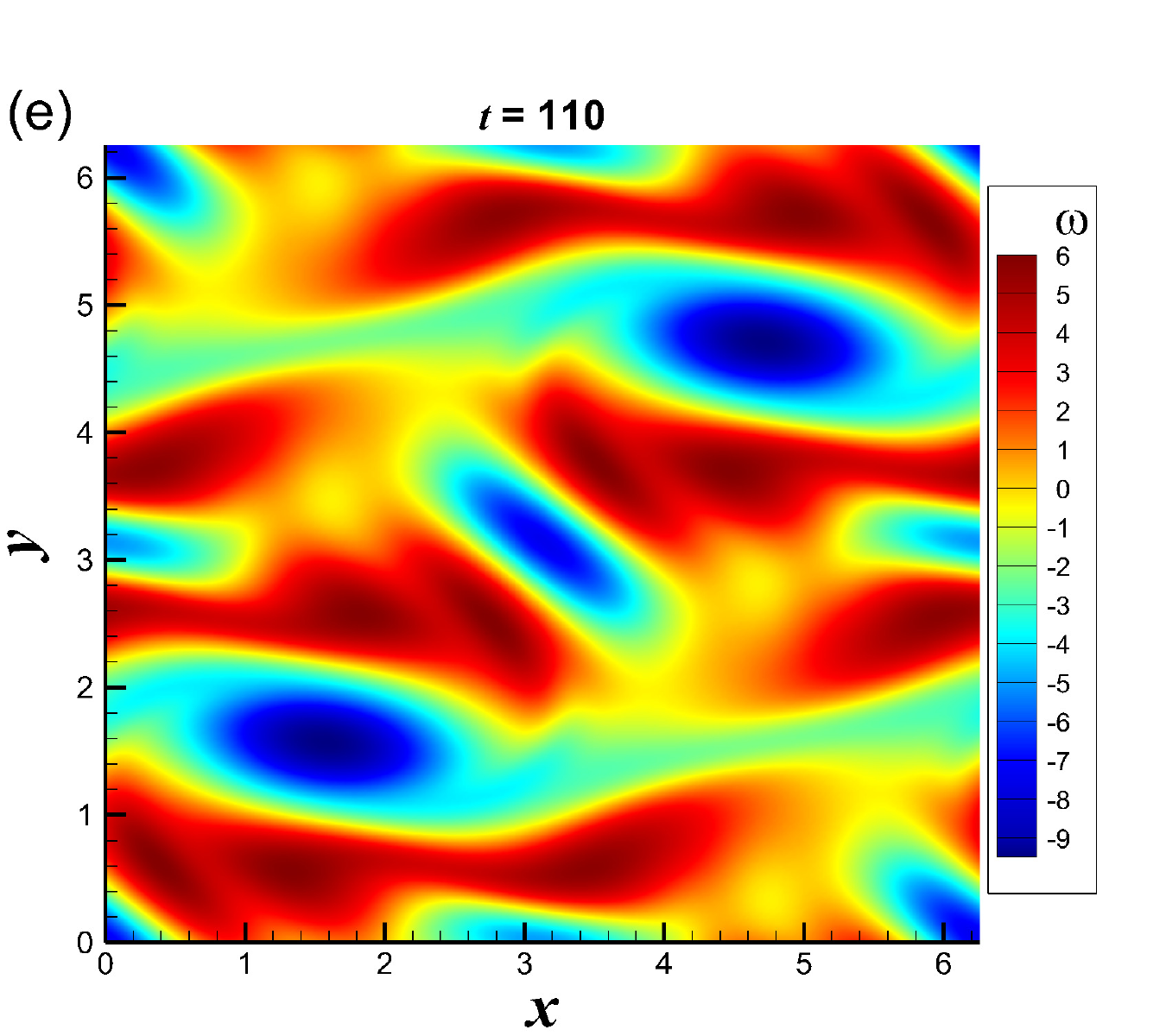}
             \includegraphics[width=2.2in]{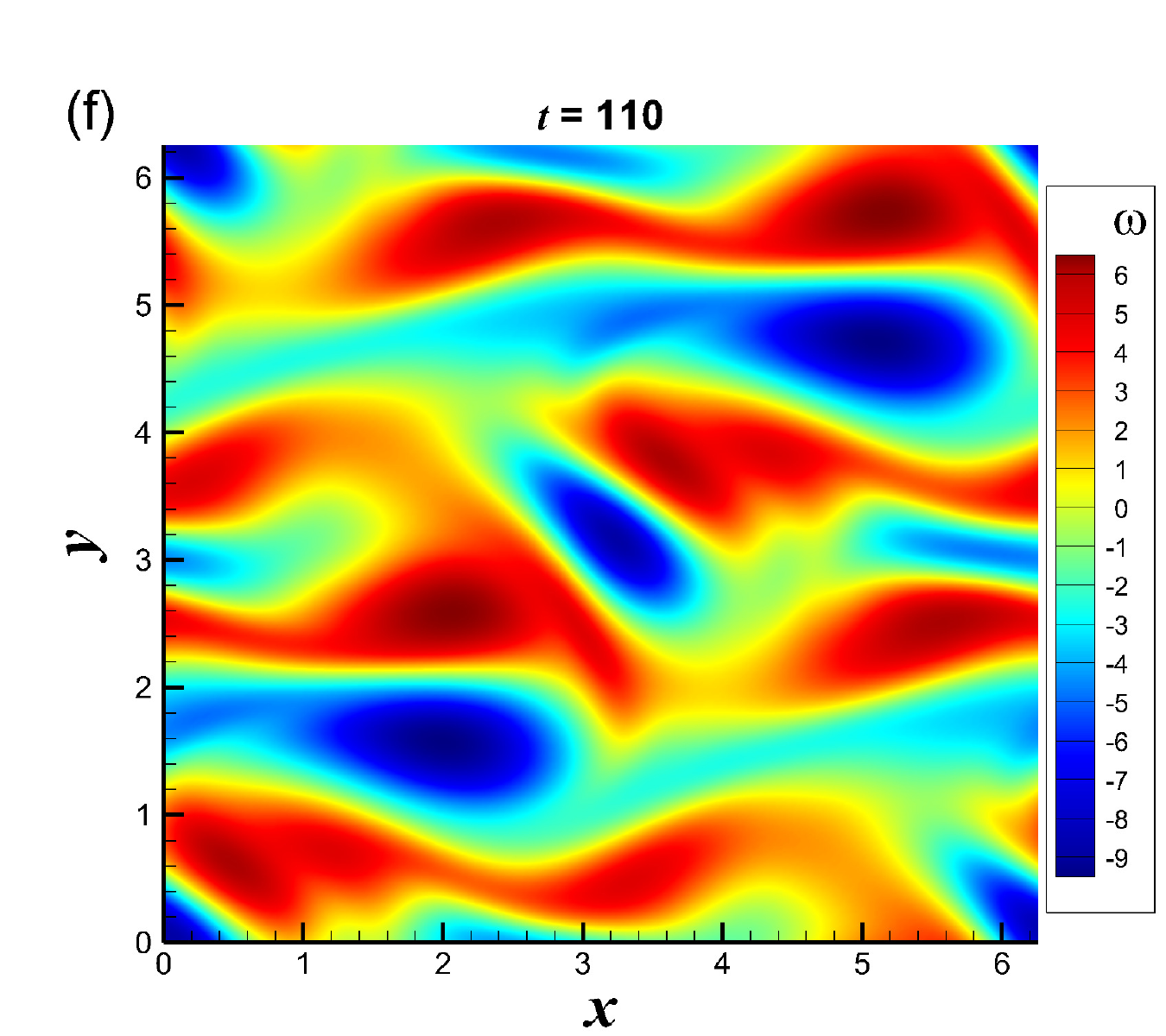}    \\
             \includegraphics[width=2.2in]{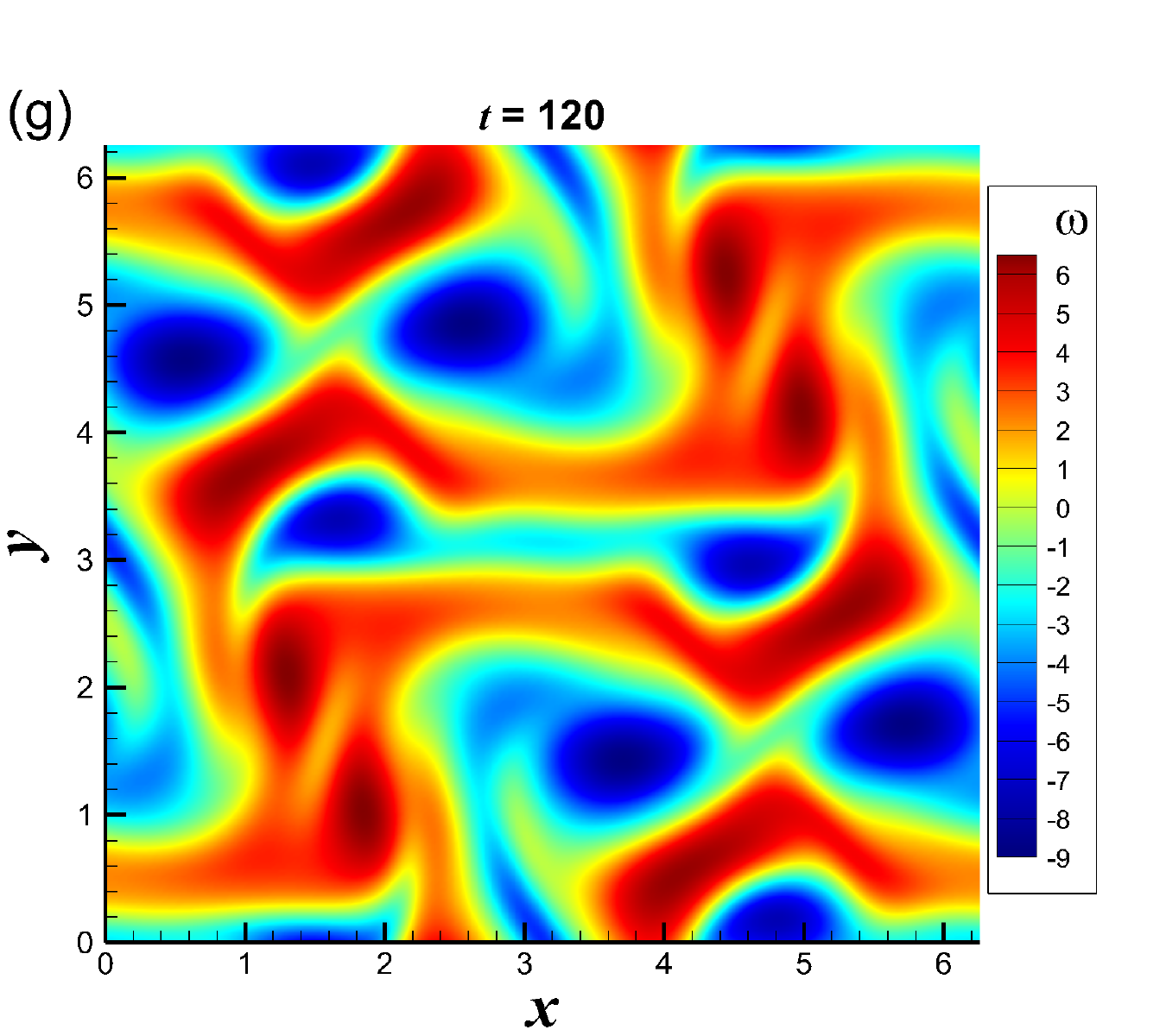}
             \includegraphics[width=2.2in]{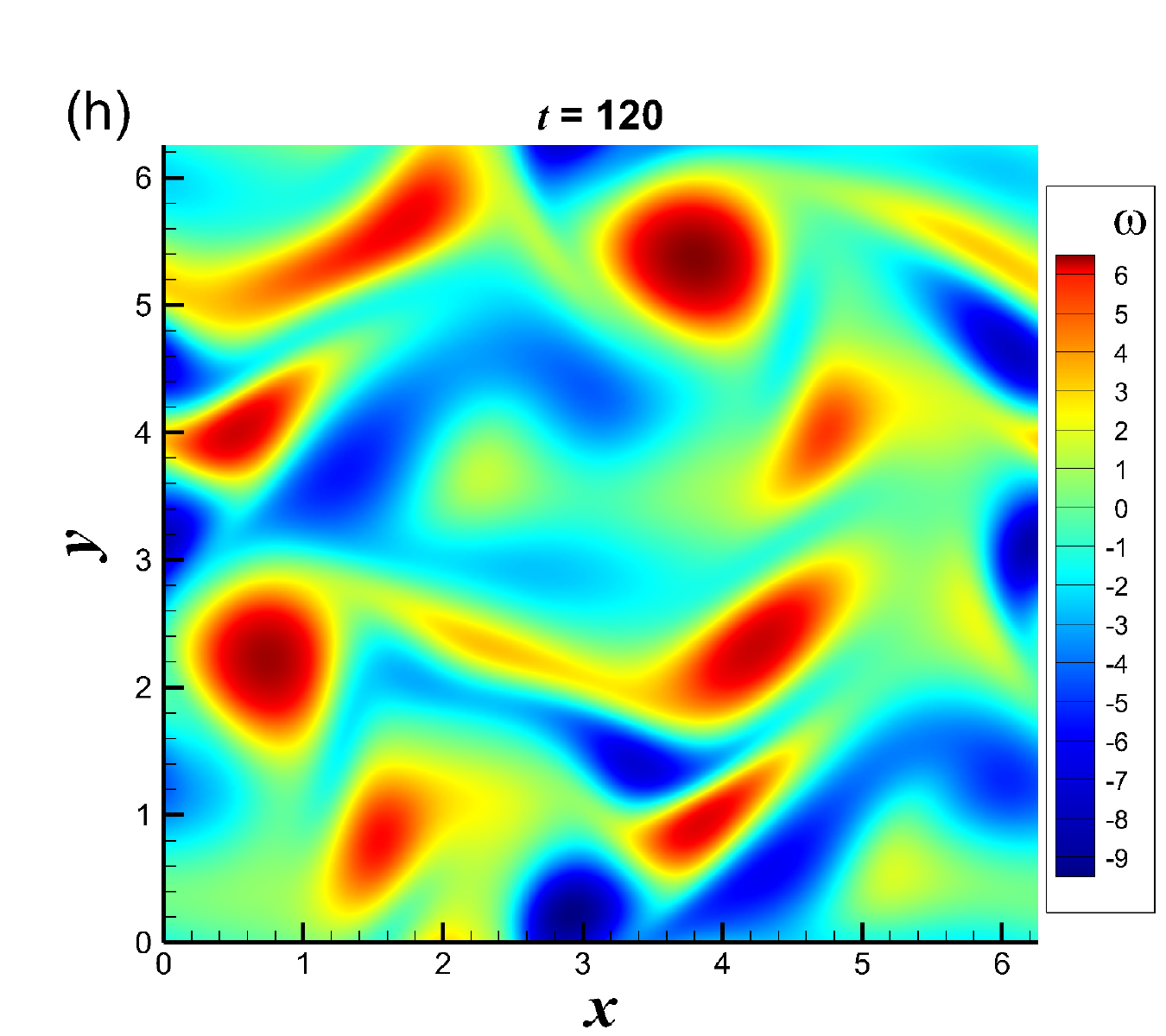}    \\
        \end{tabular}
    \caption{Comparisons of the vorticity field $\omega$ of the two-dimensional Kolmogorov turbulence governed by (\ref{eq_psi}) in the case of $n_K=4$ and $Re=40$ with the same initial condition (\ref{initial_condition})  given by the CNS (left) and DNS (right), respectively, at the different times.  (a)-(b):$t=90$; (c)-(d): $t=100$; (e)-(f):  $t=110$; (g)-(h): $t=120$.}     \label{Vor_Evolutions}
    \end{center}
\end{figure}

\begin{figure}
    \begin{center}
        \begin{tabular}{cc}
             \includegraphics[width=2.2in]{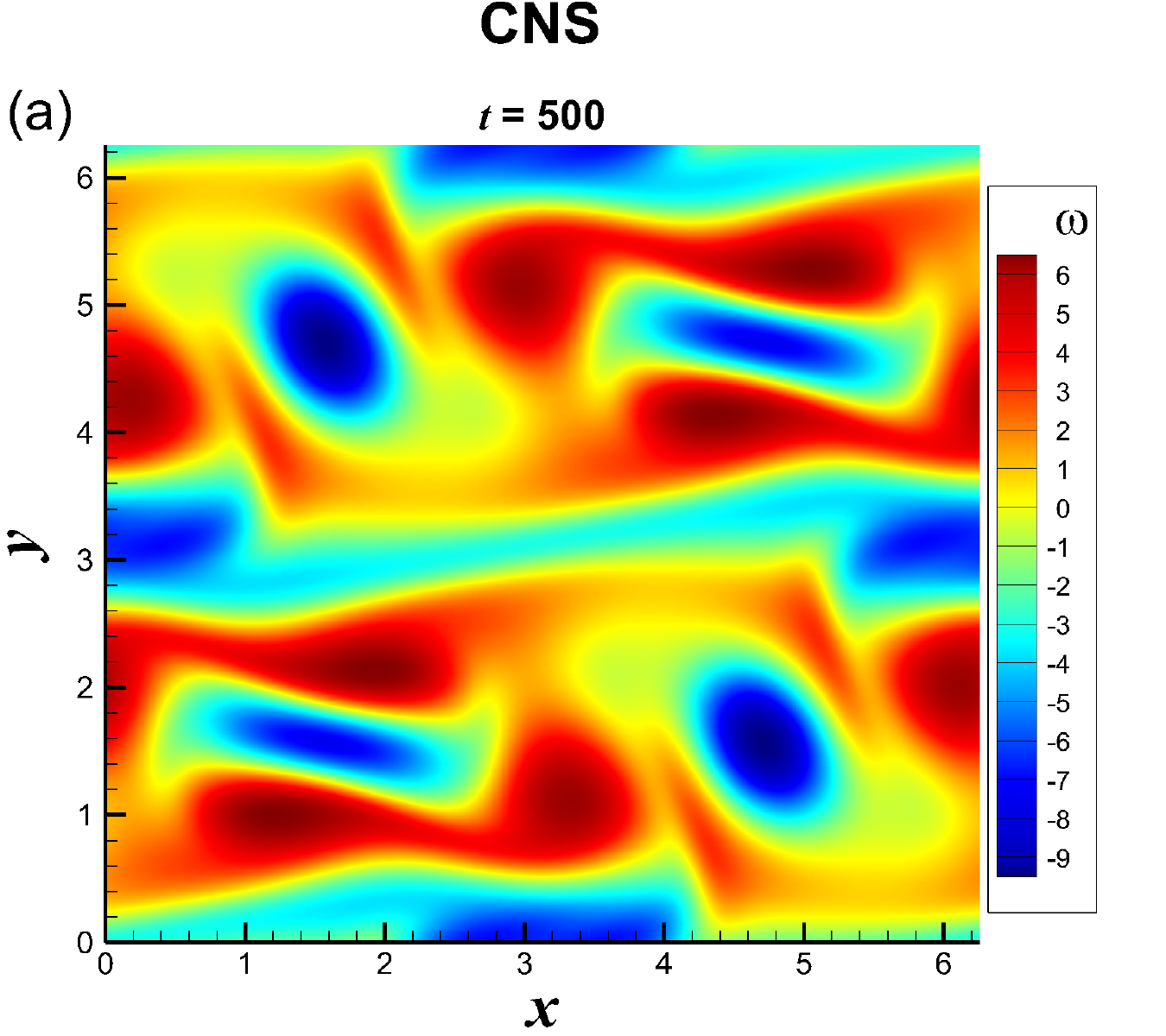}
             \includegraphics[width=2.2in]{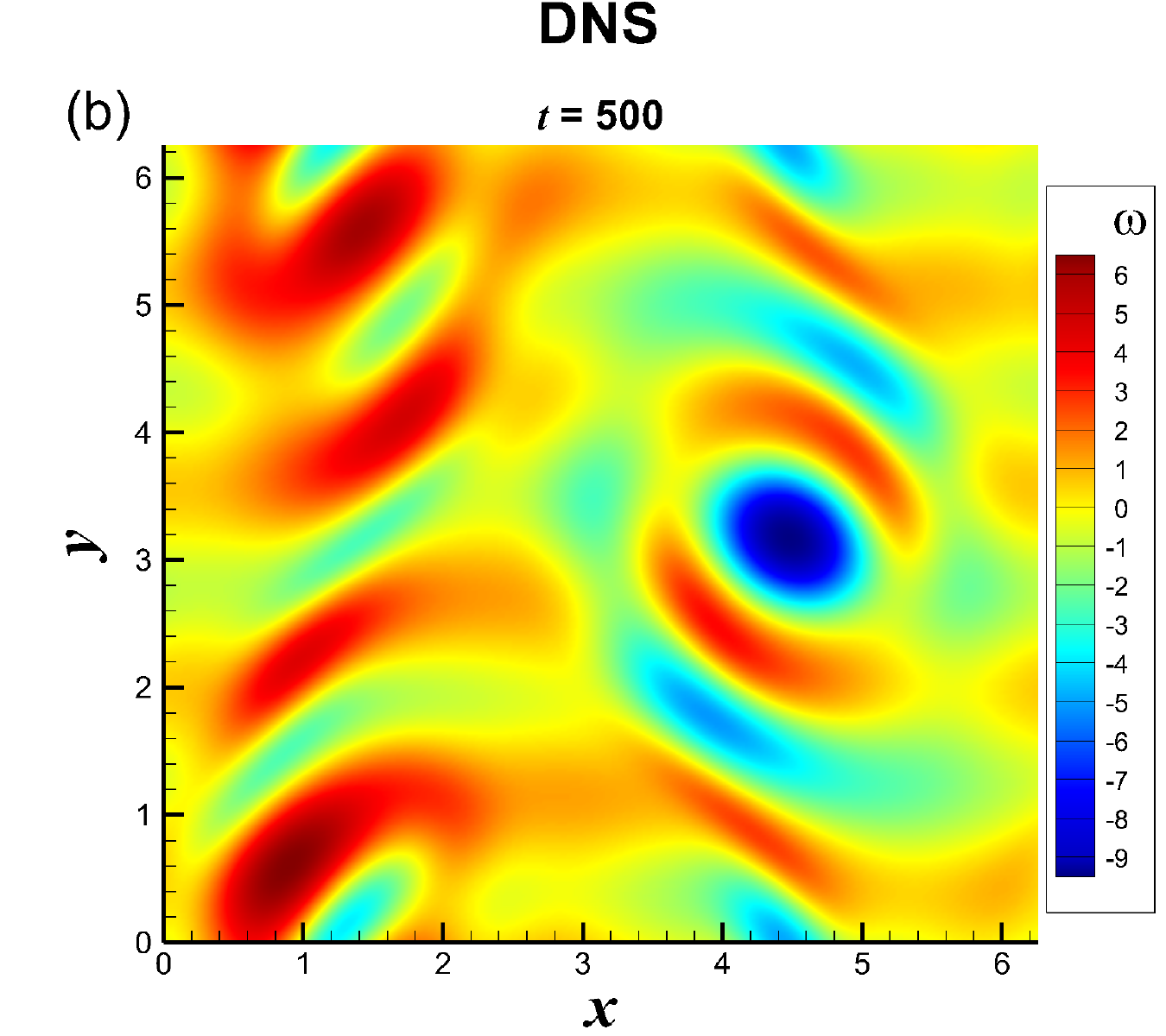}    \\
             \includegraphics[width=2.2in]{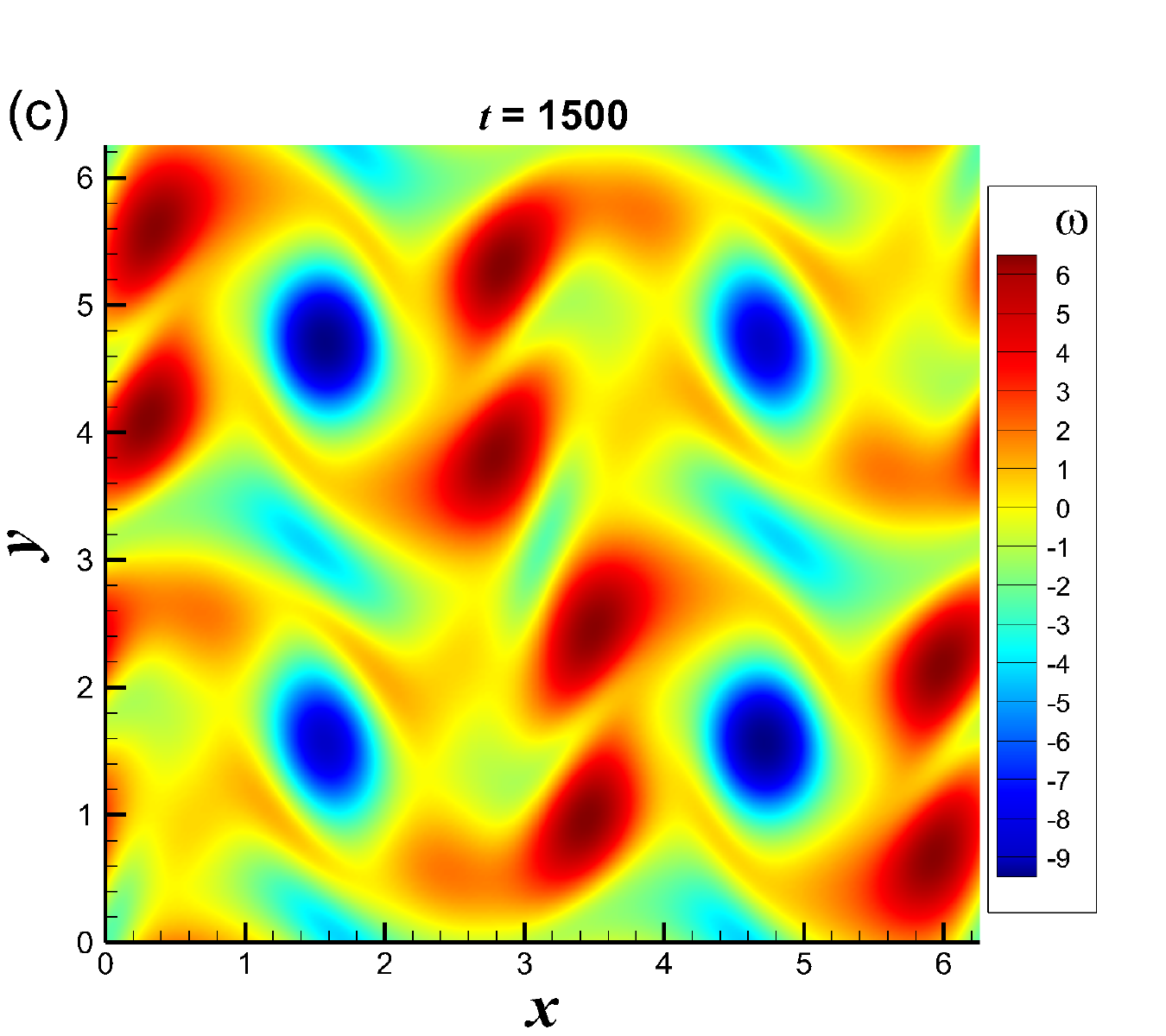}
             \includegraphics[width=2.2in]{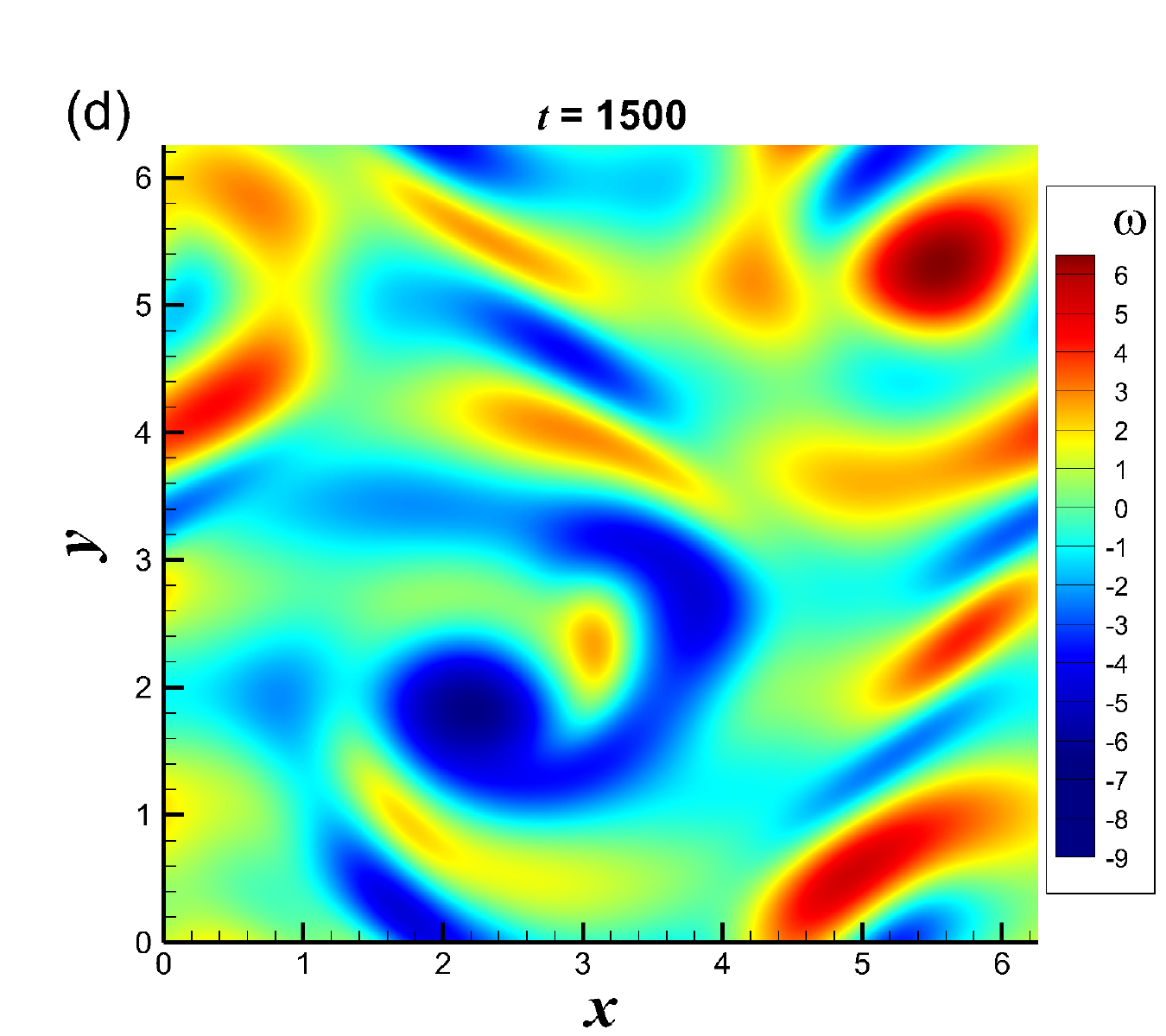}    \\
        \end{tabular}
    \caption{Comparisons of the vorticity field $\omega$ of the two-dimensional Kolmogorov turbulence governed by (\ref{eq_psi}) in the case of $n_K=4$ and $Re=40$ with the same initial condition (\ref{initial_condition}) given by the CNS (left) and DNS (right), respectively, at the different times.   (a)-(b):~$t=500$; (c)-(d): $t=1500$. }     \label{Vor_Evolutions_More}
    \end{center}
\end{figure}

\begin{figure}
    \begin{center}
        \begin{tabular}{cc}
             \includegraphics[width=2.5in]{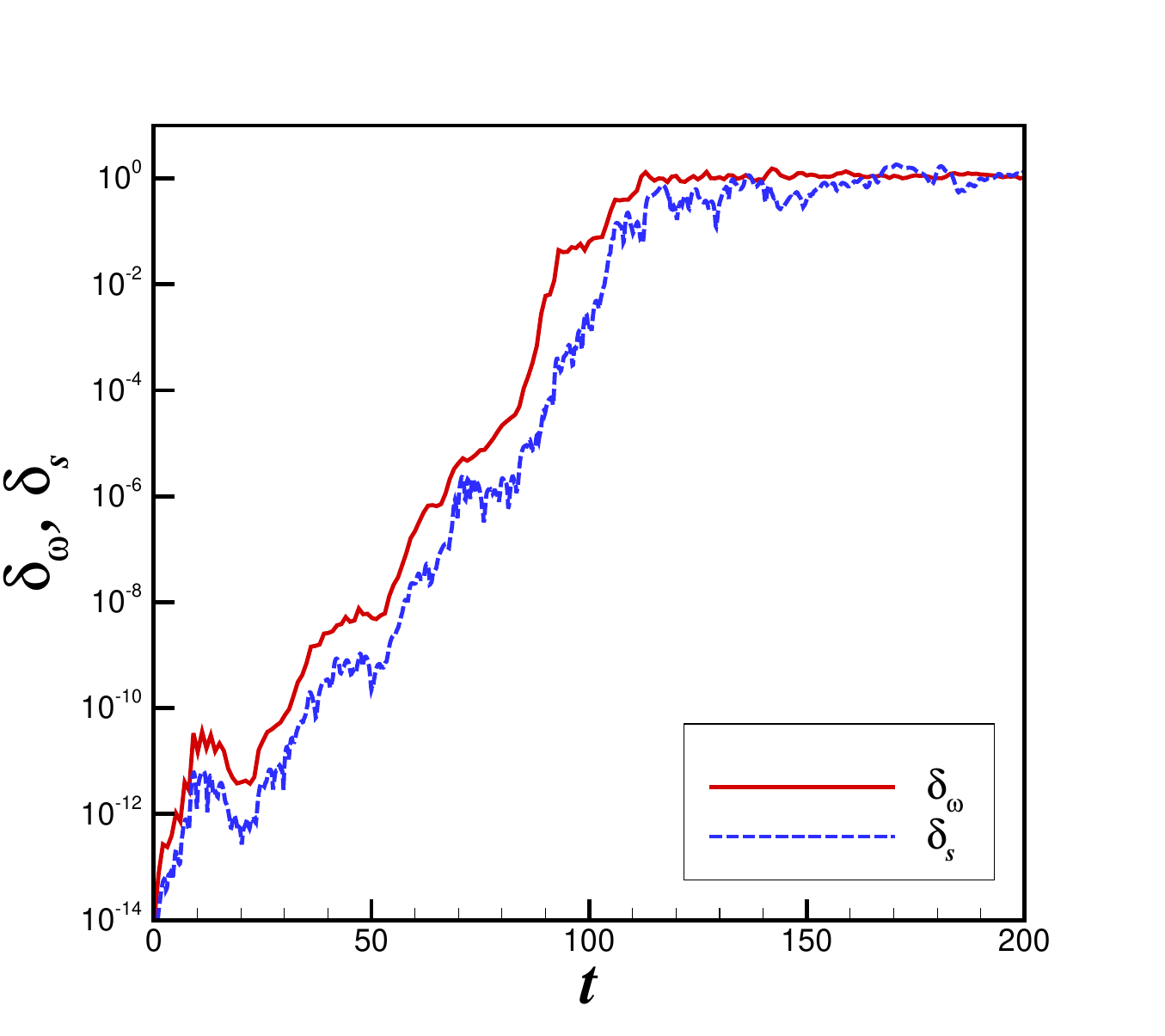}
        \end{tabular}
    \caption{The normalized absolute error $\delta_{\omega}$,  defined by (\ref{def:delta-omega}), and the spectrum-deviation $\delta_{s}$, defined by (\ref{delta_s}), of the vorticity field given by the DNS of the two-dimensional Kolmogorov turbulence governed by (\ref{eq_psi}) in the case of $n_K=4$ and $Re=40$ with the initial condition (\ref{initial_condition}), compared to the CNS benchmark solution. Red solid line: the normalized absolute error $\delta_{\omega}$; blue dash line: the spectrum-deviation $\delta_{s}$.}     \label{delta-w}
    \end{center}
\end{figure}

As shown in Figure~\ref{Vor_Evolutions} and \ref{Vor_Evolutions_More}, the DNS results of vorticity $\omega$ defined by (\ref{vorticity}) at different time $t$ are compared with the corresponding `clean'  benchmark solution given by the CNS. 
Note that the flow states given by DNS and CNS are completely the same at the beginning times such as $t=90$, as shown in Figure~\ref{Vor_Evolutions}(a) and (b), when the numerical noise of the DNS result has not yet increased to the same order of magnitude of the `true' physical solution so that the macroscopic deviation between DNS and CNS cannot be observed by eyes.  
It should be emphasized that, for {\em both} results given by the  DNS and the CNS up to $t=90$, there exists a kind of spatial symmetry,  as shown in Figure~\ref{Vor_Evolutions}(a) and (b).   So, the DNS result in $t\in[0,90]$ should be close enough to the `true' physical solution.  Unfortunately, the corresponding  interval of time is too short for calculations of statistics.  As the time increases, for instances at $t=110$ and $t=120$, the deviation between the DNS result from the CNS benchmark solution becomes more and more obvious,  as shown in Figure~\ref{Vor_Evolutions}(e)-(h).  To clearly indicate this deviation, let us consider  the normalized absolute error of the vorticity between  DNS result and the CNS benchmark solution, defined by 
\begin{equation}
\delta_{\omega} = \frac{\int_{0}^{2\pi}\int_{0}^{2\pi} \left| \omega_{DNS}-\omega_{CNS}\right| dx dy}{\int_{0}^{2\pi}\int_{0}^{2\pi} \left|\omega_{CNS}\right| dx dy}, \label{def:delta-omega}
\end{equation}
 where $\omega_{DNS}$ and $\omega_{CNS}$ denote the vorticity given by the DNS and CNS, respectively. 
 It is found that  the normalized absolute errors of the vorticity, denoted by $\delta_{\omega}$,  are 0.006 at $t=90$, 0.06 at $t=100$, 0.49  at $t=110$ and 1.12  at $t=120$, respectively.  
  Note that, when $t \ge120$, the vorticity field given by the DNS completely losses the spatial symmetry, but the  vorticity field given by the CNS benchmark solution always keeps such kind of spatial symmetry even at $t=1500$, as shown in Figure~\ref{Vor_Evolutions_More}.   Note that the `false' numerical noise of the CNS benchmark result is much smaller than its  `true'  physical  solution and thus is negligible compared to the `true' physical solution in the {\em whole} interval of time $t\in[0,1500]$.  However, when $t
\ge 120$, the `false' numerical noise of the DNS result is at the {\em same} order of magnitude as the `true' physical solution, as shown in Figure~\ref{delta-w}, say, thereafter the DNS result is {\em badly} polluted,  which naturally leads to the loss of the spatial symmetry and a large-scale illustrative deviation in the flow field  from the `true' physical solution!  The quantitative comparisons of the field geometrical structures will be given in \S~3.2.  

As mentioned in \S~2.2,  the exact solution of the Navier-Stokes equations considered in this paper have the spatial symmetry (\ref{symmetry}). 
Note that the result given by the CNS indeed has this kind of spatial symmetry in the {\em whole} interval of time $t\in[0,1500]$.  But, unfortunately, the result given by the DNS {\em loses} such kind of spatial symmetry when $t > 120$, as shown in Figs.~\ref{Vor_Evolutions} and \ref{Vor_Evolutions_More}, because the numerical noises, which have {\em no} spatial symmetry, have been enlarged to the same order of magnitude as the `true' solution and thus finally destroy this kind of spatial symmetry of the flow field.    This is a {\em clear} and {\em rigorous} evidence that the numerical results given by the DNS are indeed  ``badly polluted'' when $t>120$.    

\begin{figure}
    \begin{center}
        \begin{tabular}{cc}
             \includegraphics[width=3.0in]{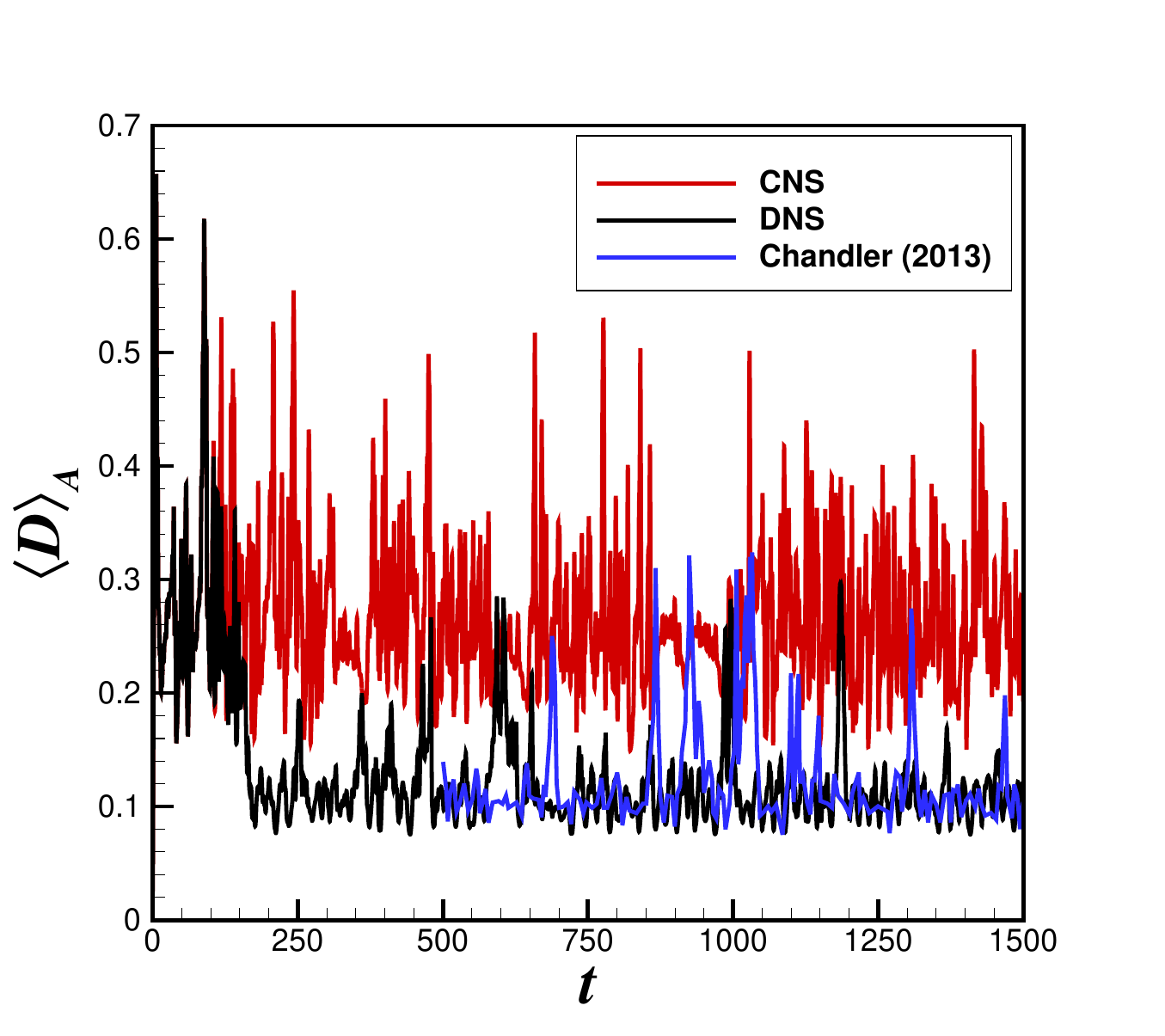}
        \end{tabular}
    \caption{Comparison of the spatially averaged kinetic energy dissipation rate $\langle D\rangle_A$ of the two-dimensional Kolmogorov turbulence governed by (\ref{eq_psi}) in the case of $n_K=4$ and $Re=40$ with the initial condition (\ref{initial_condition}) given by the CNS benchmark result (solid line in red), the DNS result (solid line in black), the previous study by Chandler and Kerswell \cite{chandler2013invariant} (solid line in blue).}     \label{D_t}
    \end{center}
\end{figure}

How about the influences of the numerical noises as a kind of artificial  tiny disturbances in statistics?  
First of all, let us consider  the spatially averaged kinetic energy dissipation rate $\langle D\rangle_A$, where  $\langle  \; \rangle_A$ is an operator of statistics defined by (\ref{average_A}),  and $D$ denotes the kinetic energy dissipation rate defined by (\ref{dissipation_rate}), respectively.  
As shown in Figure~\ref{D_t}, when $t>150$, $\langle D\rangle_A$ given by the CNS (with a approximate average $0.25$) is obviously higher than that given by the DNS (with a approximate average $0.1$), where the deviation of nearly $2.5$ times in magnitude clearly  reveals  the large  differences. Note that our DNS result is in accord with the DNS ones given by Chandler and Kerswell \cite{chandler2013invariant}, which is also a mixture of the `true' physical solution and the `false' numerical noise that are mostly at the same order of magnitude.

\begin{figure}
    \begin{center}
        \begin{tabular}{cc}
             \subfigure[]{\includegraphics[width=2.55in]{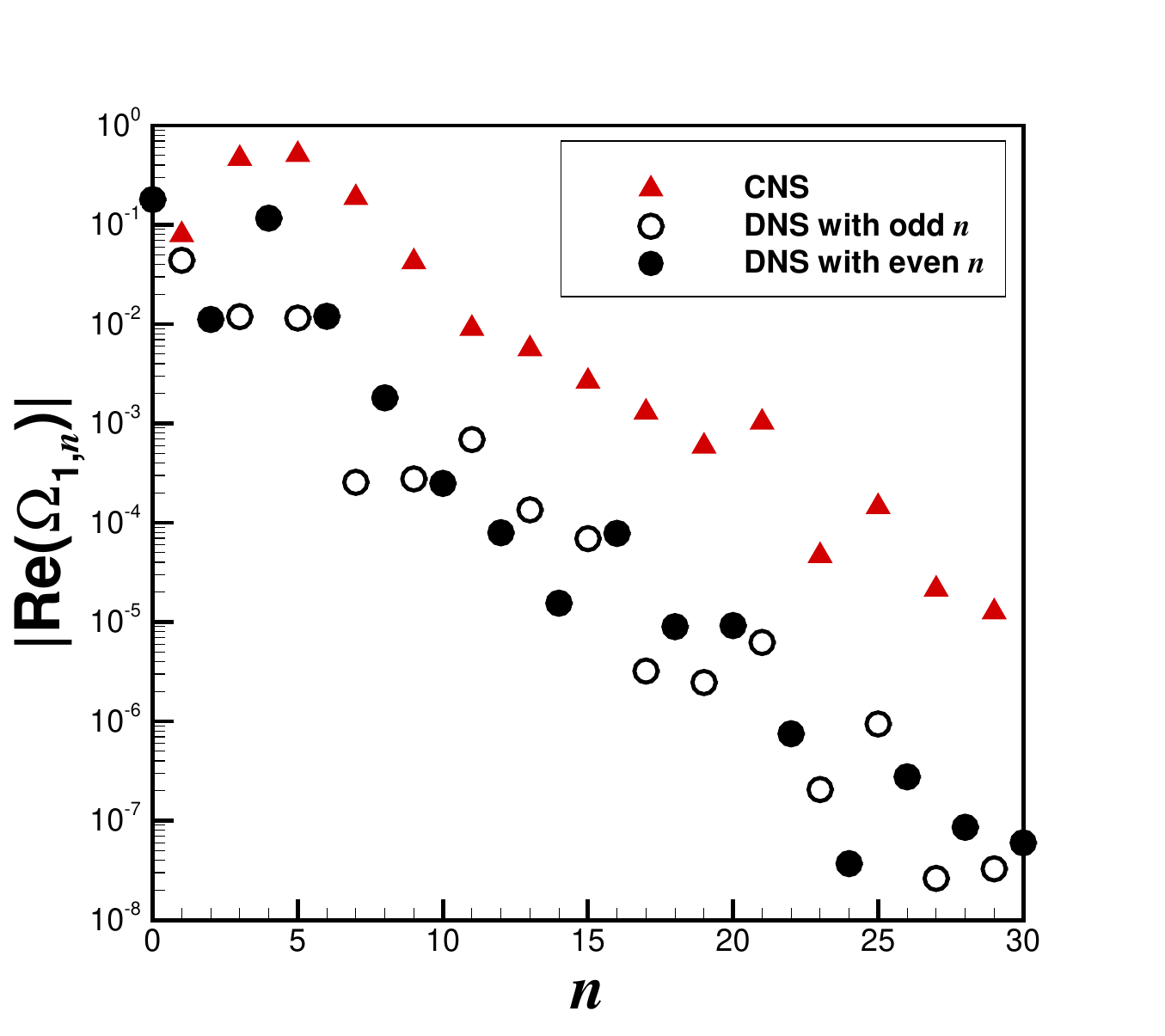}}
             \subfigure[]{\includegraphics[width=2.55in]{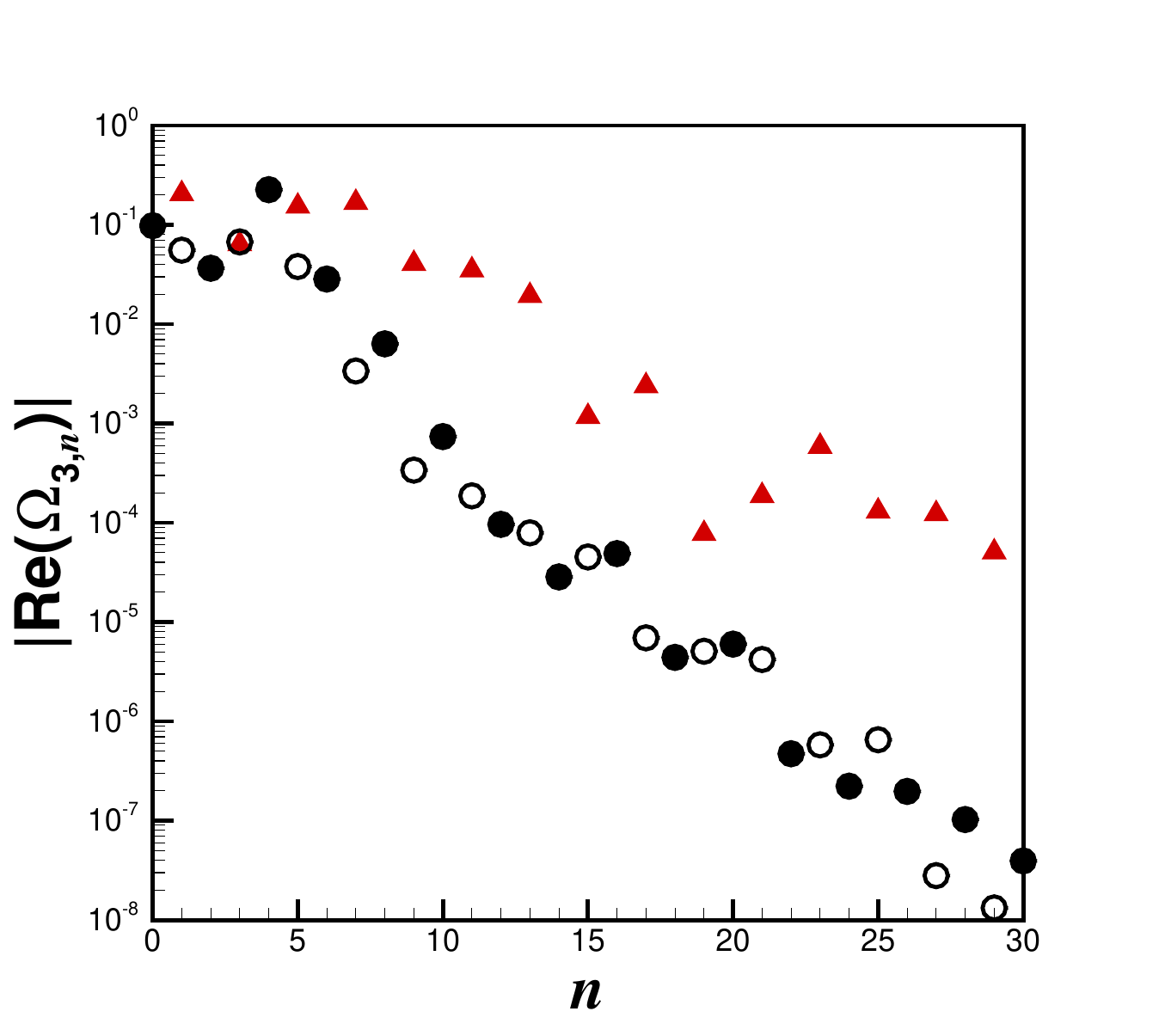}}
        \end{tabular}
    \caption{Comparisons of (a) $\mid$Re($\Omega_{1,n}$)$\mid$ and (b) $\mid$Re($\Omega_{3,n}$)$\mid$ of the two-dimensional Kolmogorov turbulence  at $t=500$ in the case of $n_K=4$ and $Re=40$ with the the initial condition (\ref{initial_condition}), where the Fourier coefficient $\Omega_{m,n}$ of the vorticity $\omega$ defined by (\ref{vorticity}) is given either by the CNS benchmark solution (solid triangle in red) or the DNS result (odd $n$: empty circle in black, even $n$: solid circle in black).  Here Re($a$) denotes the real part of the complex number $a$,  $| a |$ represents the absolute value of $a$, respectively.}     \label{RSp_x13}
    \end{center}

    \begin{center}
        \begin{tabular}{cc}
             \subfigure[]{\includegraphics[width=2.55in]{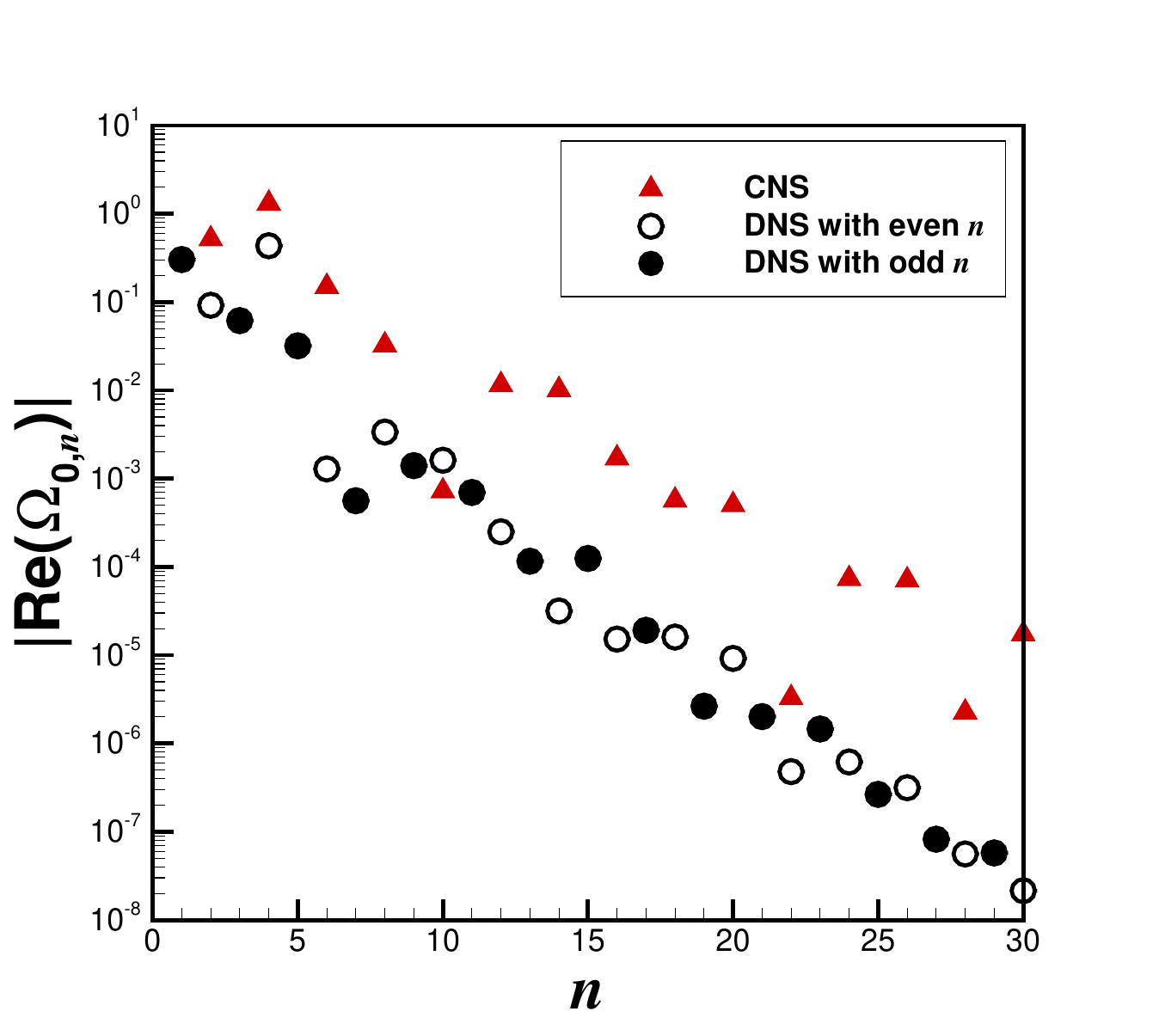}}
             \subfigure[]{\includegraphics[width=2.55in]{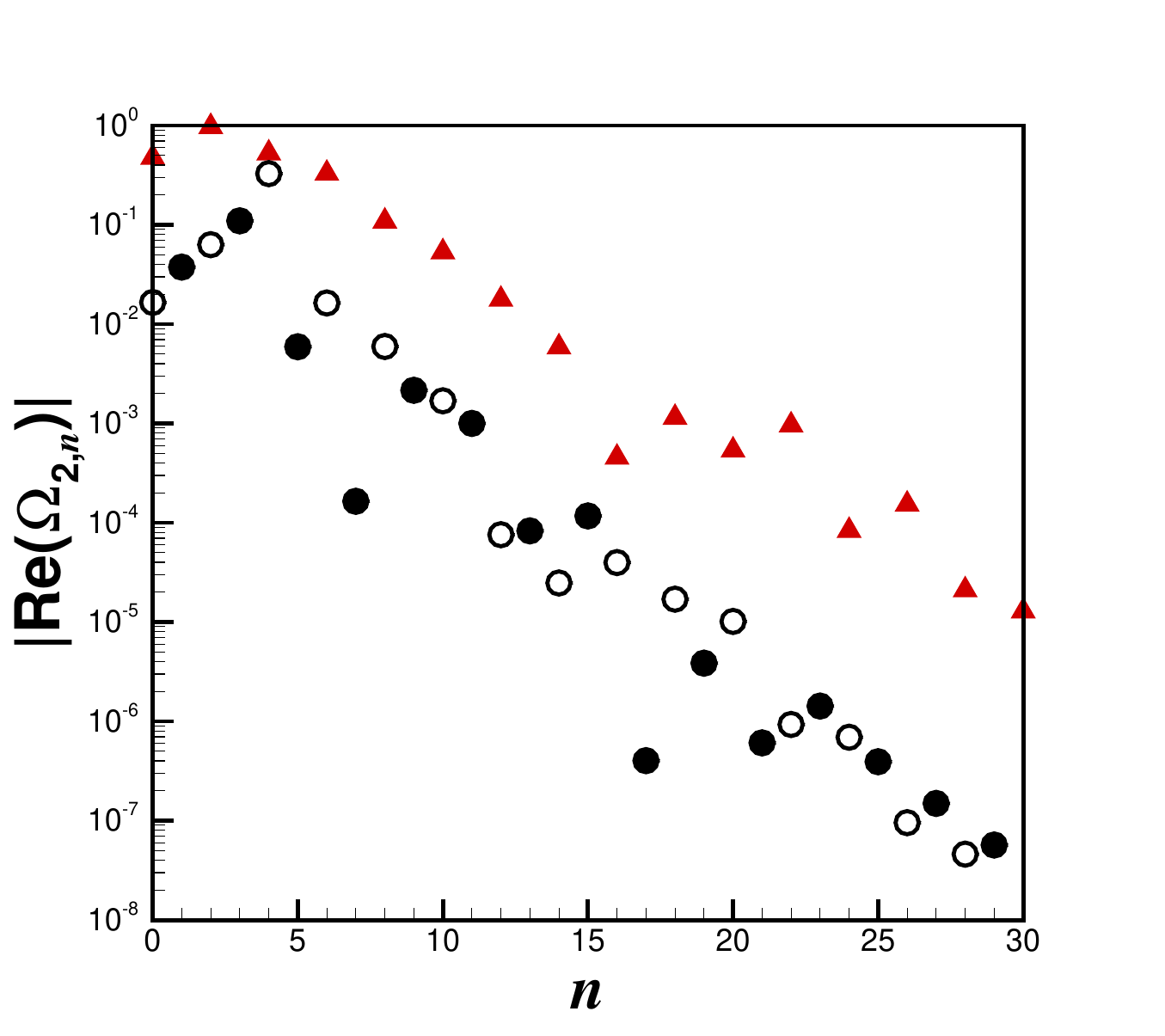}}
        \end{tabular}
    \caption{Comparisons of (a) $\mid$Re($\Omega_{0,n}$)$\mid$ and (b) $\mid$Re($\Omega_{2,n}$)$\mid$  of the two-dimensional Kolmogorov turbulence at $t=500$  in the case of $n_K=4$ and $Re=40$ with the initial condition (\ref{initial_condition}), where $\Omega_{m,n}$ of the vorticity $\omega$ defined by (\ref{vorticity}) is  given either by the CNS benchmark solution (solid triangle in red) or the DNS result (even $n$: empty circle in black, odd $n$: solid circle in black).}     \label{RSp_x02}
    \end{center}
\end{figure}

Let us further consider the Fourier coefficient $\Omega_{m,n}$ of the vorticity field $\omega$ defined by (\ref{vorticity})  at the time $t=500$, corresponding to Figure~\ref{Vor_Evolutions_More} (a) and (b).  
As shown in Figure~\ref{RSp_x13},
when $m$ is an odd number such as $m=1$ or $m=3$,  Re($\Omega_{m,n}$) given by the CNS benchmark solution  is {\em always} zero when $n$ is an even number, but Re($\Omega_{m,n}$) given by the DNS is {\em not} zero for {\em all} $n$, where Re($\;$) is an operator that takes the real part of a complex number.
On the other hand,  when $m$ is an even number such as $m=0$ or $m=2$, Re($\Omega_{m,n}$) given by the CNS is zero when $n$ is an odd number, but Re($\Omega_{m,n}$) given by the DNS is {\em not} zero nearly for {\em all} $n$  (except $\Omega_{0,0}=0$), as shown in Figure~\ref{RSp_x02}.
Besides, it is found that, when Re($\Omega_{m,n}$) given by the CNS are not zero, their values are usually an order of magnitude larger than those given by the DNS.
In addition, the imaginary part Im($\Omega_{m,n}$) given by the CNS is {\em always} zero for {\em all} values of $m$ and $n$, but this  is  {\em not} found in the DNS result, where Im($\;$) is an operator that takes the imaginary part of a complex number.  These are mainly because the `false' numerical noises of the DNS quickly increase to the same order of magnitude of the `true' physical solution of the vorticity field, which certainly leads to the large-scale  huge deviation from the CNS benchmark solution!  All of these can clearly explain why the DNS result losses the  spatial symmetry when $t>120$.  

\begin{figure}

    \begin{center}
        \begin{tabular}{cc}
             \subfigure[]{\includegraphics[width=2.55in]{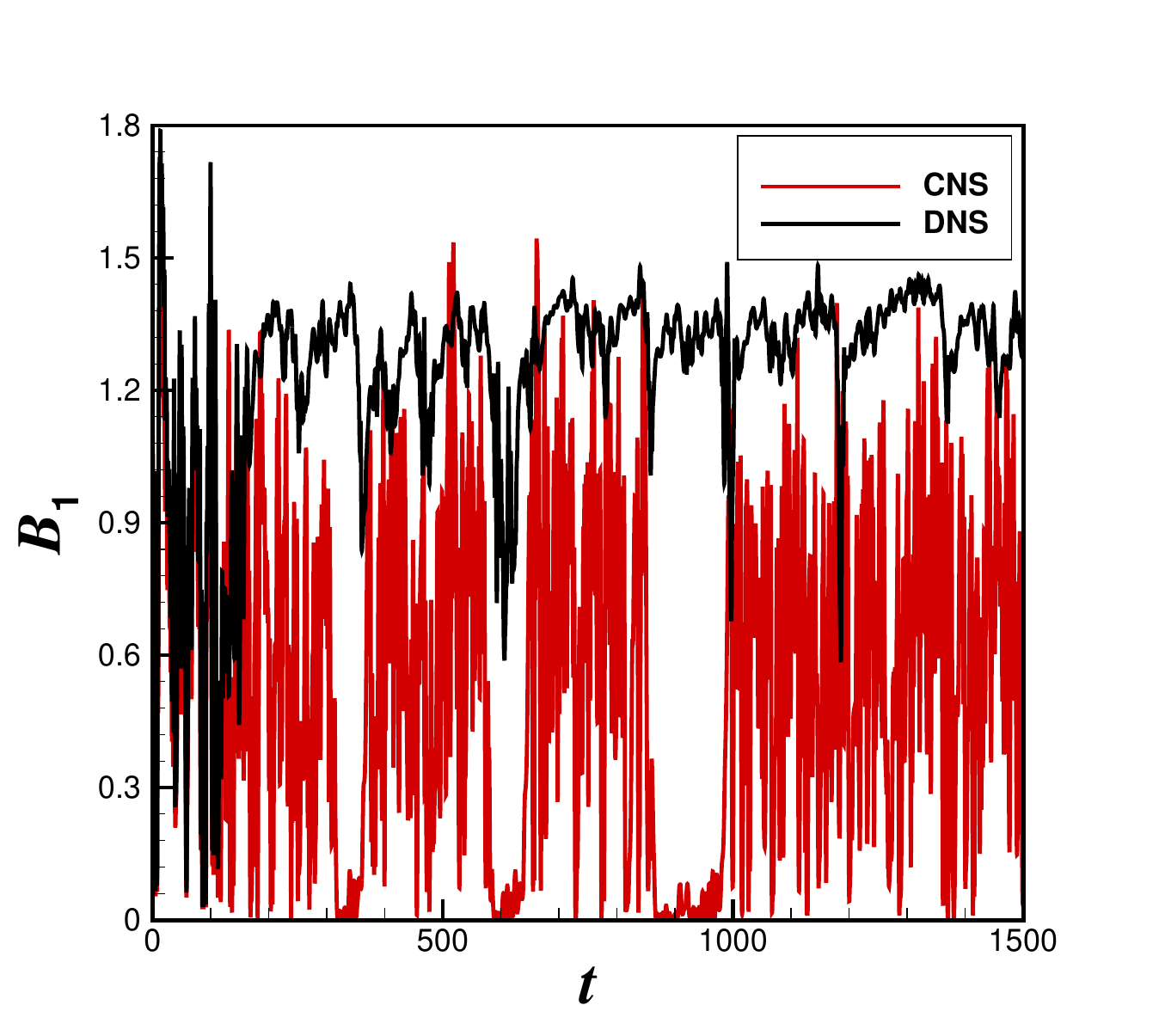}}
             \subfigure[]{\includegraphics[width=2.55in]{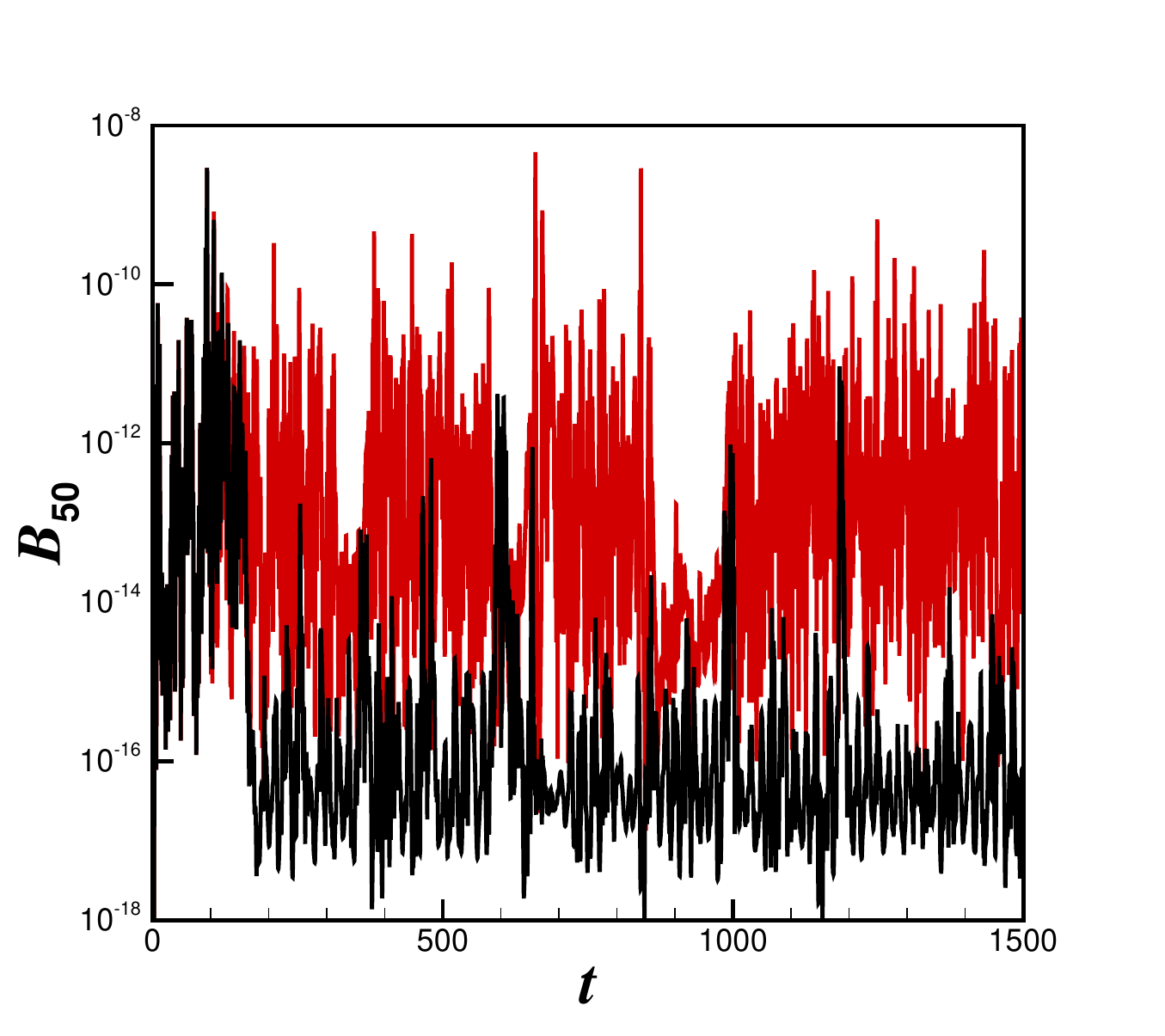}}
        \end{tabular}
    \caption{Comparisons of the enstrophy spectrums at certain wavenumbers, i.e. (a) $B_1$ and (b) $B_{50}$, of the two-dimensional Kolmogorov turbulence in the case of $n_K=4$ and $Re=40$ with the initial condition (\ref{initial_condition}), given by the CNS  (red) and the DNS (black), respectively.}     \label{B_t}
    \end{center}
    

    
    \begin{center}
        \begin{tabular}{cc}
             \subfigure[]{\includegraphics[width=2.55in]{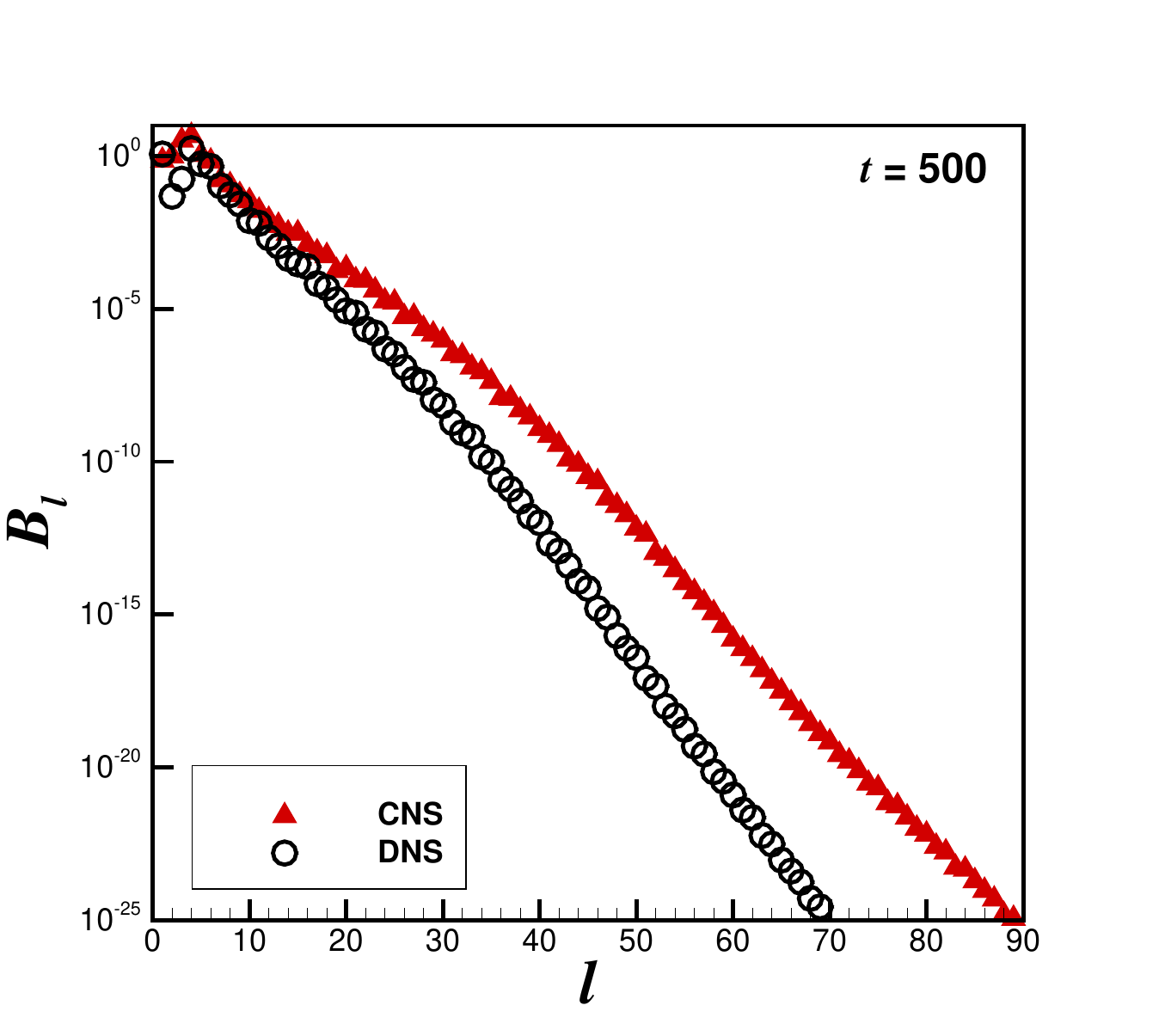}}
             \subfigure[]{\includegraphics[width=2.55in]{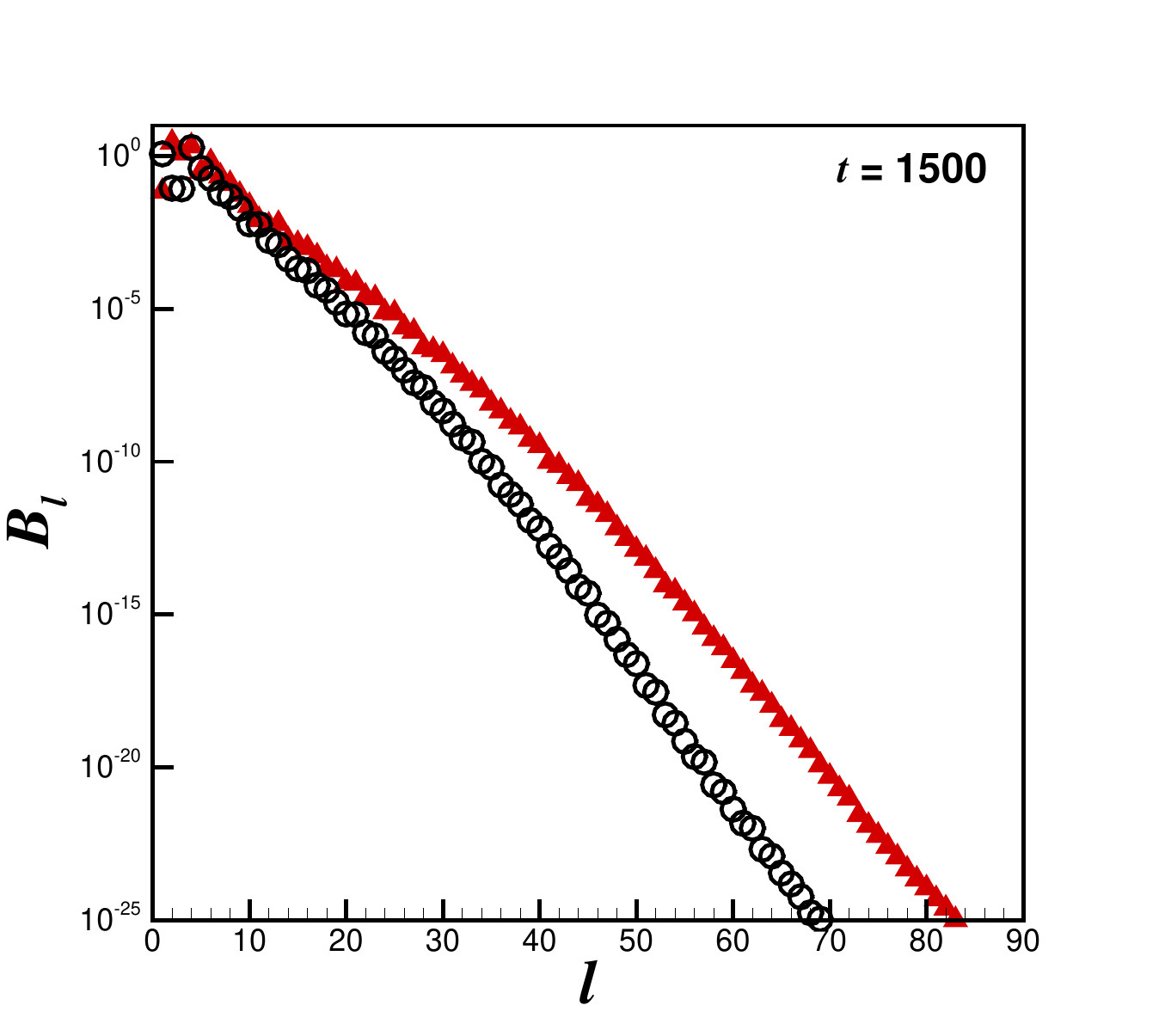}}
        \end{tabular}
    \caption{Comparisons of the enstrophy spectrums $B_l$ at (a) $t=500$ and (b) $t=1500$ of the two-dimensional Kolmogorov turbulence in the case of $n_K=4$ and $Re=40$ with the initial condition (\ref{initial_condition}), given by the CNS (triangle in red) and the DNS  (circle in black), respectively.}     \label{B_l}
    \end{center}

\end{figure}

\begin{figure}
    \begin{center}
        \begin{tabular}{cc}
             \subfigure[]{\includegraphics[width=2.55in]{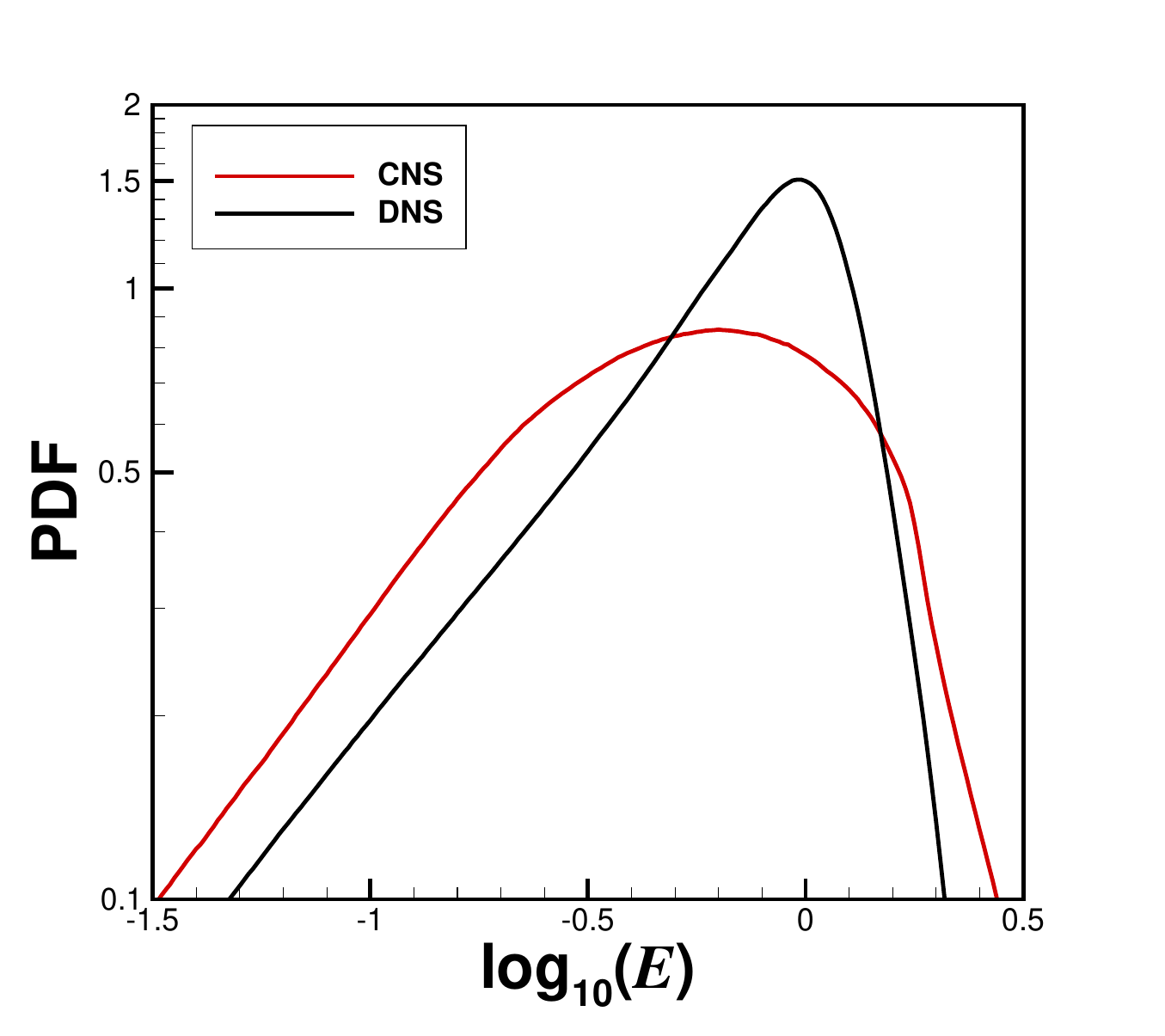}}
             \subfigure[]{\includegraphics[width=2.55in]{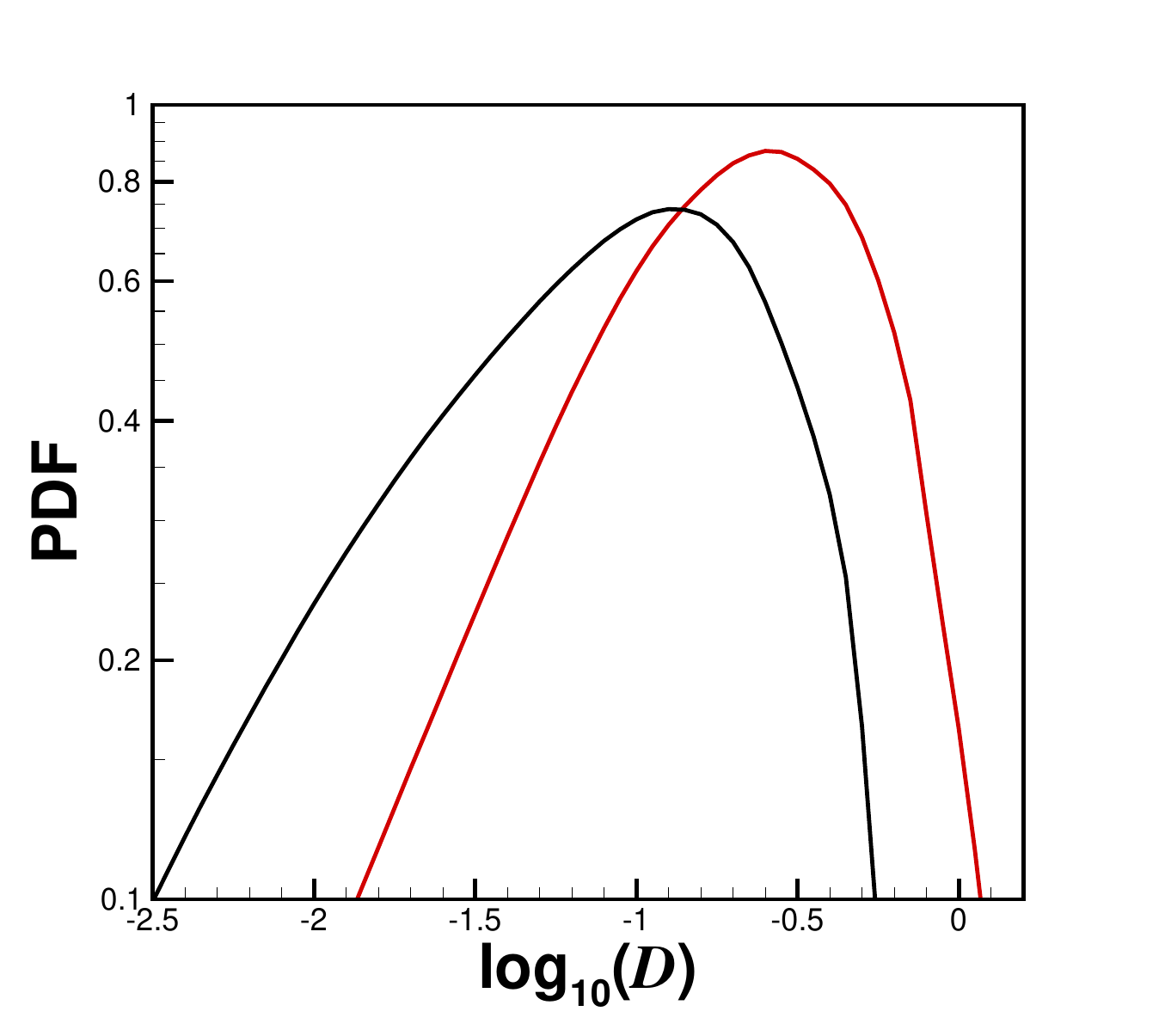}}
        \end{tabular}
    \caption{Comparisons of the probability density function (PDF) of (a) the kinetic energy $E(x,y,t)$ and (b) the kinetic energy dissipation rate $D(x,y,t)$ of the two-dimensional Kolmogorov turbulence  in the case of $n_K=4$ and $Re=40$ with the initial condition (\ref{initial_condition}), given by  the CNS (red) and the DNS (black), respectively, where PDFs are integrated in $x,y\in[0,2\pi]$ and $t \in [200, 1500]$.}     \label{PDF_E_D}
    \end{center}

    \begin{center}
        \begin{tabular}{cc}
             \subfigure[]{\includegraphics[width=2.55in]{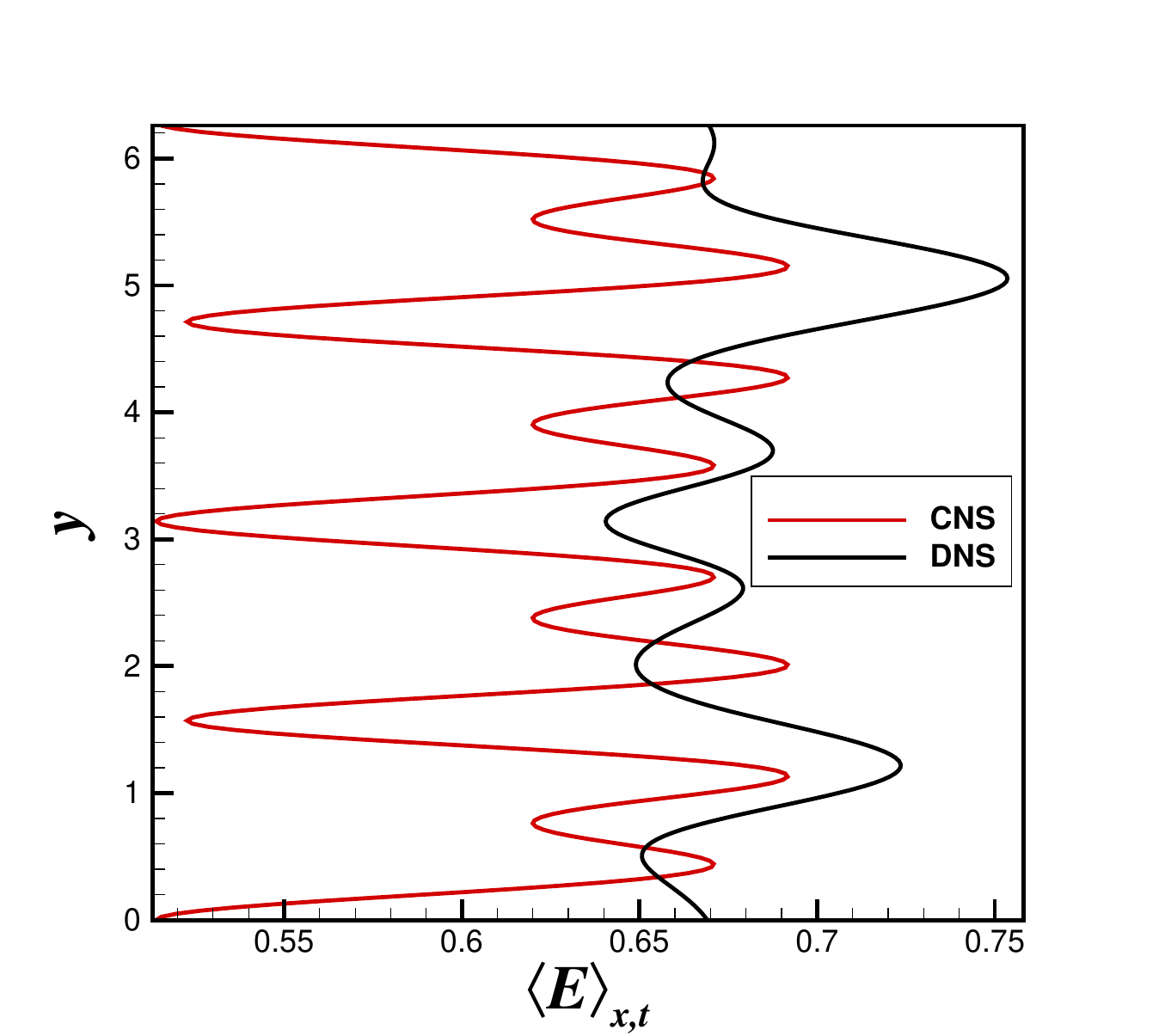}}
             \subfigure[]{\includegraphics[width=2.55in]{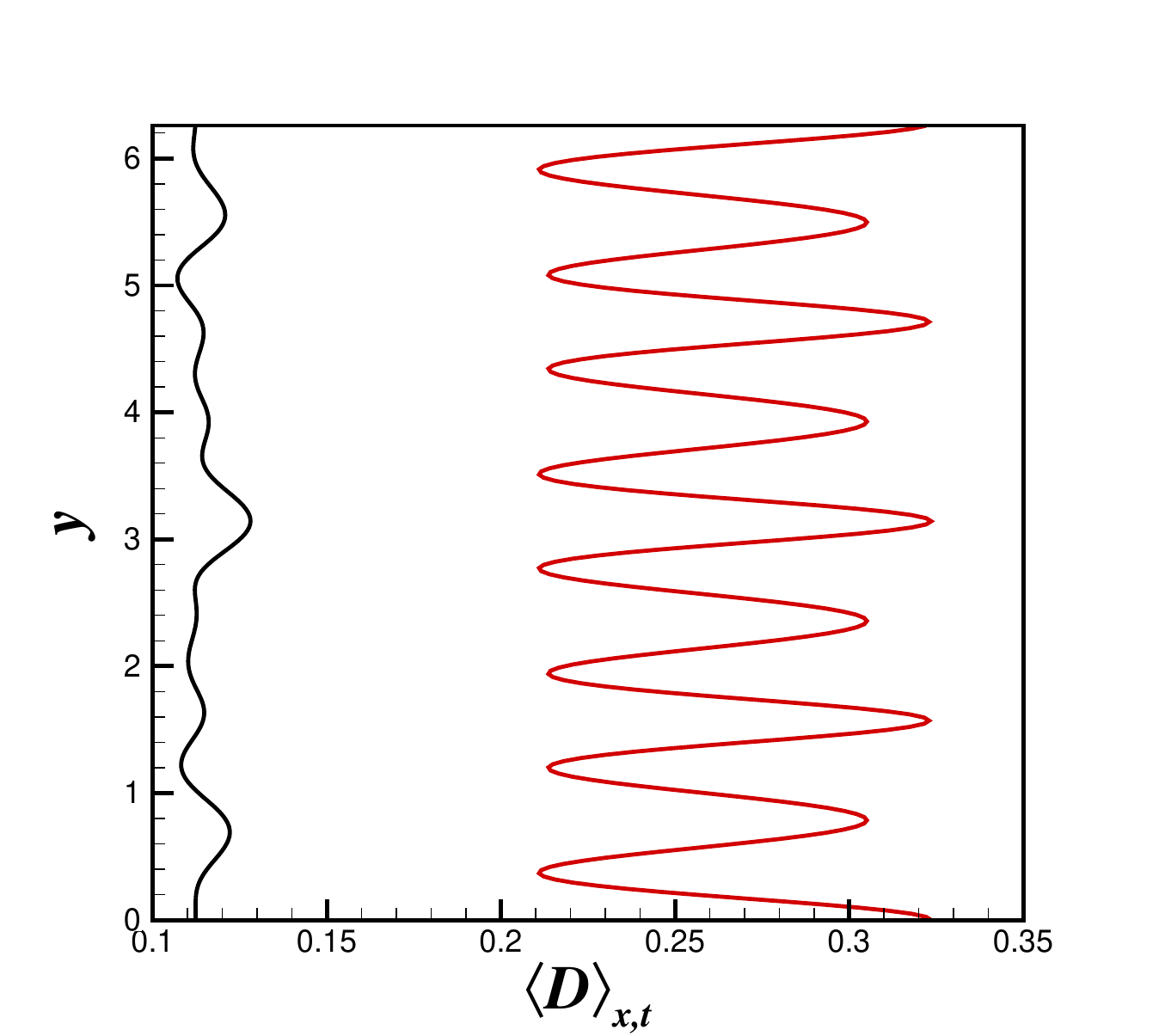}}
        \end{tabular}
    \caption{Comparisons of the vertical distributions of (a) the spatio-temporal averaged kinetic energy $\langle E\rangle_{x,t}$ and (b) the spatio-temporal averaged kinetic energy dissipation rate $\langle D\rangle_{x,t}$ of the two-dimensional Kolmogorov turbulence in the case of $n_K=4$ and  $Re=40$ with the initial condition (\ref{initial_condition}), respectively, given by the CNS  (red) and the DNS  (black).}     \label{STA_E_D}
    \end{center}
\end{figure}

\begin{figure}
    \begin{center}
        \begin{tabular}{cc}
             \subfigure[]{\includegraphics[width=2.55in]{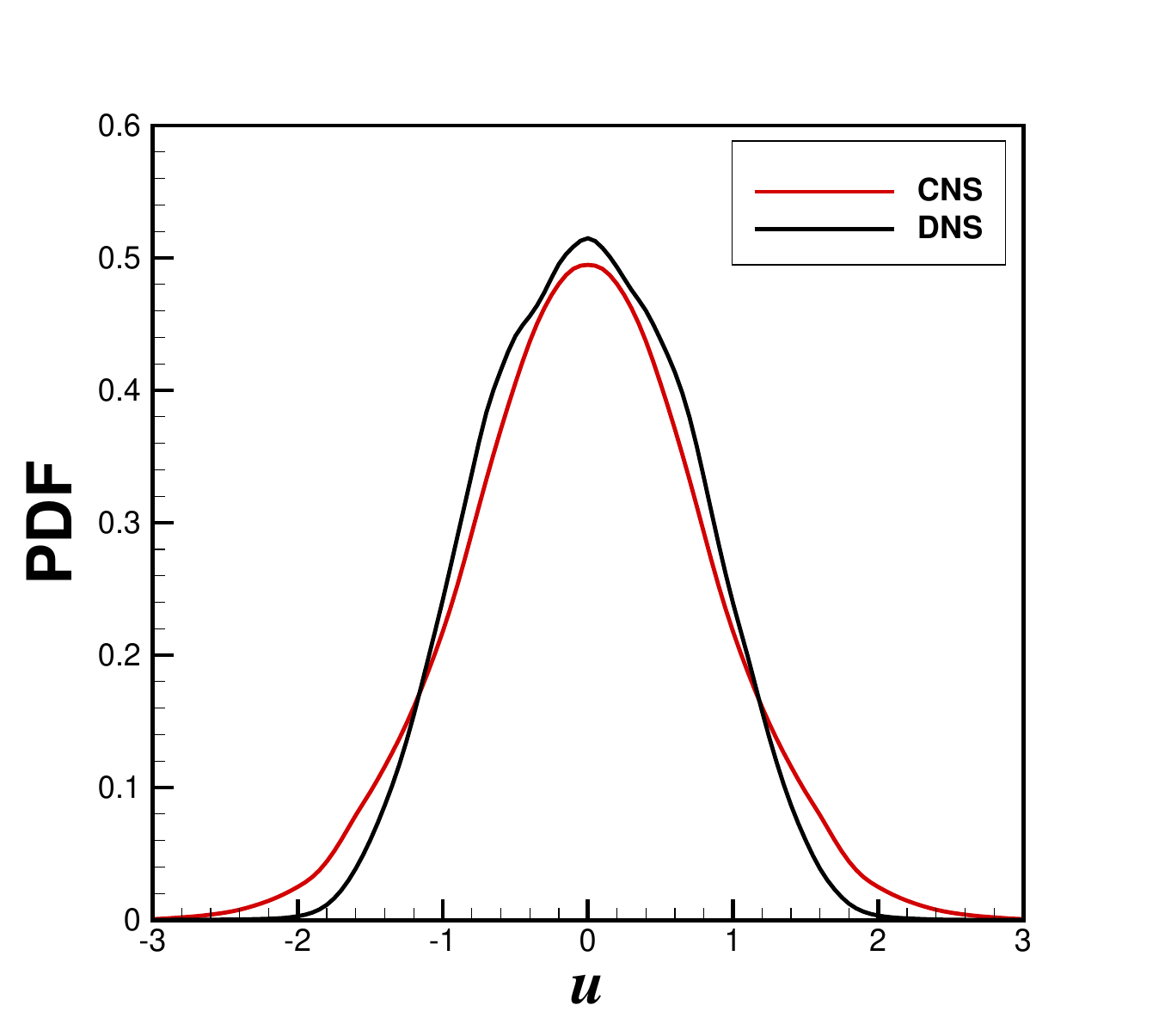}}
             \subfigure[]{\includegraphics[width=2.55in]{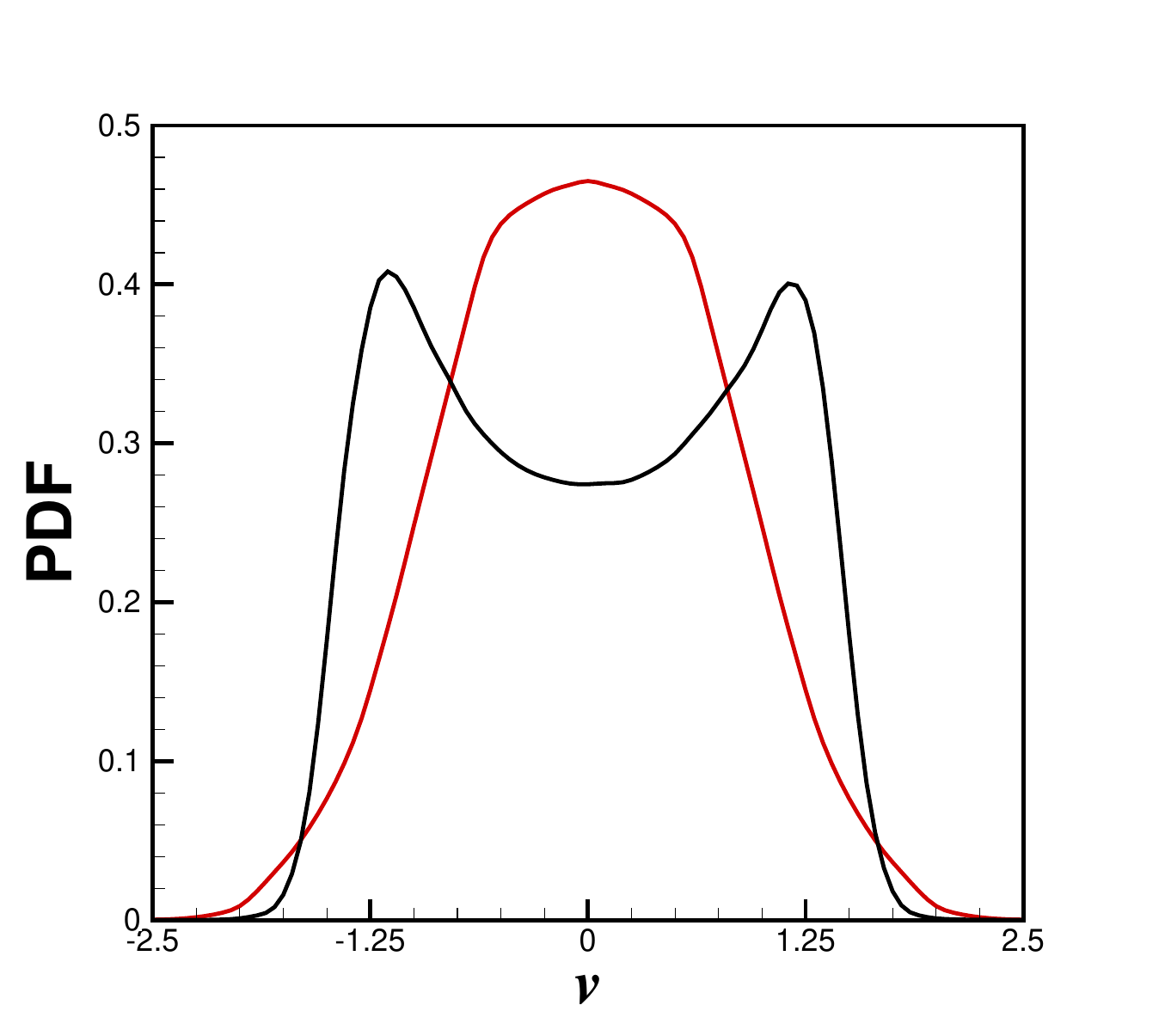}}
        \end{tabular}
    \caption{Comparisons of the PDF of (a) the horizontal velocity $u(x,y,t)$ and (b) the vertical velocity $v(x,y,t)$ of the two-dimensional Kolmogorov turbulence  in the case of $n_K=4$ and $Re=40$ with the initial condition (\ref{initial_condition}), respectively, given by the CNS  (red) and the DNS (black).}     \label{PDF_uv}
    \end{center}

    \begin{center}
        \begin{tabular}{cc}
             \includegraphics[width=2.55in]{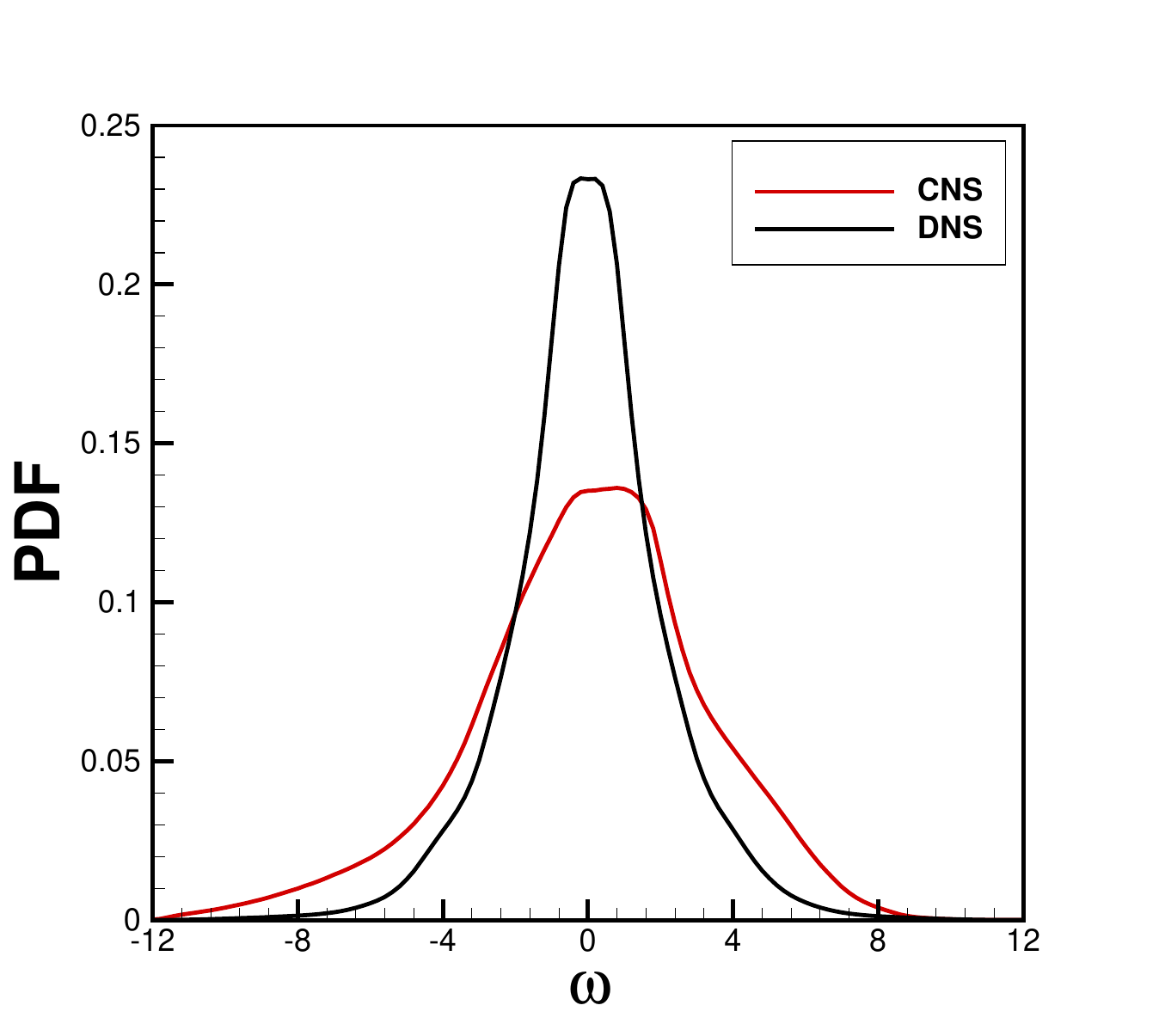}
        \end{tabular}
    \caption{Comparison of the PDF of the vorticity $\omega$ of the two-dimensional Kolmogorov turbulence  in the case of $n_K=4$ and $Re=40$ with the initial condition (\ref{initial_condition}), respectively, given by the CNS  (red) and the DNS (black).}     \label{PDF_w}
    \end{center}
\end{figure}

Furthermore, the DNS results of the enstrophy spectrum $B_l$ defined by (\ref{enstrophy_spectrum}) of the two-dimensional Kolmogorov turbulence have also huge deviation from the CNS benchmark solution,  as shown in Figure~\ref{B_t} and Figure~\ref{B_l}. 
The time histories of the enstrophy spectrums at some wave numbers, such as $B_1(t)$ and $B_{50}(t)$ as shown in Figure~\ref{B_t} (a) and (b), respectively, present the apparent sensitivity to numerical noises as a kind of artificial tiny disturbances:  the average value (about 1.13) of $B_1(t)$ given by the DNS is larger than that (about 0.60) given by the CNS, but $B_1(t)$ given by the CNS has larger variation range, which means that the flow given by the DNS contains more kinetic energy at the macroscopical scale and has a higher auto-correlation (that will be shown later in detail);
On the contrary, the average value (about $6.0 \times 10^{-12}$) of $B_{50}(t)$ given by the CNS are about 4 times larger than that (about $1.5 \times 10^{-12}$) given by the DNS, indicating that the flow given by the CNS has more energy at the microscopical scale. These phenomena can be well explained by the obvious difference between the kinetic energy dissipation rates given by the CNS and DNS, as shown in Figure~\ref{D_t}.
Besides, at some typical times such as $t=500$ and $t=1500$ as shown in Figure~\ref{B_l} (a) and (b), respectively, as $l$ increases, deviation between the spectrum $B_l$ given by the  DNS and CNS benchmark solution becomes larger and larger, and when $l > 10$  the values of $B_l$ given by the CNS are obviously higher than that given by the DNS.  Note that these agree with the results in Figure~\ref{RSp_x13} and Figure~\ref{RSp_x02}.  It suggests that, for the relatively larger wave number (i.e. smaller turbulent structure), the flow given by the CNS  should contain more kinetic energy than the DNS, which is reasonable considering the apparently higher kinetic energy dissipation rate of the CNS as shown in Figure~\ref{D_t}.  In summary, the numerical noises as a kind of artificial tiny disturbances lead to huge deviations of the enstrophy spectrum given by the DNS from the CNS benchmark solution, on {\em both} of large and small scales,  as shown in Figure~\ref{B_t} and Figure~\ref{B_l}.  

 The comparisons of the probability density function (PDF) of the kinetic energy $E$ defined by (\ref{kinetic_energy})  and the  kinetic energy dissipation rate $D$ defined by (\ref{dissipation_rate}) given by the CNS and DNS are as shown in Figure~\ref{PDF_E_D} (a) and (b), respectively.  Note that in the whole paper $x,y\in[0,2\pi]$ and  $t \in [200, 1500]$ are used to calculate the PDFs.  
 Near $E\approx 1$, the DNS result has a larger probability of the distribution  than the CNS benchmark solution, as shown in Figure~\ref{PDF_E_D} (a).
 For large $D$, the CNS benchmark solution has a larger probability of the distribution than the DNS result, as shown in Figure~\ref{PDF_E_D} (b).    
The comparisons of the vertical distributions of  the spatio-temporal averaged kinetic energy $\langle E\rangle_{x,t}$ and the spatio-temporal averaged kinetic energy dissipation rate $\langle D\rangle_{x,t}$ given by the CNS and DNS, respectively, are as shown in Figure~\ref{STA_E_D}, where $\langle \; \rangle_{x,t}$ is an operator of statistics defined by (\ref{average_xt}).
Note that the DNS result  mostly  has larger $\langle E\rangle_{x,t}$ than the CNS benchmark solution, but $\langle D\rangle_{x,t}$ given by the CNS is several times larger than that of the DNS.  
Besides, although the PDF of the horizontal velocity $u(x,y,t)$ given by the DNS is close to that given by the CNS, the PDF of the vertical velocity $v(x,y,t)$ given by the DNS is quite different from that given by the CNS, as shown in Figure~\ref{PDF_uv}.   
In addition,  the PDF of the vorticity $\omega(x,y,t)$ given by the CNS and DNS are also distinctly different:   the vorticity $\omega$ given by the DNS is more concentrated on zero, as shown in Figure~\ref{PDF_w}.  

All of the above-mentioned  comparisons between the CNS benchmark solution and the DNS result clearly indicate that tiny disturbances, resulting from the micro-level background numerical noise, can lead to huge deviations on {\em both} of large and small scales not only in spatio-temporal trajectory of the flow field but also in spatial symmetry of the vorticity field as well as statistics of many physical quantities, such as the enstrophy spectrum, the spatio-temporal average and PDFs of the velocity, vorticity, kinetic energy, and kinetic energy dissipation rate of the two-dimensional Kolmogorov turbulence considered in this paper.

\subsection{Field geometrical structures}

\begin{figure}
    \begin{center}
        \begin{tabular}{cc}
             \subfigure[]{\includegraphics[width=2.55in]{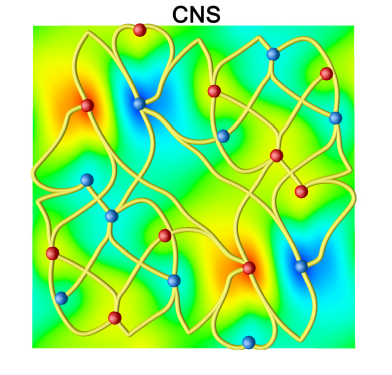}}
             \subfigure[]{\includegraphics[width=2.55in]{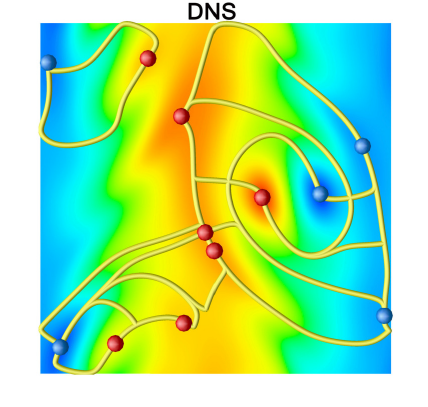}}
        \end{tabular}
    \caption{Illustration of the dissipation element (DE) structures on the instantaneous vertical velocity field $v(x, y)$ at $t=500$ for (a) the CNS benchmark solution or (b) the DNS result, where color represents the magnitude of $v$, the DE boundaries are shown in yellow solid lines, and the extremal points are marked by red (maximum) and blue (minimum) dots.}     \label{DE_t500}
    \end{center}

    \begin{center}
        \begin{tabular}{cc}
             \subfigure[]{\includegraphics[width=2.55in]{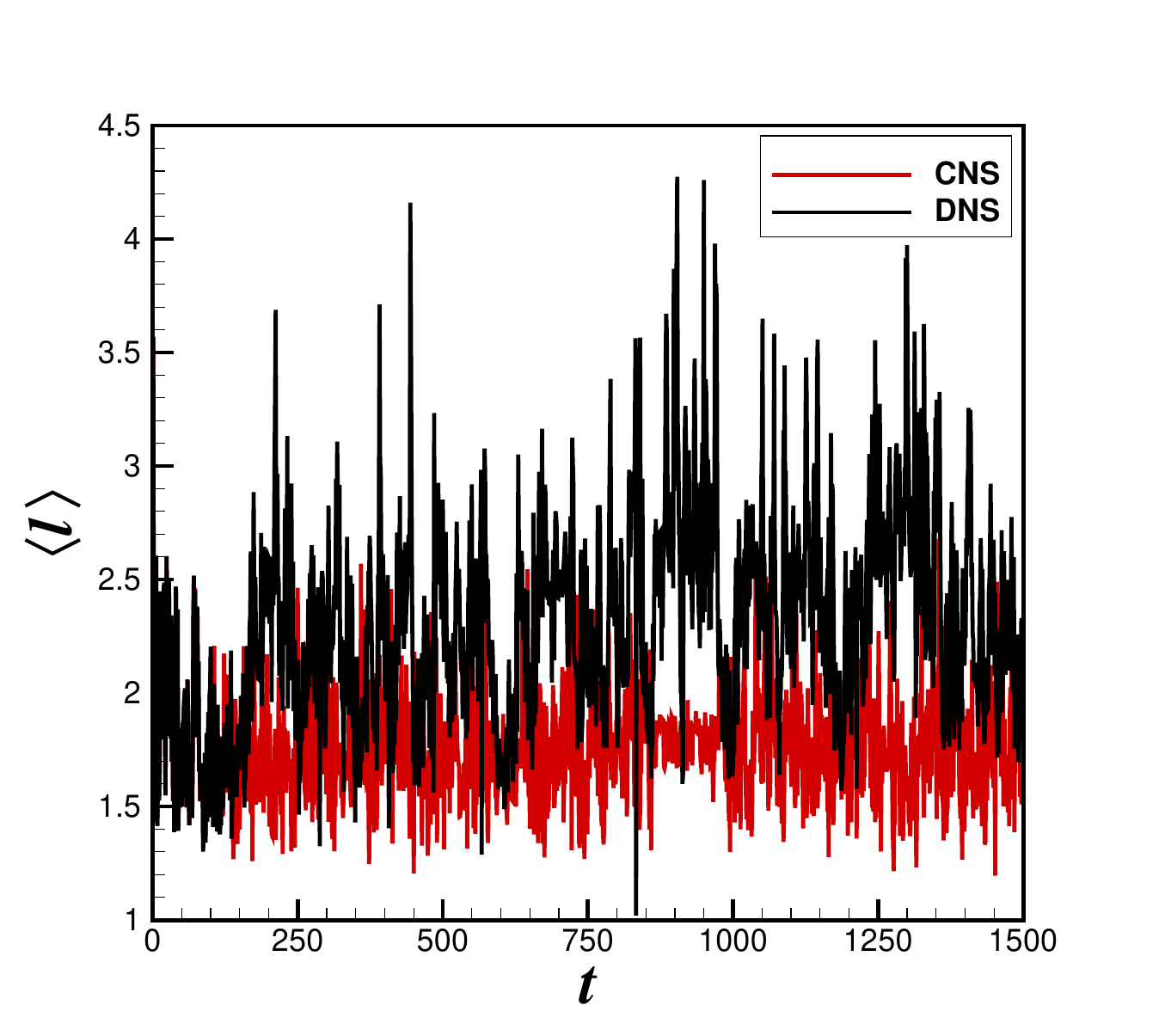}}
             \subfigure[]{\includegraphics[width=2.55in]{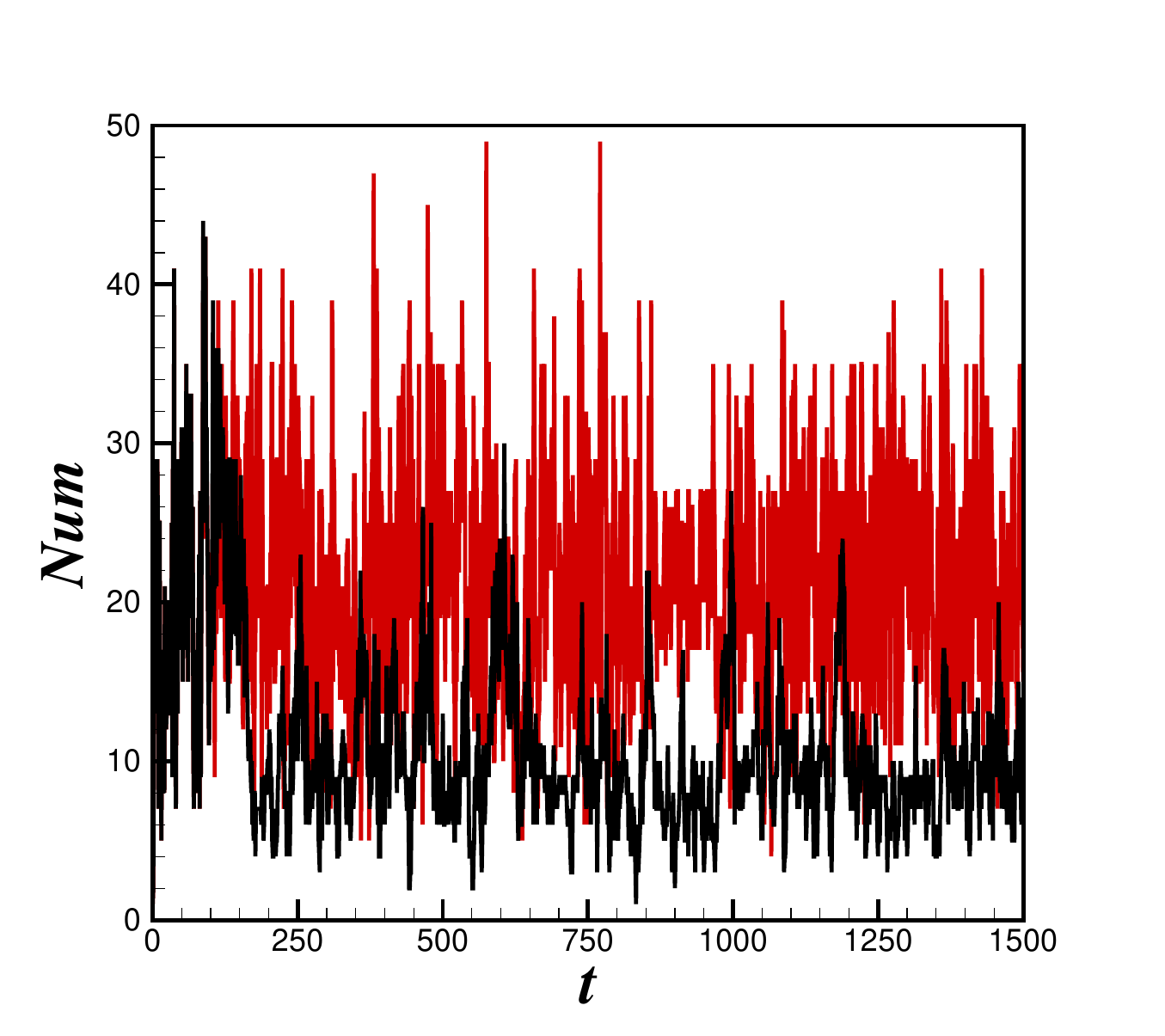}}
        \end{tabular}
    \caption{Comparison of the time histories of (a) averaged length scale $\langle l\rangle$ of the DE units or (b) total number $Num$ of the DE units, where the corresponding DE structures are obtained by means of the instantaneous vertical velocity field given by the CNS (red) or the DNS (black).}     \label{DE-t}
    \end{center}
\end{figure}

Based on the aforementioned investigation, the evolution of the flow field can also be observed through the changes in field geometrical structures.
For the non-local statistics related to the topology of flow field, the main challenge lies in the quantitative (rather than illustrative) structure identification, which is relatively rare in the existing literature. To overcome this challenge, a novel attempt for the natural decomposition of scalar fields is the dissipation element (DE) analysis \cite{jfm06, wang_peters_2008, wang2013new}.
The corresponding DE structures, as a representation of the whole flow field, are space-filling and able to effectively capture and portray the flow dynamics across the entire field. Additionally, the utilization of the DE analysis approach can avoid the defect of the so-called scale-mixing effect \cite{jfm06}, which is often encountered in traditional structure function analyses.

Specifically, starting from any spatial point in a (large enough) scalar field, along with its descending and ascending directions of the scalar gradient trajectory, one can inevitably reach a local minimum point and a local maximum point, respectively. The ensemble of spatial points whose gradient trajectories share the same pair of extremal points define a DE unit, and thus the DE structures consisting of several DE units can represent the whole flow field.
Typically, the length scale of each DE unit can be parameterized with $l$, say, the linear distance between the two extremal points. For a deeper understanding of the properties and applications of the DE analysis, please refer to  \cite{jfm06, wang_peters_2008,wang2013new,denker_2020}.

Applying the DE analysis on both the CNS benchmark solution and the DNS result, without loss of generality, for the instantaneous vertical velocity field $v(x, y)$ at $t=500$, the difference between topological properties of these two flow fields can be revealed, as shown in Figure~\ref{DE_t500}: the local maximal and minimal points are represented by the red and blue dots respectively and the DE boundaries are shown in yellow solid lines. 
Note that here we choose the vertical velocity field rather than the horizontal one mainly for exhibiting an obvious deviation between the CNS benchmark solution and the DNS result, which corresponds to the results shown in Figure~\ref{PDF_uv}.
At this moment, the sizes of the DE structures for the DNS result are generally larger than those given by the CNS in the same fixed computational domain. Quantitatively, on the one hand, the evolution of the averaged length scale, i.e. $\langle l\rangle$, of the DE units (obtained by means of each instantaneous vertical velocity field) is shown in Figure~\ref{DE-t}(a), which indicates that the averaged length scales at different time moments given by the DNS are generally larger than those of the CNS benchmark solution after an initial period of time. On the other hand, Figure~\ref{DE-t}(b) illustrates that the total number (denoted by $Num$) of the DE units (obtained likewise via each instantaneous vertical velocity field) given by the DNS is less than that given by the CNS most of the time, which is a natural result corresponding to the difference between length scales mentioned above. Note that all of these results can be explained from a perspective of the kinetic energy dissipation rate (as shown in Figure~\ref{D_t} and Figure~\ref{PDF_E_D}(b)): considering the definition of a DE unit, the lower kinetic energy dissipation rate given by the DNS usually corresponds to the smoother and larger DE structures.

In summary, the numerical noises as a kind of artificial tiny disturbances leads to  quantitatively large deviations on the field geometrical structures of the two-dimensional Kolmogorov turbulence under consideration.  

\subsection{Intermittent stability}

\begin{figure}
    \begin{center}
        \begin{tabular}{cc}
             \subfigure[]{\includegraphics[width=2.55in]{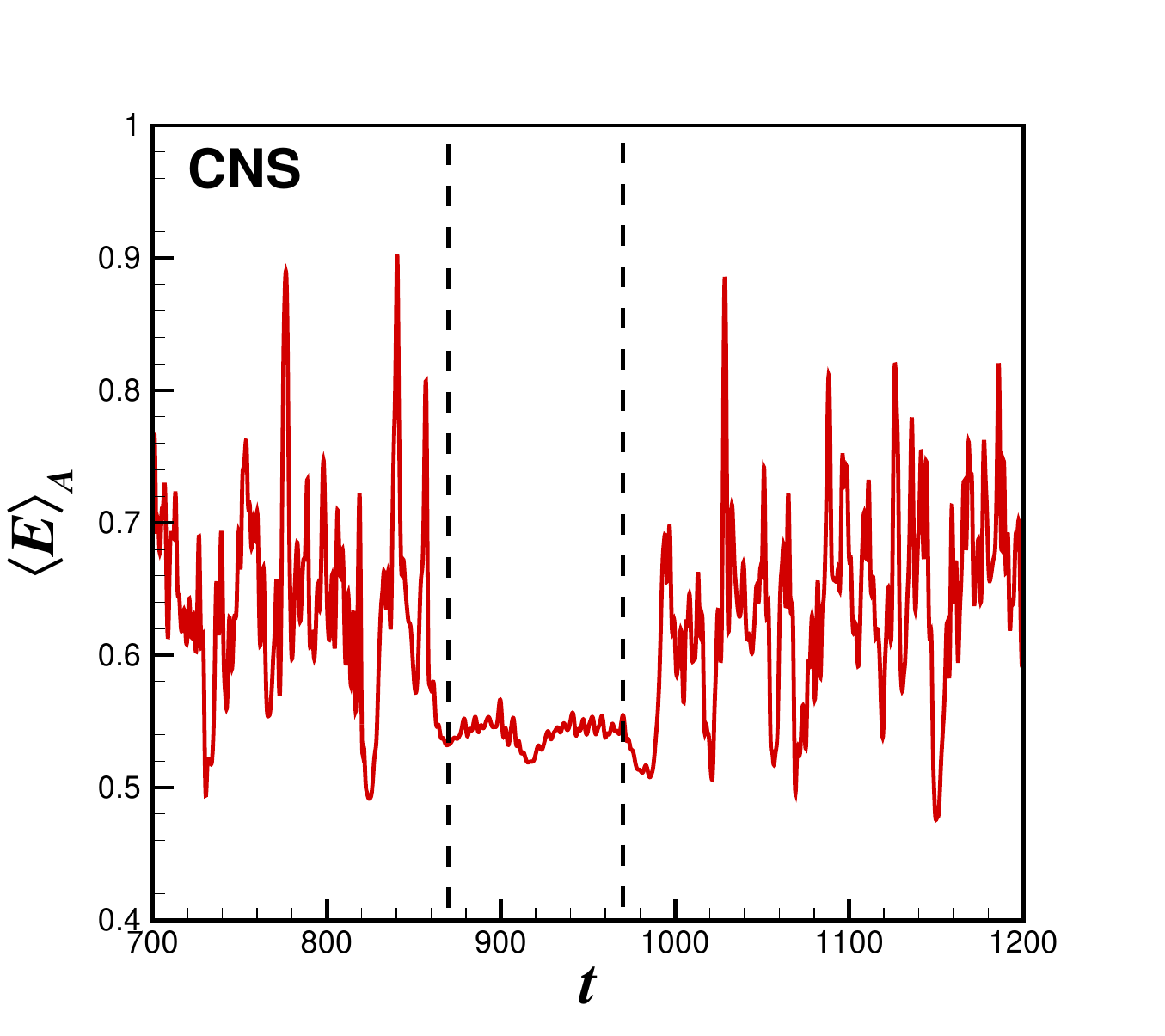}}
             \subfigure[]{\includegraphics[width=2.55in]{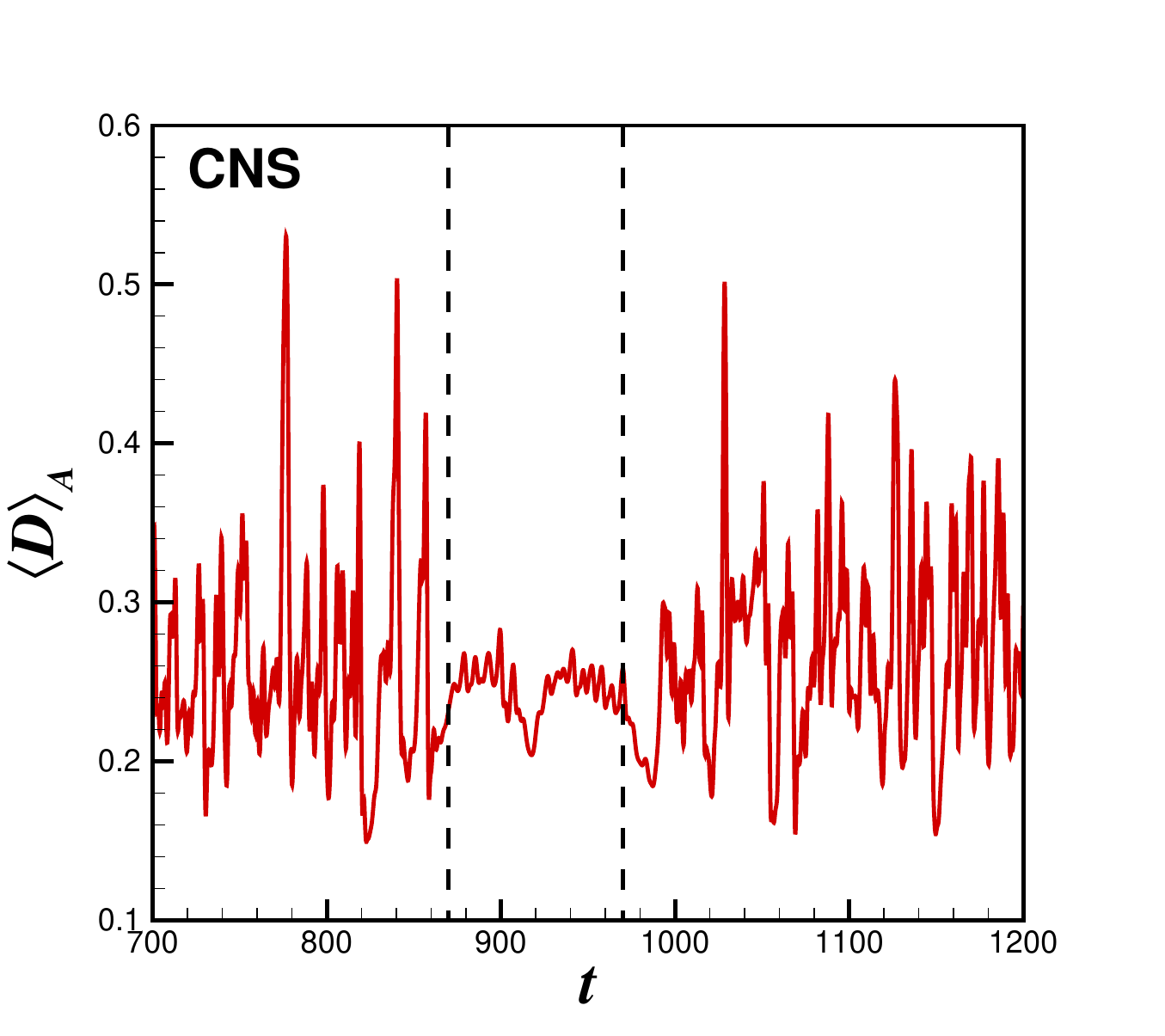}}    \\
             \subfigure[]{\includegraphics[width=2.55in]{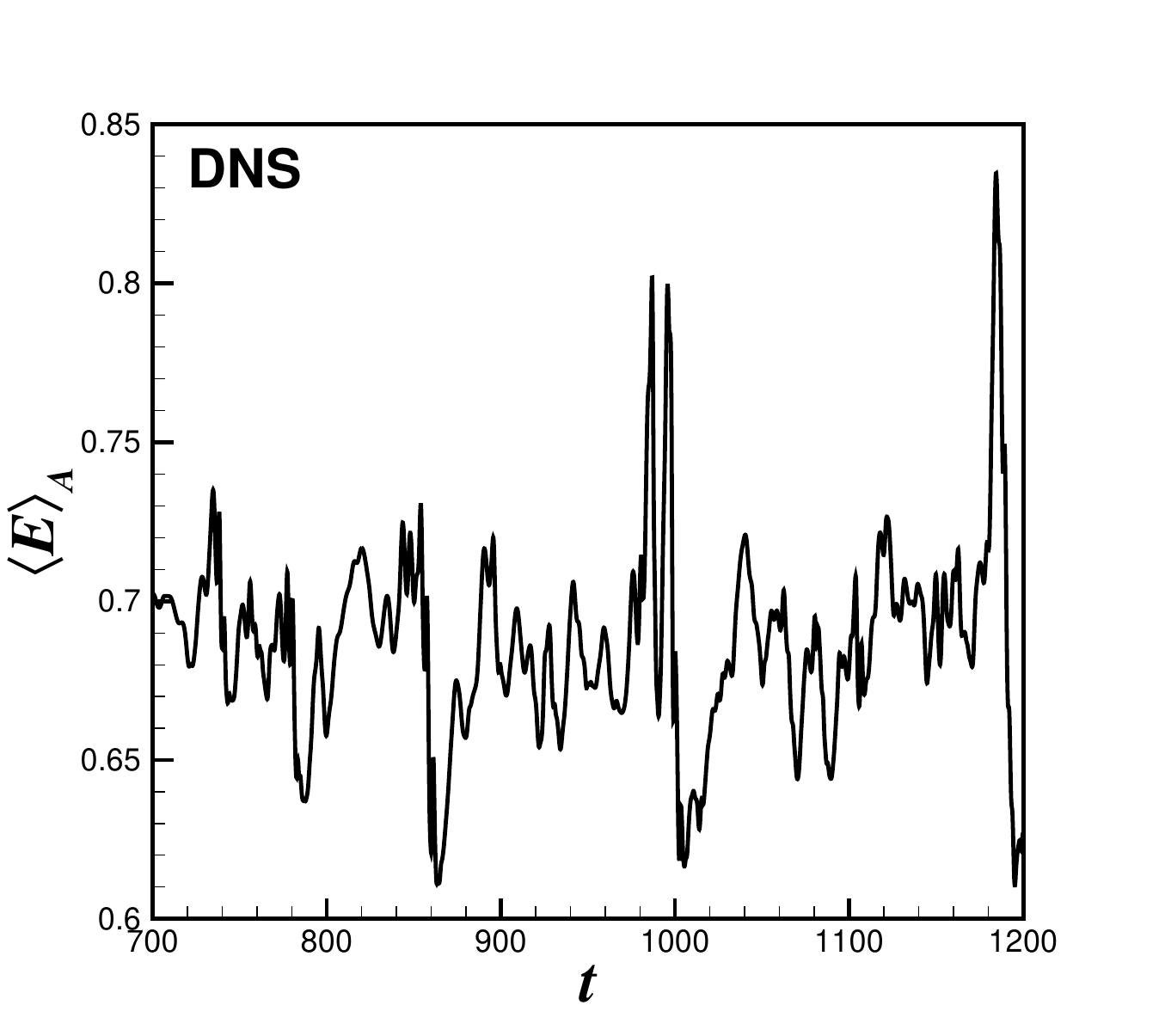}}
             \subfigure[]{\includegraphics[width=2.55in]{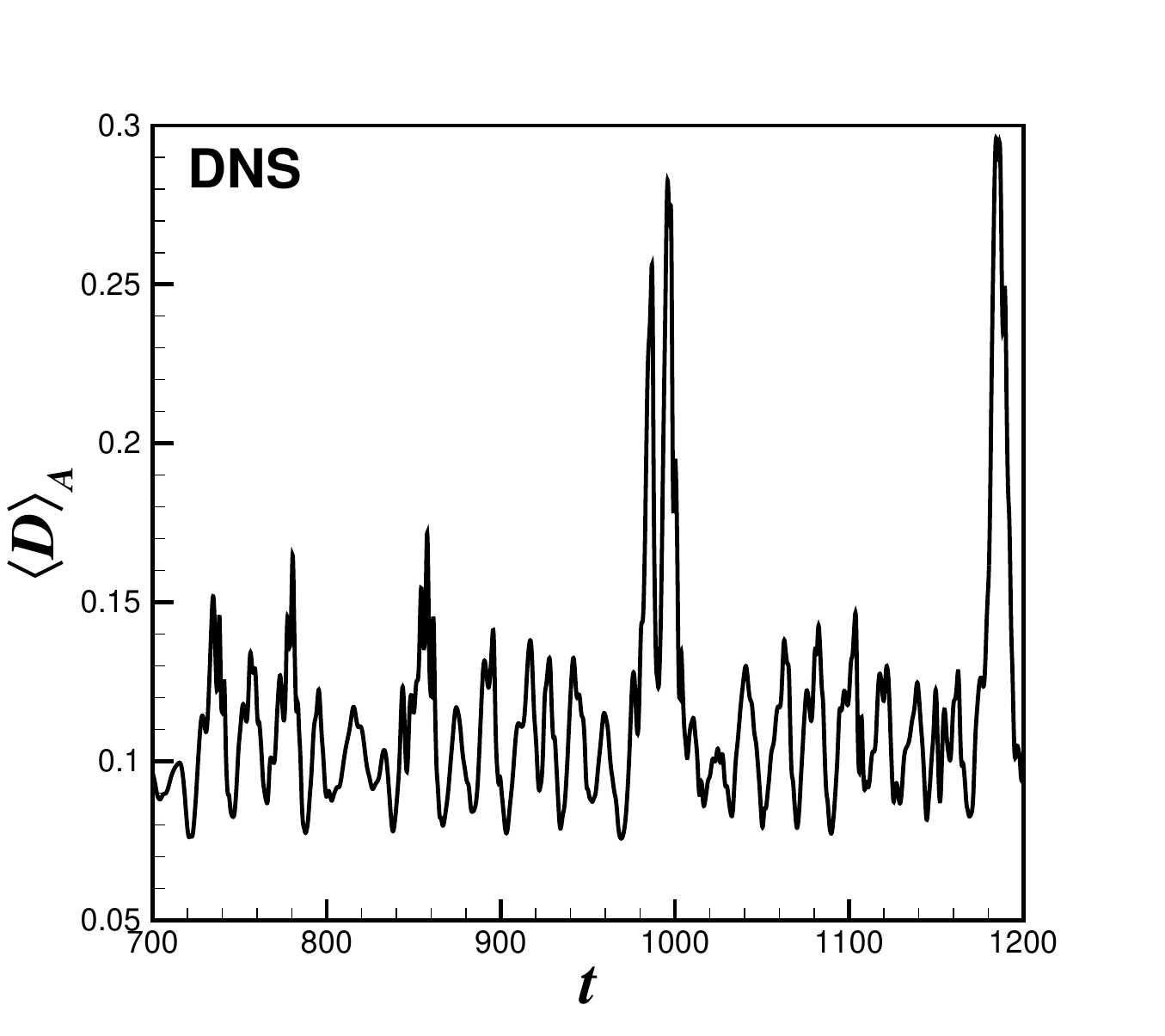}}    \\
        \end{tabular}
    \caption{Time histories of $\langle E\rangle_A$ and  $\langle D\rangle_A$ of the two-dimensional Kolmogorov turbulence  in the case of $n_K=4$ and $Re=40$ with the initial condition (\ref{initial_condition}) given by the (a)-(b) CNS and (c)-(d) DNS in $t \in [700, 1200]$, respectively, where the period of time $870<t<970$ corresponds to the intermittent stability of Kolmogorov turbulence.}     \label{ED_t}
    \end{center}
\end{figure}

\begin{figure}
    \begin{center}
        \begin{tabular}{cc}
             \subfigure[]{\includegraphics[width=2.55in]{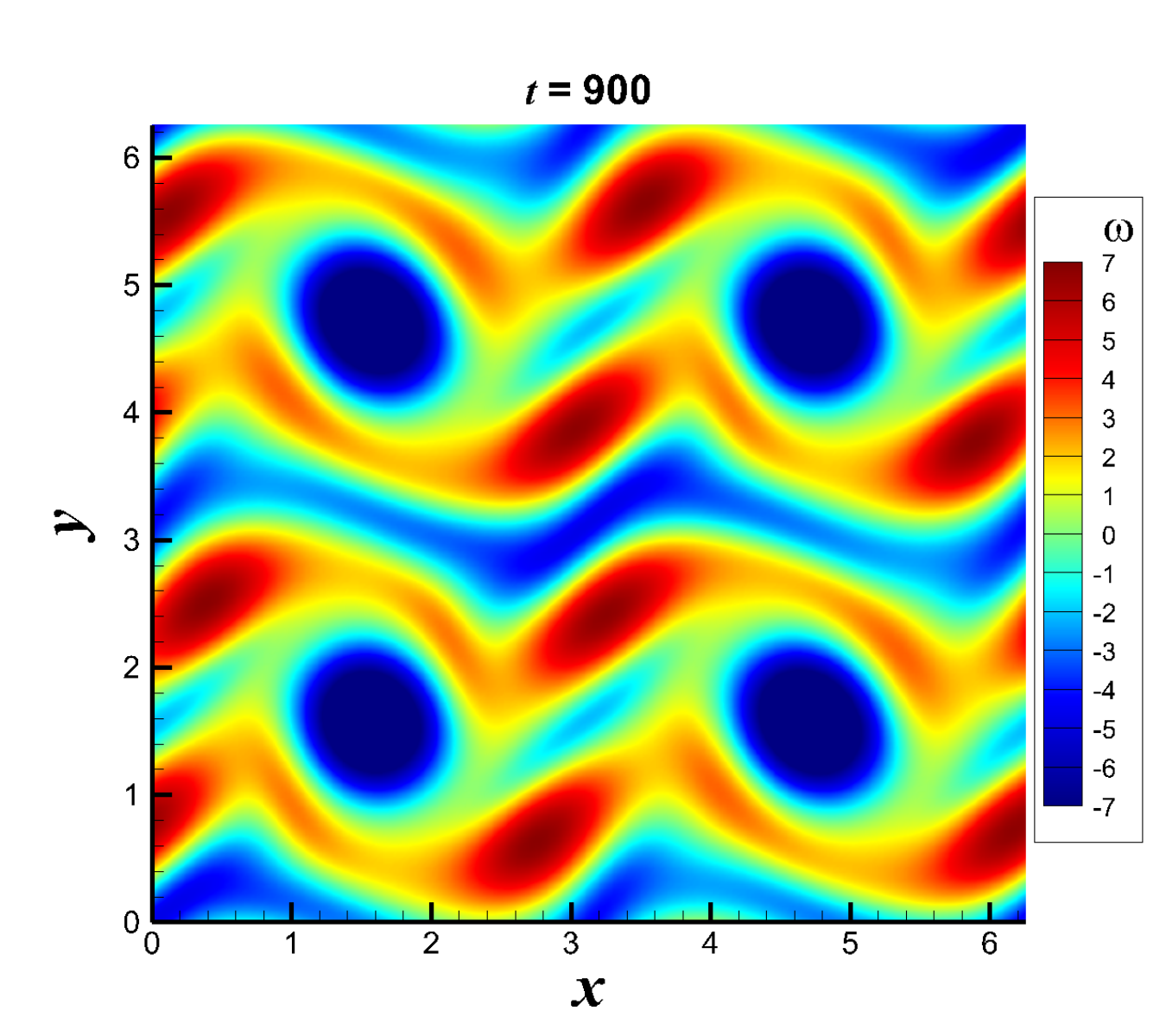}}
             \subfigure[]{\includegraphics[width=2.55in]{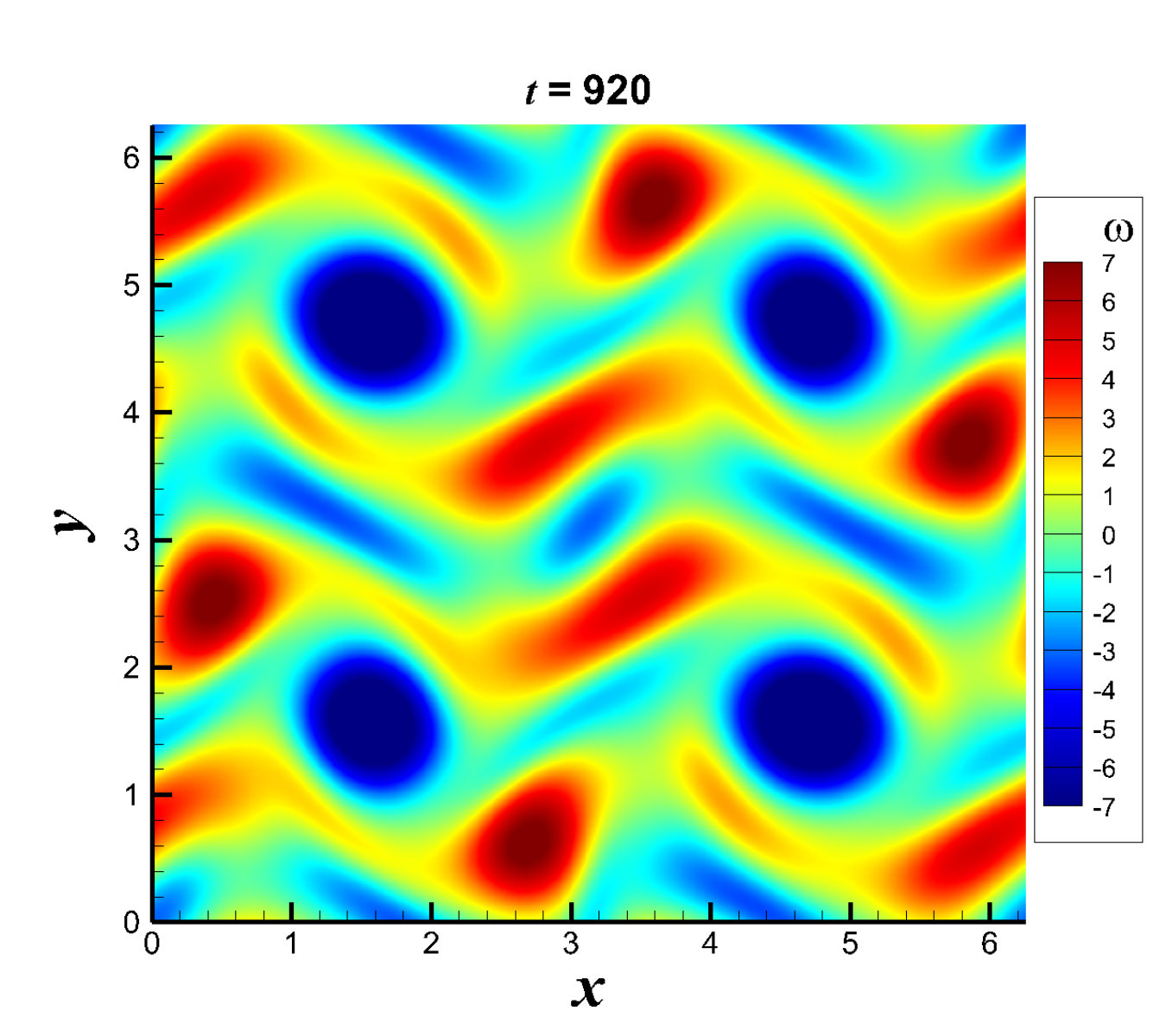}}    \\
             \subfigure[]{\includegraphics[width=2.55in]{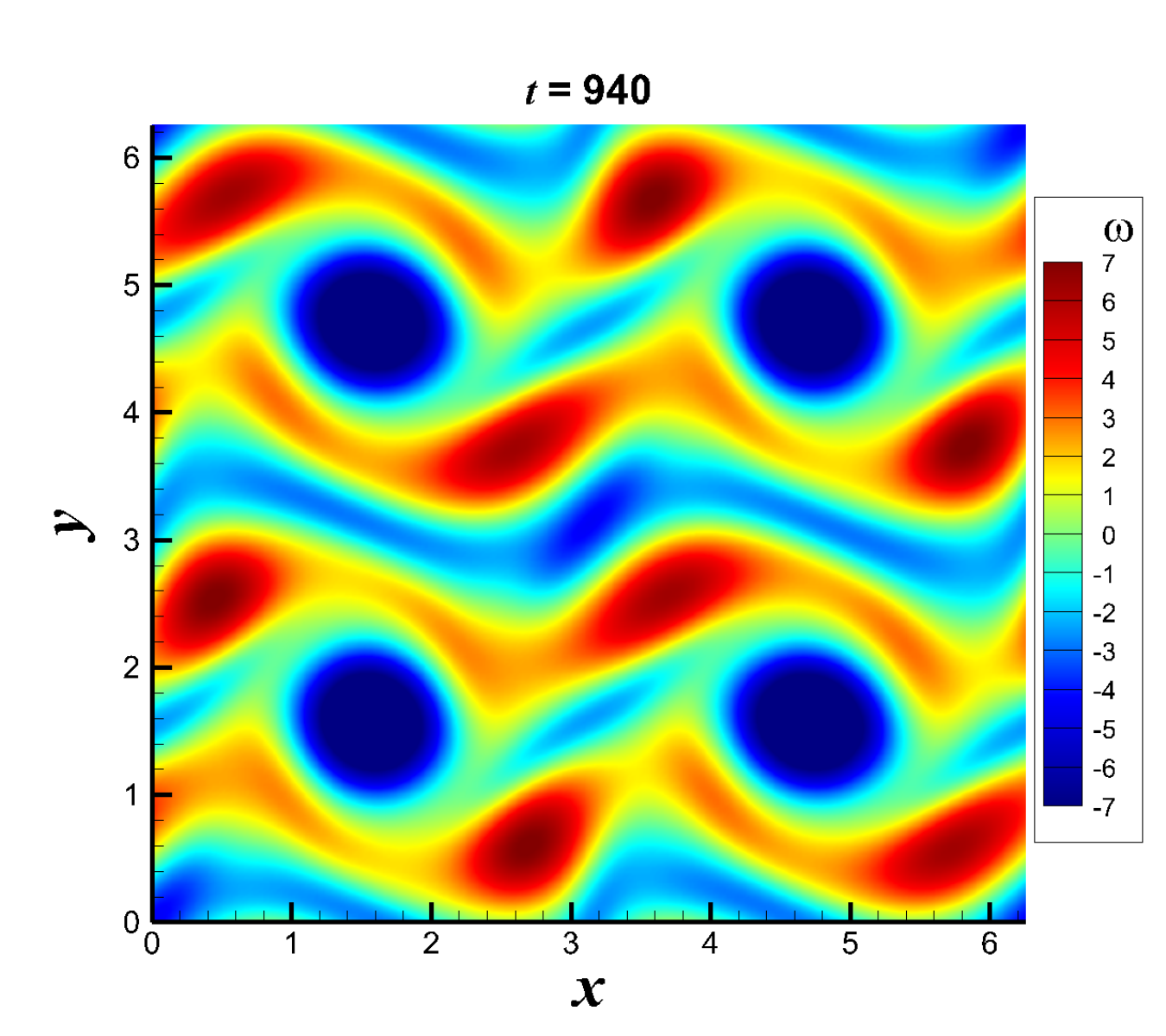}}
             \subfigure[]{\includegraphics[width=2.55in]{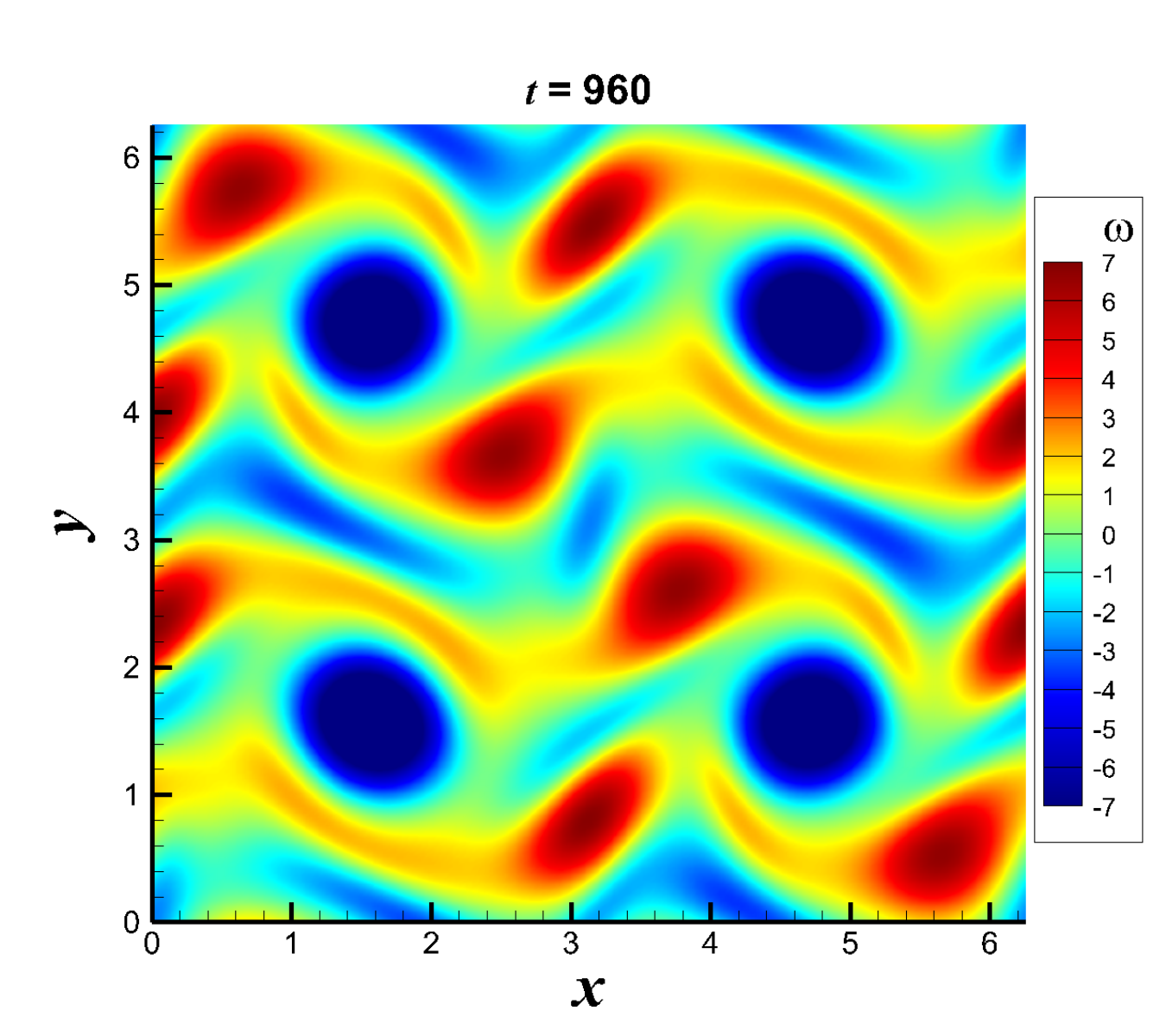}}    \\
        \end{tabular}
    \caption{Vorticity field $\omega$ of the two-dimensional Kolmogorov turbulence in the case of $n_K=4$ and $Re=40$ with the initial condition (\ref{initial_condition}) given by the CNS in the state of the intermittent stability at (a) $t=900$, (b) $t=920$, (c) $t=940$, and (d) $t=960$.}     \label{Vor_stable}
    \end{center}
\end{figure}

According to the time histories of $\langle E\rangle_A$ and $\langle D\rangle_A$ in $t \in [700, 1200]$ given by the CNS benchmark solution, as shown in Figure~\ref{ED_t}(a) and (b), there exists a short period of time, for example,  $870<t<970$, in which the two-dimensional Kolmogorov turbulence is in a relatively stable state, corresponding  to an intermittent stability of turbulence \cite{costa2011simplified, gogia2017emergent, gogia2020intermittent}.    Note that the vorticity fields $\omega$ given by the CNS benchmark solution at $t=900$, $t=920$, $t=940$, and $t=960$  are almost unchanged, as shown in Figure~\ref{Vor_stable}(a)-(d).  By contrast, the corresponding vorticity fields given by the DNS has {\em no} such kind of  intermittent stability of turbulence in the {\em whole} period of simulation!   

It should be emphasized that the DNS result quickly becomes a mixture of the `false' numerical noise $\delta$ and the `true' physical solution $p$, which are at the same order of magnitude mostly, i.e. $\delta \sim p$.  
On the contrary, the numerical noise $\delta$ of the CNS benchmark solution is much smaller than the `true' physical solution $p$, i.e. $|\delta| \ll |p|$, and thus negligible, so that the CNS benchmark solution is very close to the `true' physical solution $p$.   
As illustrated above in this section, the DNS result of the two-dimensional  Kolmogorov turbulence has significant deviations from the CNS benchmark solution not only in trajectory but also in statistics of  many physical variables.  

Note that the intermittent stability of turbulence in $870<t<970$ can be successfully found by the CNS, but is unfortunately lost by the DNS.  This indicates once again that CNS should be a powerful tool to investigate turbulent flows more accurately than DNS.  Hopefully, CNS as a new method could bring us some new discoveries in turbulence.     

\subsection{Scale-to-scale energy flux}

To delve into the underlying physical mechanisms, we employ the Filter-Space-Technique (FST) to extract the scale-to-scale energy flux, denoted as $\Pi^{[l]}$ (see definition below). FST, initially developed for large eddy simulation in the 1970s \cite{Leonard1975AG}, involves applying a low-pass filter to the velocity field. Mathematically, this is expressed as:
\begin{equation}
f^{[l]}(\mathbf{x},t) = \int G^{[l]}(\mathbf{x}-\mathbf{x}') f(\mathbf{x}',t) d \mathbf{x}',
\end{equation}
where $f=\mathbf{u}(\mathbf{x},t)$ represents a vector of the two-dimensional velocity field, and the Gaussian kernel $G^{[l]}(r)=\exp(-r^2/2l^2)$ is commonly used as the filter \cite{Chen2003PRL, Boffetta2012ARFM, zhou2016JFM}.
For the incompressible Navier-Stokes equation, the scale-to-scale energy flux can be derived analytically as:
\begin{equation}
\Pi^{[l]}(\mathbf{x},t) = -\sum_{i,j=1,2} \left[ \left( u_iu_j \right)^{[l]} -u_i^{[l]}u_j^{[l]}\right]\frac{\partial u_i^{[l]}}{\partial x_j}
\end{equation}
where $u_1=u$, $u_2=v$ are velocity components and $x_1=x$, $x_2=y$ denote spatial coordinates. 
This term emerges from filtering the nonlinear terms in Navier-Stokes equations, capturing the interaction between scales removed by the filter and those that remain. Crucially, the sign of $\Pi^{[l]}$ reveals the direction of energy transfer: a positive value indicates a cascade from larger scales ($>l$) to smaller scales ($<l$), while a negative value signifies the reverse. Thus, $\Pi^{[l]}$ provides valuable insights into both the direction and intensity of energy transfer across scales in turbulent flows.

\begin{figure}
    \begin{center}
        \begin{tabular}{cc}
             \subfigure[]{\includegraphics[width=2.55in]{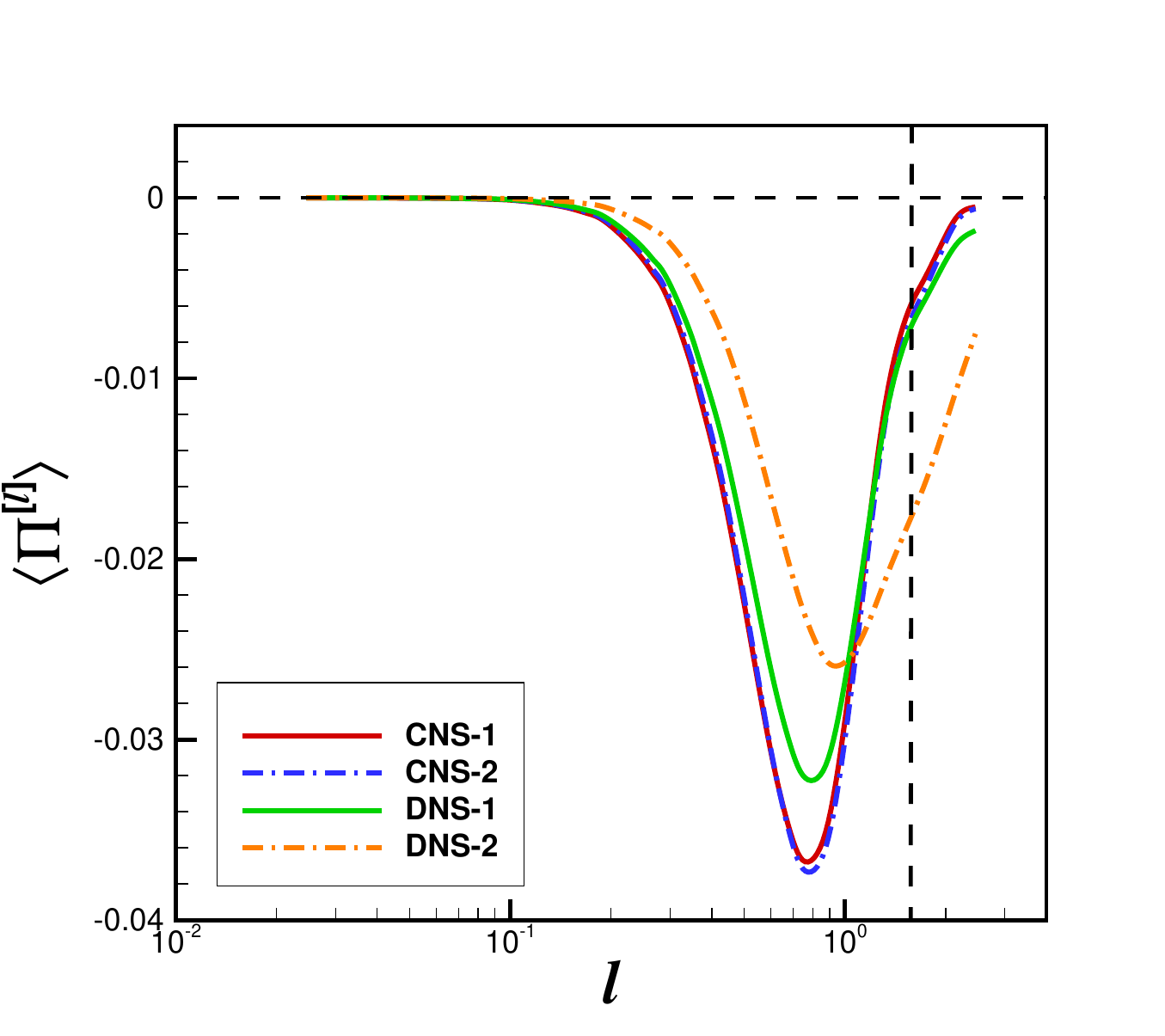}}
             \subfigure[]{\includegraphics[width=2.55in]{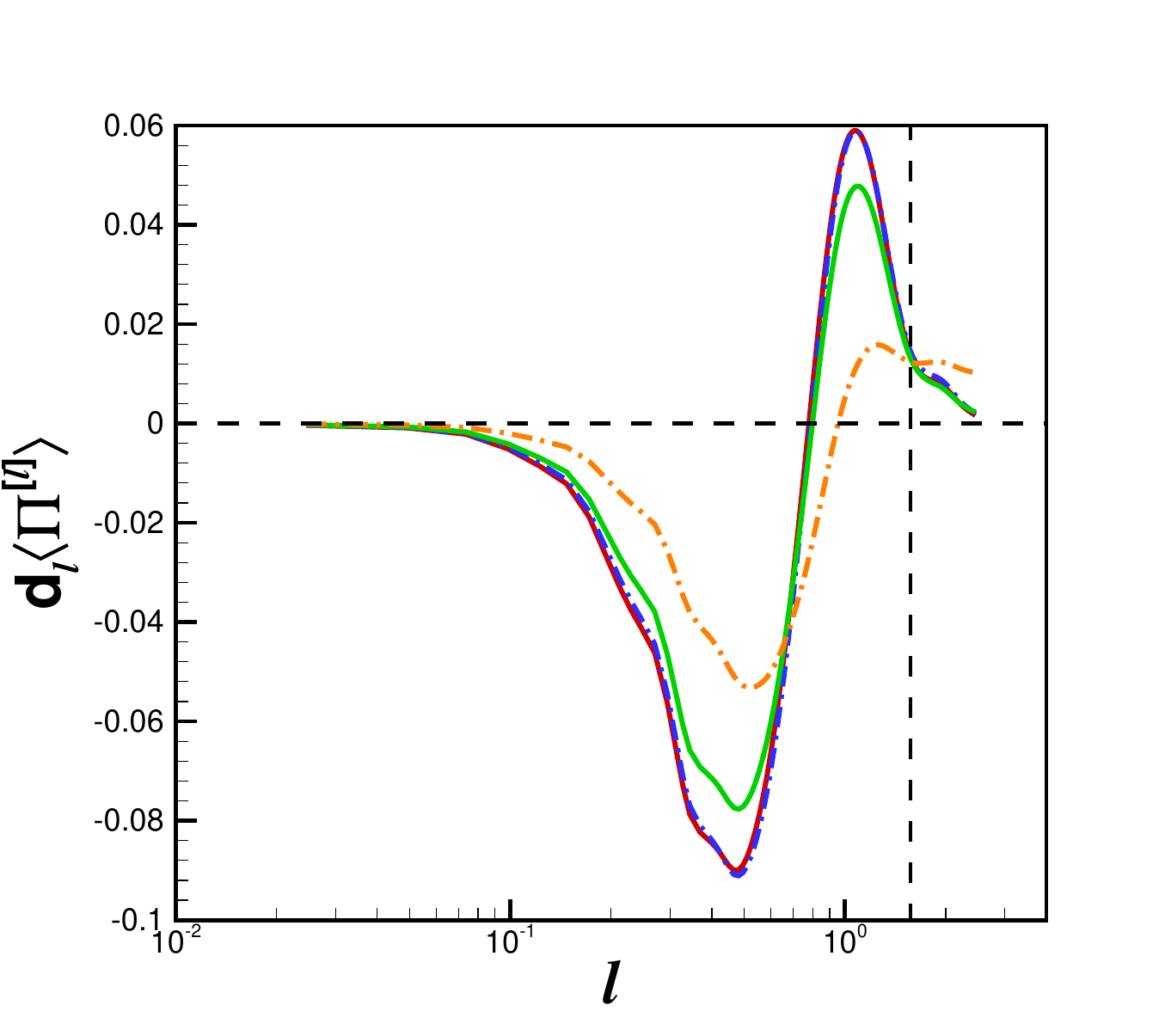}}    \\
        \end{tabular}
    \caption{(a) Averaged scale-to-scale energy flux $\langle \Pi^{[l]} \rangle$ and (b) the corresponding derivative $\mathrm{d}_{l} \langle \Pi^{[l]} \rangle$ versus the scale $l$ of the two-dimensional Kolmogorov turbulence investigated in this paper, where the vertical dash line in black corresponds to the forcing scale $l_f=2\pi /n_K=1.57$ and the horizontal one denotes the zero value of vertical coordinate. These averaged results are integrated in $t \in [50, 100]$ (solid lines, i.e. `-1') and $t \in [110, 250]$ (dash and dot lines, i.e. `-2') by means of the CNS benchmark solution (red and blue lines) and the DNS result (green and orange lines).}     \label{fig:scale2scale}
    \end{center}
\end{figure}

To quantify the energy transfer across different scales, we apply FST to the velocity field in this investigation, extracting $\Pi^{[l]}$ for scales ranging from $0$ to $2.45$. Figure~\ref{fig:scale2scale}(a) depicts the averaged scale-to-scale energy flux $\langle \Pi^{[l]} \rangle$ as a function of the scale $l$ for two distinct time periods: $50\leq t\leq 100$ and $110\leq t\leq 250$, where the spatio-temporal average $\langle \, \rangle$ is defined by (\ref{average_all}).
For the two-dimensional Kolmogorov turbulence investigated in this paper, the consistently negative values of $\langle \Pi^{[l]} \rangle$ across all scales confirm the presence of an inverse energy cascade that is a characteristic of two-dimensional turbulence.
Notably, while the CNS curves remain {\em unchanged} between these two time periods, a marked deviation is observed between the DNS curves, which corresponds to the emergence of the above-mentioned qualitative difference of the flow field:  when $t \geq120$, the DNS result is badly polluted by numerical noises, as shown in Figure~\ref{Vor_Evolutions} and Figure~\ref{Vor_Evolutions_More}. Furthermore, FST reveals a significant discrepancy in $\langle \Pi^{[l]} \rangle$ at scales around $l \approx l_f/2$ (where $l_f=2\pi /n_K=1.57$ denotes the forcing scale) between the results obtained by the CNS and DNS, even in the early stages (such as $50\leq t\leq 100$) when the visible difference between these two cases' velocity fields is absent, which is particularly intriguing and of course caused by the numerical noise.

Considering that the CNS has greatly reduced the background numerical noise so that the corresponding result can be regarded as a `clean' benchmark solution, these findings, coupled with the observed inverse energy cascade, provides rigorous evidence that the deviation caused by the  numerical noise of the DNS first manifest at the smallest scales and subsequently propagate upwards to larger scales along with the inverse energy cascade in this two-dimensional Kolmogorov turbulence. This amplification process eventually leads to the emergence of significant differences at larger scales as time progresses.

To delve deeper into the intricacies of the energy cascade loop, we turn our attention to the local balance of energy flux at a given scale $l$, as described by the following equation \cite{frisch1995turbulence}:
\begin{equation}
\mathrm{d}_{l} \Pi^{[l]} = \epsilon_{\nu}(l) - E_{\mathrm{in}}(l).
\end{equation}
In this equation, $\mathrm{d}_{l}$ stands for the derivation with respect to $l$, $\epsilon_{\nu}(l)$ represents the energy dissipation density function due to fluid viscosity, while $E_{\mathrm{in}}(l)$ denotes the energy injection density function resulting from external forces. 
It is worth noting that $\epsilon_{\nu}(l)$ must be greater than or equal to zero, and it approaches zero for scales significantly larger than the Kolmogorov scale (such as its minimum $\eta_{min}\approx 0.042$ mentioned in \S~3.5). This allows us to simplify the equation as follows:
\begin{equation}
E_{\mathrm{in}}(l) \approx -\mathrm{d}_{l} \Pi^{[l]}.
\end{equation}
Figure~\ref{fig:scale2scale}(b) presents the measured $\mathrm{d}_{l} \langle \Pi^{[l]} \rangle$ versus the scale $l$ in this two-dimensional Kolmogorov turbulence, providing insights into energy injection and extraction processes. The results reveal that kinetic energy is injected into the system within the scale ranges of $0.1\lesssim l \lesssim 0.7$ for the CNS and $0.1\lesssim l \lesssim 1.0$ for the DNS. However, when scales exceed these thresholds ($0.7$ for the CNS and $1.0$ for the DNS), kinetic energy is extracted by external forcing. This extraction is attributed to the alignment of the velocity field and the external force with two opposite phases at large scale.  It is worth noting that this pattern of energy injection and extraction has also been observed in other two- and quasi-two-dimensional systems of turbulence, which is a topic that we plan to explore in greater detail in the future.

The inverse energy cascade has been observed in various (quasi) 2D flow systems, including Kraichnan turbulence  \cite{Kraichnan1967, Boffetta2012ARFM}, bacterial turbulence \cite{Wang2017PRE}, oceanic flows  \cite{Vallis2017Book, Lovejoy2019Book, Zhang2024FMS}, and the dynamics of Jupiter's and Saturn's weather layers \cite{Read2024}. Surprisingly, these 2D systems retain intricate dynamics despite dimensionality reduction. Our findings here {\em challenge} the perception of reduced complexity in lower dimensions and suggests that energy transfer pathways can be {\em more} intricate than often assumed. For example, an external force may not always inject energy; it could pump energy into the cascade at an intermediate scale while simultaneously extracting energy at larger scales to balance the inverse cascade.  A similar pattern was observed in oceanic surface flows under the Coriolis force  \cite{Zhang2024FMS}. Ultimately,  the final status may depend on intricate small-scale interactions, as our results demonstrate.  A systematical study of other two- and quasi-two-dimensional systems of turbulence will be reported in future.


\subsection{Validity of direct numerical simulation (DNS)}    

Traditionally, DNS results of Navier-Stokes equations  are widely regarded as `reliable' benchmark solutions of turbulence, as long as grid spacing is fine enough (i.e. less than the minimum Kolmogorov scale) \cite{pope2001turbulent} and time-step is small enough, say, satisfying the Courant-Friedrichs-Lewy condition (Courant number $<$ 1)  \cite{Courant1928}.   Is this {\em definitely} true?  

According to our CNS  result, the maximum kinetic energy dissipation rate $D_{max}=5.095$ appears at $t=659.6$, corresponding to the minimum Kolmogorov scale \cite{pope2001turbulent}, say, 
\begin{align}
& \eta_{min}\approx Re^{-3/4}(D_{max})^{-1/4}=0.042.    \label{Kolmogorov_scale}
\end{align}
Note that  $\Delta_x=\Delta_y=2\pi/N_{x}  \approx 0.0245$.  Thus, 
the famous criterion on the grid spacing \cite{pope2001turbulent}, i.e.    
\begin{equation}
\Delta_x = \Delta_y <  \eta_{min},    \label{criterion-grid}
\end{equation} 
is satisfied for the CNS.  It should be emphasized that, for the DNS result,  the maximum kinetic energy dissipation rate $D_{max}$ is about 4.372, corresponding to the minimum Kolmogorov scale $ \eta_{min} = 0.043$,  so that the grid spacing  criterion (\ref{criterion-grid}) is also satisfied.  Therefore,  the spatial resolution adopted in this paper is fine enough for {\em both} of the CNS and DNS of the two-dimensional Kolmogorov turbulence under consideration. 

Besides, it is well-known that a small enough time-step is needed for DNS of Navier-Stokes equations, say, the so-called Courant-Friedrichs-Lewy condition, i.e. Courant number $<$ 1,   must be satisfied \cite{Courant1928}.   For the two-dimensional Kolmogorov turbulence under consideration, the corresponding Courant number is 0.16 for CNS and 0.016 for DNS: {\em both} of them satisfy Courant-Friedrichs-Lewy condition!  Therefore, the time-step used for {\em both} of DNS and CNS are small enough, too.      

Unfortunately,   even so,  the DNS result has a huge deviation on both of small and large scales from the CNS benchmark solution {\em not only} in trajectory and spatial symmetry {\em but also} even in statistics, as reported above.  Especially, as mentioned in \S~\ref{geq-initial},  the true solution of the NS equations under consideration has the spatial symmetry (\ref{symmetry}), but unfortunately the DNS loses this kind of spatial symmetry when $t>120$, while the CNS remains this kind of spatial symmetry in the whole interval of time $t\in[0,1500]$: this clearly indicates the large deviation of the DNS results (when $t>120$) from the true solution of the NS equations!     
This highly suggests that even the fine enough grid spacing satisfying the spacing criterion  (\ref{criterion-grid})  with small enough time-step satisfying the Courant-Friedrichs-Lewy condition \cite{Courant1928}  could {\em not} guarantee the reliability and correction of DNS results: it is only a necessary condition but {\em not} a sufficient condition for the validity of the DNS!   This also provides us a theoretical evidence that the hypothesis $\langle p + \delta \rangle = \langle p \rangle$ does not stand up for the two-dimensional Kolmogorov turbulence under consideration, where $\langle \; \rangle$ denotes an operator of statistics, $p + \delta$ is the result of the NS equations given by the DNS, $p$ is the `true' physical solution of the NS equations, respectively.    
 
 \begin{figure}
    \begin{center}
        \begin{tabular}{cc}
             \subfigure[]{\includegraphics[width=2.55in]{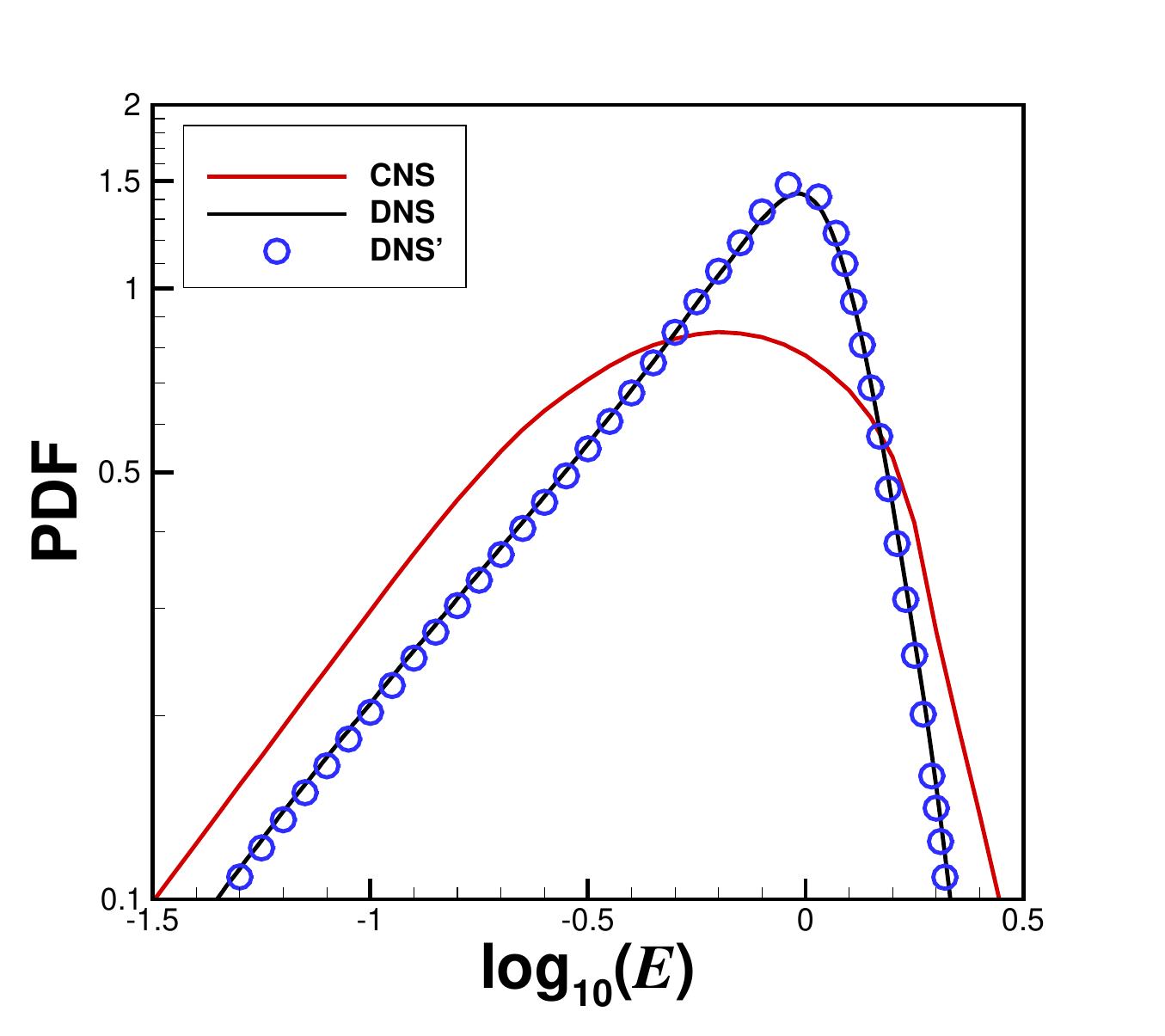}}
             \subfigure[]{\includegraphics[width=2.55in]{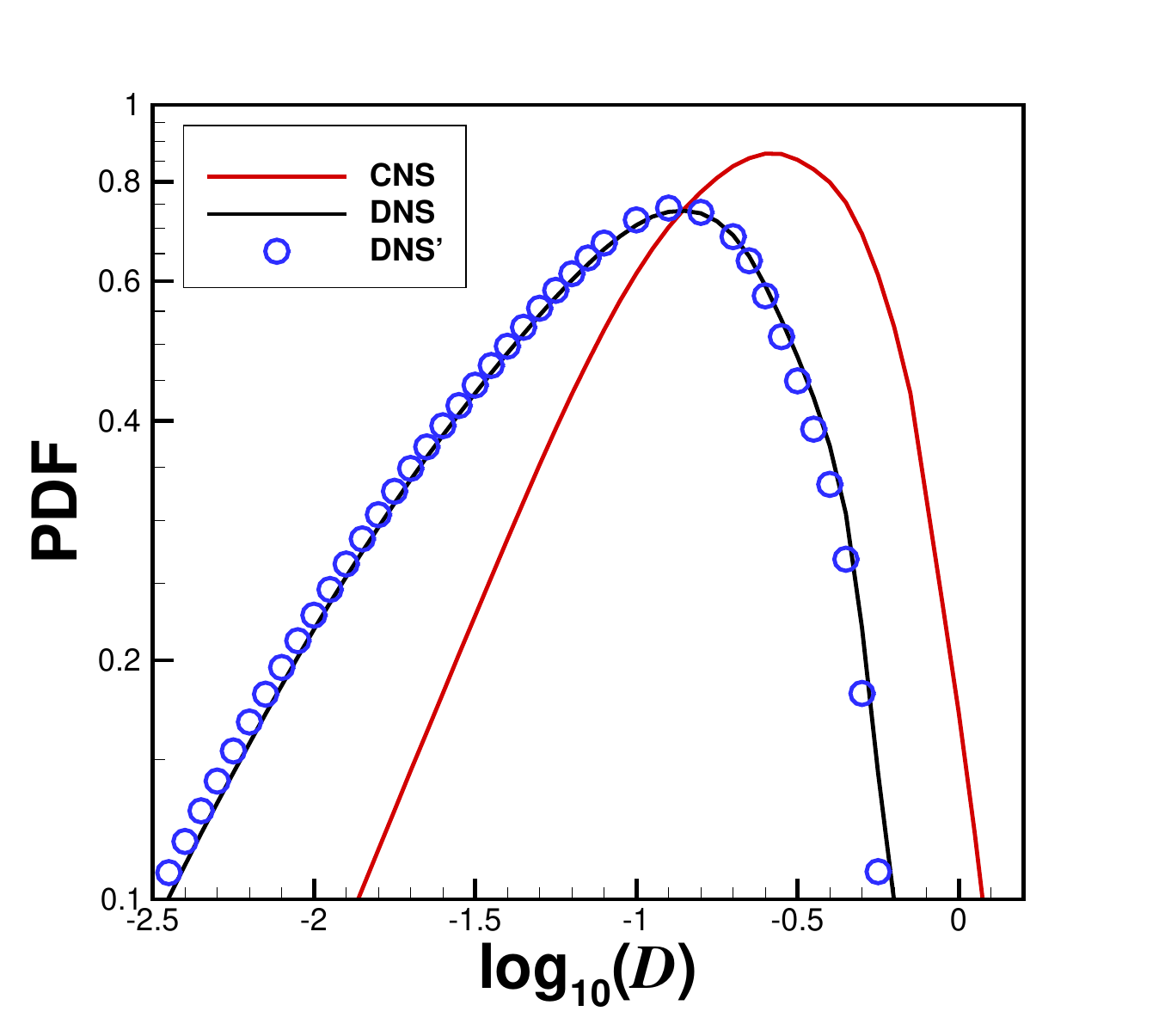}}
        \end{tabular}
    \caption{Comparisons of the probability density function (PDF) of (a) the kinetic energy $E(x,y,t)$ and (b) the kinetic energy dissipation rate $D(x,y,t)$ of the two-dimensional Kolmogorov turbulence  in the case of $n_K=4$ and $Re=40$ with the initial condition (\ref{initial_condition}), given by  the CNS (red line), the DNS (black line) with $\Delta t = 10^{-4}$ and the DNS (blue symbol) with $\Delta t = 2\times 10^{-4}$, respectively, where PDFs are integrated in $x,y\in[0,2\pi]$ and $t \in [200, 1500]$.}     \label{PDF_E_D_2}
    \end{center}
\end{figure}

In addition,  it is traditionally assumed  that DNS results are correct and reliable if they are obtained using different time-step and/or grid spacing but match well each other in statistics.  
However, it is found that the statistic results given by the DNS using different time-steps match perfectly with each other, as shown in Figure~\ref{PDF_E_D_2}.  But unfortunately, even such kind of perfect match still  {\em cannot} guarantee the validity of the DNS results:   they all quickly lose the spatial symmetry (\ref{symmetry}) and are quite different from those given by the CNS  benchmark solution even in statistics!    All of these challenge our traditional beliefs.  

\section{Concluding remarks and discussions}

Traditionally, results given by the direct numerical simulation (DNS) of Navier-Stokes equations are widely assumed as `reliable' benchmark solutions of turbulence, as long as grid spacing is fine enough (i.e. less than the minimum Kolmogorov scale) \cite{pope2001turbulent} and time-step is small enough, say, satisfying the Courant-Friedrichs-Lewy condition (Courant number $<$ 1)  \cite{Courant1928}.   Is this definitely true?        

To answer this important fundamental question, a two-dimensional  sustained turbulent Kolmogorov flow  driven by an external body force is solved by means of the two numerical methods with detailed comparisons:  one is the traditional `direct numerical simulation' (DNS), the other is the `clean numerical simulation' (CNS).  The results given by the DNS are a kind of mixture of the `false' numerical noise and the `true'  solution of the NS equations, which are mostly at the same order of magnitude, as shown in Figure~\ref{Vor_Evolutions}, Figure~\ref{Vor_Evolutions_More} and Figure~\ref{delta-w}.   On the contrary, the numerical noise of the result given by the CNS is much smaller than the `true' solution of the NS equations  in a finite but long enough interval of time, i.e. $t\in[0,1500]$, so that the CNS result is very close to the `true'  solution and thus  can be used as a benchmark solution of the NS equations for comparison.       
It is found that numerical noise as a kind of small-scale disturbances can lead to huge deviations in both of large and small scales on the two-dimensional Kolmogorov turbulent flow, not only  qualitatively but also  quantitatively (even in statistics), including the spatial symmetry of the flow field, the enstrophy spectrum as well as many statistics of physical quantities such as the velocity, vorticity, kinetic energy, kinetic energy dissipation rate and so on, as reported in \S3.1.   
Besides, the field geometrical structures are also sensitive to the background numerical noise as a kind of artificial tiny disturbance, as illustrated in \S~3.2.  In addition, it should be emphasized that the so-called intermittent stability of turbulence can be successfully found by the CNS, but is unfortunately lost by the DNS, as shown in \S~3.3.    
It is found  in \S~3.4 that the averaged scale-to-scale energy flux given by the DNS has a large deviation from that given by the CNS benchmark solution.    
Furthermore, it is illustrated in \S~3.5 that a fine enough spatial grid with a small enough time-step alone  {\em cannot} guarantee the correction/reliability and validity of the DNS: it is only a necessary condition but {\em not} sufficient.  This also provides us a theoretical evidence that the hypothesis $\langle p + \delta \rangle = \langle p \rangle$ does {\em not} stand up for the two-dimensional Kolmogorov turbulent flow under consideration, where $\langle \; \rangle$ denotes an operator of statistics, $p + \delta$ is the result given by the DNS that might be ``badly polluted'', $p$ is the `true' solution of the Navier-Stokes equations, respectively.  So,  unfortunately,  DNS results of sustained turbulent flows, especially those  having energy exchange with external systems, might have huge deviation  from the `true' solution of Navier-Stokes equations.   Hopefully, CNS as a new tool to investigate turbulent flows more accurately than DNS could bring us some new discoveries.    Of course, more investigations for various types of turbulent flows are needed in the future.

It should be emphasized that all results reported above is for the case of the Reynolds number  $Re=40$.  Similarly, we investigated the two-dimensional Kolmogorov turbulent flow under the same initial condition  (\ref{initial_condition}) in ten different Reynolds numbers $Re$ that are randomly chosen in the domain $Re\in[35,50]$, and obtained the similar results described above.  Thus, all of our conclusions mentioned above have  general meanings.  

In summary,  using a two-dimensional  sustained turbulent Kolmogorov flow  driven by an external body force as an example, we illustrated that a DNS result of the Navier-Stokes equations is {\em not} definitely correct/reliable even in statistics, even if the grid spacing and time-step are small enough!  
This finding might {\em challenge} some traditional beliefs and assumptions about the NS equations and turbulence, although they are worldwide accepted nowadays.    
Our finding highly suggests that one had better  check very carefully  DNS results of some complicated turbulent flows governed by the Navier-Stokes equations, especially those having energy exchange with external systems.    

Certainly, a lots of investigations should be done in future.  Below are some of our viewpoints for discussions.

First of all,  it should be emphasized that the CNS result has a kind of spatial symmetry (\ref{symmetry}), which comes from the spatial symmetry of the initial condition (\ref{initial_condition}) and the governing equation (\ref{eq_psi}).   As shown in Figs.~1 and 2, the flow field keeps such kind of spatial symmetry (\ref{symmetry}) for CNS in the {\em whole} interval of time $t\in[0,1500]$, but for the DNS only in a {\em short} interval of time from the beginning, i.e. $t\in[0,120)$, beyond which the random numerical noises of the DNS, which have {\em no} spatial symmetry at all,  have been enlarged to the same order of magnitude as the ``true'' solution of the NS equations and thus finally  destroy this kind of spatial symmetry  (\ref{symmetry}) of the flow field.  So, the loss of the spatial  symmetry  (\ref{symmetry})  of the DNS clearly indicates its large deviation from the ``true'' solution of the NS equations.  In mathematics this provides us a {\em rigorous} evidence that small disturbances of the NS equations might be increased to a macroscopic level, say, at the same level of magnitude as the macroscopic physical variables under considerations.  This confirms once again the previous conclusions reported by Lin, Wang \& Liao \cite{lin2017origin} and Qin \& Liao  \cite{qin_liao_2022} for a Rayleigh-B\'{e}nard convection turbulent flow.   However, this is  {\em contrary} to the general belief that small disturbances of turbulent flow should be dissipated by the viscosity of fluid: it is based on this general belief  that  NS equations generally do {\em not} contain the unavoidable tiny noises such as thermal fluctuations, environmental disturbances, and so on.   However, if NS equations are indeed an excellent  model for turbulence,  then our results given by the CNS reveals that its spatiotemporal trajectories are very sensitive to small disturbances, so that these small disturbances, no matter they are artificial or physical,  {\em must} be considered, say, should {\em not} be neglected.  Obviously, this leads to a logical {\em paradox}!   Therefore, it should be {\em wrong} in physics for NS equations to neglect unavoidable small disturbances! 

Note that nearly all DNS results of NS equations quickly become badly polluted.  This is obvious from the well-known fact that,  even for  exactly the same NS equations with the same boundary/initial conditions and the same physical parameters, all DNS results quickly become quite different in spatiotemporal trajectories, although their animations of flow field  look quite similar.  Here, the obvious difference of spatiotemporal trajectories  implies  the `badly' polluted, denoted by $p+\delta$.    However, if the statistics of the corresponding turbulent flows are {\em stable} in statistics, say, the hypothesis $\langle p  + \delta\rangle =  \langle p \rangle$ holds,  then the badly polluted DNS results $p+\delta$ agree well with the true solution in {\em statistics}!   This is the reason why many DNS results agree well with the experiments:  the key point is the {\em statistical stability}, which is more important than the accuracy of numerical simulations of DNS.

Secondly,   how can we know in general cases whether a DNS result of turbulent flow governed by the NS equations  is correct/reliable in statistics or not?  Although the two-dimensional  sustained turbulent Kolmogorov flow considered in this paper is just a simple example,  it highly suggests that the statistics stability, say,  $\langle p  + \delta\rangle =  \langle p \rangle$,  should be very important  for  turbulent  flows.  This is exactly the reason why Liao \cite{Liao2023book} proposed the so-called  `{\bf modified fourth Clay millennium problem}':   
  \begin{enumerate}
 \item[]  {\em ``The existence, smoothness and  {\bf statistic stability} of the Navier-Stokes equation: Can we prove the existence and smoothness of the solution of the Navier-Stokes equation  with  physically proper boundary and initial conditions, whose statistics are   stable (or unstable), i.e., insensitive (or  sensitive) to small disturbances? ''}   
   \end{enumerate}
Note that  the statistic stability is emphasized here, since unavoidable {\em physical} disturbances (such as thermal fluctuations)  always exist in practice.   Statistic stability of turbulent flows is also the first problem of ``Eight Key Open Questions in Ocean Engineering'' \cite{SKLOE2023} issued in 2023 by State Key Laboratory of Ocean Engineering, China.   The key point is to find out the necessary and sufficient condition of such kind of statistic stability.    From practical view-point of CFD, statistic stability of NS equations is more important than the existence and smoothness of its solution.  

Thirdly, it should be emphasized that, from physical point of view, the NS equations (as a model of turbulence)  neglect the influence of the unavoidable small {\em physical} disturbances such as thermal fluctuations and environmental noises.  Note that physical thermal fluctuation is often larger than the artificial  numerical noise of the DNS.  Thus, if artificial numerical noises could be regarded as physical thermal fluctuations (see Eckmann and Ruelle \cite{Eckmann1985RMP}), our results imply that small but unavoidable physical disturbances sometimes might have huge influences on the long-term statistics of turbulent flow and thus should {\em not} be neglected in physics.  Therefore,  from {\em physical} point of view, it should be better to use the Landau-Lifshitz-Navier-Stokes (LLNS) equations \cite{LLNS1959, Bandak2022PRE}, which consider the influence of unavoidable thermal fluctuations, instead of the NS equations, to model turbulent flows.                                
So, although the NS equations might be {\em physically} wrong since small unavoidable physical disturbances are {\em not} considered,  {\em artificial} numerical noises in computational fluid dynamics (CFD) practically play a role like {\em physical} thermal fluctuations \cite{Eckmann1985RMP}, which might ``correct'' ,  more or less,  this kind of {\em physical} mistakes of the NS equations.  Therefore, from physical viewpoint, artificial numerical noises might have {\em positive} meanings  for the NS equations, although they  have no mathematical meanings  at all.    It should be emphasized that the NS equations are only a mathematical model of turbulent flows in physics: they are quite different things.   Possibly,  the unavoidable but nasty numerical noises might be very important in correctly modelling turbulence.   It is a pity that the statistic stability and the influences of artificial numerical noises on modeling turbulence have been  neglected.

Finally, let us point out the differences and relationships between the CNS and DNS.   First of all, CNS contains an important concept, namely the ``critical predictable time'' $T_{c}$, within which numerical noise of numerical simulation (given by CNS) is much smaller than its true solution and thus is reliable (i.e. ``clean'') and can be used as a benchmark solution.  Secondly, the reliability of result in $t\in[0,T_{c}]$ given by CNS with a numerical noise ${s}_{1}$ can be verified by another result given by CNS with a smaller numerical noise $s_{2} < s_{1}$.    However, different from the CNS, DNS has {\em not} the concept  ``critical predictable time'' $T_{c}$ at all, since it is a traditional {\em belief} that small disturbances of NS equations should {\em not} enlarge to a macroscopic level.  In other words, $T_{c}$ of DNS  is {\em assumed} to be {\em infinite} from the viewpoint of CNS.  Unfortunately,  our results reported in this paper  indicate that this {\em assumption} is wrong!   This is a fundamental difference between CNS and DNS.   On the other side, the traditional DNS that uses data in the double precision and the 4th-order Runge-Kutta method in temporal dimension can be regarded as a special case of the CNS that adopts the multiple-precision and high-order Taylor expansion in temporal dimension.   Note that the DNS result reported in this paper loses the spatial symmetry (\ref{symmetry}) and thus deviates largely from its true solution when $t > 120$, mainly because its ``critical predictable time'' $T_c$ is only about 120, say, $T_c \approx 120$, which is too short for calculating statistics.   However, unlike DNS, the CNS can greatly enlarge the so-called  ``critical predictable time'' $T_c$ by reducing the numerical noises.   For  example, the ``critical predictable time'' $T_c$ of the CNS result reported in this paper is greater than 1500, say, $T_c > 1500$,  by means of the multiple precision with 200 significant digits for all physical/numerical variables and parameters and the 60th-order Taylor expansion with the time-step $\Delta t =10^{-3}$.   In this way, one can gain a `clean', reliable, benchmark solution of NS equations by means of the CNS.    Shortly speaking, DNS can be regarded as a {\em special} case of the CNS:  the CNS has much less numerical noises and thus much larger ``critical predictable time'' $T_c$ than the DNS. In other words, CNS is more general than DNS.    Indeed, for the very first time,  CNS provides us a way to rigorously verify  DNS results:  any a CNS result with a ``critical predictable time'' $T_{c}$ can be verified by another CNS result with even smaller numerical noises.  
Without doubt, CNS as a more {\em general} and more {\em precise} tool than DNS can provide us a  totally new way to  investigate  turbulent  flows.

\section*{Funding}
{This work is partly supported by National Natural Science Foundation of China (Grant No. 12302288; 12272230) and State Key Laboratory of Ocean Engineering.}

\section*{Declaration of Interests}
{The authors report no conflict of interest.}

\section*{Data availability statement}
{The data that support the findings of this study are available on request from the corresponding author.}

\section*{Author ORCID}
{Shijie QIN, https://orcid.org/0000-0002-0809-1766;  Yongxiang HUANG, https://orcid.org/0000-0002-2011-9215; Shijun LIAO, https://orcid.org/0000-0002-2372-9502}

\section*{Author contributions}
{Qin performed the analysis and wrote the draft.  Yang  performed the analysis.  Mei and Wang analysed the results especially related to the field geometrical structures.  Huang analysed the results, especially related to the scale-to-scale energy flux and the energy injection/extraction processes. Liao conceived/designed the analysis, and wrote the manuscript.  }

\appendix

\section{The CNS algorithm for two-dimensional Kolmogorov flow}

In 2020, in order to overcome the huge computation requirements of the CNS algorithm in spectral space,  an efficient CNS algorithm  \cite{lin2017origin,  hu2020risks, qin2020influence} was proposed in physical space for spatio-temporal chaos.  Recently, Qin and Liao \cite{qin_liao_2022} have successfully applied the CNS algorithm in physical space to a two-dimensional turbulent Rayleigh-B{\'e}nard convection.
Here, the basic idea of this kind of efficient CNS algorithm is briefly described by using the two-dimensional Kolmogorov flow as an example.

We rewrite the governing equation (\ref{eq_psi}), which is based on Navier-Stokes equations, as following:
\begin{eqnarray}
\frac{\partial}{\partial t}(\psi_{xx}+\psi_{yy})=\psi_{y}(\psi_{xxx}+\psi_{xyy})
-\psi_{x}(\psi_{xxy}+\psi_{yyy}) \nonumber \\
+\frac{1}{Re}(\psi_{xxxx}+2\psi_{xxyy}+\psi_{yyyy})-n_K\cos(n_Ky),       \label{psi_0}
\end{eqnarray}
where $x$ and $y$ as subscripts denote the spatial derivatives corresponding to horizontal and vertical directions, respectively.

The computational domain $x,y\in[0,2\pi]$ is discretized by means of $N_x\times N_y$ equidistant, i.e.
\begin{equation}
x_{j}=\frac{2\pi j}{N_x},\hspace{1.0cm} y_{k}=\frac{2\pi k}{N_y},    \label{points}
\end{equation}
where $j=0, \,1, \,2, \,..., \,N_x-1$ and $k=0, \,1, \,2, \,..., \,N_y-1$.

To reduce truncation errors in the dimension of time, the high-order Taylor expansions are adopted, i.e.
\begin{equation}
\psi(x_{j},y_{k},t+\Delta t)\approx\sum^{M}_{m=0}\psi^{[m]}(x_{j},y_{k},t)(\Delta t)^{m},  \label{Taylor_psi}
\end{equation}
where $\Delta t$ is the time-step, $M$ is the order of Taylor expansion, with the definition
\begin{equation}
\psi^{[m]}(x_{j},y_{k},t)=\frac{1}{m!}\frac{\partial^{m}\psi(x_{j},y_{k},t)}{\partial t^{m}}.      \label{Taylor_details}
\end{equation}
Here, the order $M$ should be large enough so as to reduce the truncation errors to a required tiny level.

Differentiating $(m-1)$ times both sides of (\ref{psi_0}) with respect to $t$ and then dividing them by $m!$, we obtain the governing equation of $\psi^{[m]}$:
\begin{eqnarray}
\psi^{[m]}_{xx}(x_{j},y_{k},t)+\psi^{[m]}_{yy}(x_{j},y_{k},t)
=\frac{1}{m} \Big\{F_m
+\frac{1}{Re} \big[ 2\psi^{[m-1]}_{xxyy}(x_{j},y_{k},t) \nonumber \\
+\psi^{[m-1]}_{xxxx}(x_{j},y_{k},t)+\psi^{[m-1]}_{yyyy}(x_{j},y_{k},t) \big] \nonumber \\
+\sum^{m-1}_{r=0}\psi^{[r]}_{y}(x_{j},y_{k},t) \big[ \psi^{[m-1-r]}_{xxx}(x_{j},y_{k},t)
+\psi^{[m-1-r]}_{xyy}(x_{j},y_{k},t) \big ] \nonumber \\
-\sum^{m-1}_{r=0}\psi^{[r]}_{x}(x_{j},y_{k},t) \big[ \psi^{[m-1-r]}_{xxy}(x_{j},y_{k},t)
+\psi^{[m-1-r]}_{yyy}(x_{j},y_{k},t) \big] \Big\},    \label{psi_m}
\end{eqnarray}
where $m\geq1$ and
\begin{equation}
F_m=\left\{
\begin{array}{l}
-n_K\cos(n_Ky_{k}),    \hspace{1.0cm}    m=1\\
0,    \hspace{3.0cm}    m>1.
\end{array}
\right.  \label{Fm}
\end{equation}

Note that there exist some spatial partial derivatives denoted by subscripts in (\ref{psi_m}), such as $\partial^{s_1+s_2}\psi^{[r]}/(\partial x^{s_1}\partial y^{s_2})$ with $r,\,s_1,\,s_2\geq0$. In order to approximate these spatial partial derivative terms with high computational efficiency and precision from the known discrete variable $\psi^{[r]}(x_{j},y_{k},t)$, we adopt the spatial Fourier series
\begin{equation}
\psi^{[r]}(x,y,t)\approx\sum^{\lfloor N_x/3 \rfloor}_{\,n_x=-\lfloor N_x/3 \rfloor}\sum^{\lfloor N_y/3 \rfloor}_{\,n_y=-\lfloor N_y/3 \rfloor}
\Psi^{[r]}(n_x,n_y,t)
\exp(\mathbf{i}\,n_xx)\exp(\mathbf{i}\,n_yy),   \label{Appro_psi}
\end{equation}
where $n_x$, $n_y$ are integers, $\lfloor\,\,\rfloor$ stands for a floor function, $\mathbf{i}=\sqrt{-1}$ represents the imaginary unit, for dealiasing $\Psi^{[r]}=0$ is imposed for wavenumbers outside the above domain $\sum$, and
\begin{equation}
\Psi^{[r]}(n_x,n_y,t)=\frac{1}{N_x\hspace{0.03cm}N_y}\sum^{N_x-1}_{j=0}\sum^{N_y-1}_{k=0}\psi^{[r]}(x_{j},y_{k},t)
\exp(-\mathbf{i}\,n_xx_{j})\exp(-\mathbf{i}\,n_yy_{k}),   \label{Coe_psi}
\end{equation}
is determined by the known $\psi^{[r]}(x_{j},y_{k},t)$. Then, we can obtain the rather accurate approximations of the spatial partial derivative terms in (\ref{psi_m}), say,
\begin{eqnarray}
&& \frac{\partial^{s_1+s_2}\psi^{[r]}}{\partial x^{s_1}\partial y^{s_2}}(x_{j},y_{k},t)\nonumber\\
&\approx &  \mathbf{i}^{s_1+s_2} \sum^{\lfloor N_x/3 \rfloor}_{\,n_x=-\lfloor N_x/3 \rfloor}\sum^{\lfloor N_y/3 \rfloor}_{\,n_y=-\lfloor N_y/3 \rfloor} 
(n_x)^{s_1}(n_y)^{s_2} \hspace{0.03cm}
\Psi^{[r]}(n_x,n_y,t)
\exp(\mathbf{i}\,n_xx_{j})\exp(\mathbf{i}\,n_yy_{k}).        \label{Appro_psi_xy}
\end{eqnarray}
Here, the fast Fourier transform (FFT) algorithm is used to increase computational efficiency. In this way, the spatial truncation error can be decreased to a required tiny level, as long as the mode numbers $N_x$ and $N_y$ are large enough.

Note that, if the order $M$ of the Taylor expansion (\ref{Taylor_psi}) is large enough, temporal truncation errors can be reduced to a required tiny level. Besides, if the spatial discretization is fine enough, say, the mode numbers $N_x$ and $N_y$ are large enough, spatial truncation errors in Fourier expression (\ref{Appro_psi}) and the corresponding spatial derivative terms in (\ref{psi_m}) can be decreased to a required tiny level. However, this is not enough, since there always exist some round-off errors that might be larger than temporal or spatial truncation errors. Therefore, all physical/numerical variables and parameters are expressed in multiple precision (MP) with a large enough number $N_s$ of significant digits, so that the round-off errors can also be reduced to a required tiny level. In this way, the background numerical noise, i.e. truncation errors and round-off errors as a whole, can be reduced to a required tiny level. This is different from most of the other numerical algorithms, including the general DNS which has double precision at best.
In addition, note that the results given by the CNS are useful only in a limited (but long enough) interval of time $t\in[0,T_{c}]$, in which the numerical noise can be negligible.


\bibliography{kolmogorov} 

\end{document}